\documentclass[12pt,english]{article}
\pdfoutput=1
\usepackage[T1]{fontenc}
\usepackage[utf8]{inputenc}
\usepackage{subfigure,lmodern, amsmath,amssymb, graphicx, pifont, adjustbox, bm, xcolor}
\usepackage{amsfonts}
\usepackage{geometry}
\usepackage{comment}
\usepackage{mathtools}
\usepackage{float}
\usepackage{slashed}
\usepackage{cite}
\usepackage{ragged2e}
\usepackage{etoolbox}
\usepackage{array}

\apptocmd{\thebibliography}{\justifying\setlength{\leftskip}{7.4mm}}{}{}
\makeatletter\g@addto@macro\bfseries{\boldmath}\makeatother

\usepackage{stackengine}
\usepackage{esint}
\usepackage[unicode=true,pdfusetitle,
 bookmarks=true,bookmarksnumbered=false,bookmarksopen=false,
 breaklinks=false,pdfborder={0 0 1},backref=false,colorlinks=true]
 {hyperref}
\hypersetup{pdfauthor={Nima Arkani-Hamed, Carolina Figueiredo, and Grant N. Remmen},
 citecolor=black,linkcolor=black,urlcolor=black}

\newcommand{\appendixref}[1]{\hyperref[#1]{appendix~\ref{#1}}}
\def\equationautorefname~#1\null{eq.\,(#1)\null}
\usepackage{breakurl}
\usepackage[hang,flushmargin]{footmisc} 
\allowdisplaybreaks

\usepackage{changepage}

\newcommand{\be}{\begin{equation}}
\newcommand{\ee}{\end{equation}}
\newcommand{\bea}{\begin{eqnarray}}
\newcommand{\eea}{\end{eqnarray}}

\newcommand{\eq}[2]{\be\begin{aligned}#1 \label{#2}\end{aligned}\ee}

\newcommand{\Fig}[1]{Fig.~\ref{#1}}

\newcommand{\Eq}[1]{Eq.~(\ref{#1})}

\newcommand*\diff{\mathop{}\!\mathrm{d}}

\newcolumntype{P}[1]{>{\centering\arraybackslash}p{#1}}

\begin{document}

\interfootnotelinepenalty=10000
\baselineskip=18pt
\hfill 
\hfill
\vspace{1cm}
\thispagestyle{empty}
\begin{center}
{\LARGE \bf
Open String Amplitudes:\\[1.2mm] \Large Singularities, 
 Asymptotics, and New Representations
}\\
\bigskip
\begin{center}{Nima Arkani-Hamed,${}^{a}$ Carolina Figueiredo,${}^{b}$ and Grant N. Remmen${}^{c}$}\end{center}
{
\it  ${}^a$School of Natural Sciences, Institute for Advanced Study, Princeton, NJ, 08540 \\[1.5mm]
${}^b$Department of Physics, Princeton University, Princeton, NJ 08544\\[1.5mm]
${}^c$Center for Cosmology and Particle Physics, Department of Physics,\\[-1mm] New York University, New York, NY 10003}
\let\thefootnote\relax\footnote{e-mail: \\ 
\url{arkani@ias.edu}, 
\url{cfigueiredo@princeton.edu},
\url{grant.remmen@nyu.edu}}
\end{center}

\bigskip
\centerline{\large\bf Abstract}
\begin{quote} \small
Open string amplitudes at tree level have been studied for over fifty years. However, there is no known analytic form for general $n$-point amplitudes, and their conventional representation in terms of worldsheet integrals does not make many of their most basic physical properties manifest. Recently, a  formulation of these amplitudes exposing the underlying ``binary geometry'' via the use of ``$u$'' variables has given us many insights into their basic features. In this paper, we initiate a systematic exploration of fundamental aspects of open string amplitudes from this new point of view. We begin by finding explicit expressions for the factorization of amplitudes at general massive levels, which are seen to be determined by products of lower-point massless amplitudes with shifted kinematics. We then study the asymptotic behavior when subsets of kinematic variables become large, delineating regimes with exponential (generalized hard scattering) and power-law (generalized Regge) behavior. We also give precise expressions for the asymptotics, which reveal another example of the recently observed property of factorization away from poles. We derive new recursion relations for the amplitude, which when repeatedly applied reduce to infinite series representations with a wider domain of convergence than the usual integral representations. For the five-point case, we present a new closed-form expression for the amplitude that for the first time gives its analytic continuation to all of kinematic space. We also discuss novel relations between amplitudes at different kinematic points following from the recently observed ``split'' factorizations. 

\end{quote}
	
\setcounter{footnote}{0}

\setcounter{tocdepth}{2}
\newpage
\tableofcontents

\newpage

\section{What Are String Amplitudes?}

Before string theory was a theory of strings~\cite{Nambu, Nielsen, Susskind}, it was a model for the S-matrix in flat space~\cite{Veneziano}.  Famously, the original inspiration for this subject had little to do with  the top-down pursuit of a quantum theory of gravity.  Rather, string amplitudes were discovered entirely inadvertently, in the process of a bottom-up search for functions exhibiting remarkable physical properties like dual resonance and softened high-energy behavior.  

By rights, we should then expect the string theory S-matrix to be the most exceptional of all possible S-matrices.
It is therefore ironic that our knowledge of what string amplitudes {\it actually are} remains frustratingly incomplete, even at tree level. This surprising and much underappreciated fact is already evident in the most familiar equation in string theory,
\begin{equation}
\mathcal{A}_4 = \int_0^1 \frac{\diff z}{z(1 - z)} z^{s} (1 - z)^{t}, \label{eq:Veneziano_worldsheet}
\end{equation}
which is the worldsheet representation of the Veneziano amplitude~\cite{Veneziano} encoding the tree-level scattering of four open strings (where we have set $\alpha^\prime\equiv 1$).
This expression only converges for $s,t>0$, which does not cover the physically relevant case of Lorentzian kinematics, corresponding to $s<0$ or $t<0$ in our sign conventions.\footnote{We work in mostly-plus signature, $(-\, + \, \cdots + \,)$, so that $p_i^2 = -m^2$, and define our Mandelstam invariants by $s = (p_1 + p_2)^2$ and $t = (p_1 + p_4)^2$, so physical $s$-channel scattering corresponds to $s<0$ and $t>0$, and poles for non-tachyonic states occur at negative values of $s$ and $t$.}  Notably, the physical singularities of the amplitude, which arise from the exchange of massive string resonances, all reside precisely in this non-convergent regime.  This simple fact highlights the yawning gap between what we can formally write down and what we can actually evaluate.

Fortunately, in the case of four-point open string scattering we can analytically continue \Eq{eq:Veneziano_worldsheet} to an expression for the amplitude that is evaluable for any $s$ and $t$, 
\begin{equation}
{\cal A}_4 = \frac{\Gamma(s) \Gamma(t)}{\Gamma(s + t)}.\label{eq:Euler}
\end{equation}

This formula is remarkably useful: it is analytic except at the poles corresponding to resonant exchanges, which occur when $s$ or $t$ is a negative integer.  Even better, it manifests the hidden zeroes of the amplitude~\cite{Zeros} when $s + t$ is a negative integer. 

This same basic deficiency of \Eq{eq:Veneziano_worldsheet} is an affliction of all $n$-point scattering, where the tree-level amplitudes for the open string are given by the Koba-Nielsen (KN) formula~\cite{Koba:1969rw},
\begin{equation}
\mathcal{A}_n(1,2,\cdots,n) = \int\limits_{z_1<z_2\cdots< z_n} \frac{\diff^n z}{{\rm SL(2,\mathbb{R})}} \frac{\prod_{i<j} (z_j - z_i)^{-2 p_i \cdot p_j}}{(z_1 -z_2)\cdots(z_n - z_1)}.\label{eq:KN}
\end{equation}

This expression is stunningly compact, but it is only well defined and convergent for unphysical kinematics far from the most interesting regions where massive string resonances can actually be produced. 

The implications of this shortcoming are far-reaching.
Granted the wish of a string-scale collider, an experimentalist who dutifully measures the cross section for $2 \to 4$ string scattering at center-of-mass energy $\sim 10\, M_{{\rm string}}$ will be sorely disappointed.  There is simply no theoretical prediction for this process, since the KN formula cannot be evaluated for physical kinematics, even numerically.  The worldsheet integrals are horribly divergent, which is especially ironic for a theory celebrated for having the softest possible behavior for high-energy amplitudes! 
As in the case of four-point scattering, our only recourse at $n$ point is to attempt to analytically continue the KN formula to physical kinematics.  However, no such construction exists for general $n$-point scattering. Even the known closed-form expression for the scattering of five strings in terms of a ${}_3 F_2$ generalized hypergeometric function~\cite{Bialas:1969jz} does not converge for all values of the Mandelstam invariants.\footnote{A similar issue should be expected for the form of the six-point amplitude written in terms of Srivastava's generalized triple hypergeometric function~\cite{Oprisa}.}

Recently, a novel formulation of string amplitudes has been actively explored, introducing a number of new ideas that hold the promise of  changing this situation. Instead of the conventional worldhseet expressions, this formulation exposes the underlying  binary geometry~\cite{BinGeom,CountProblem,Arkani-Hamed:2019mrd,Gluons} of the dynamics using so-called $u$ variables. While the $u$ variables have been known at tree level since early days~\cite{BardakciRuegg,ChanTsou,Koba:1969rw,David}, they receded from use in the ensuing decades, but have enjoyed a recent resurgence, as their conceptual underpinning has been properly understood. This has allowed an explicit ``positive parametrization'' of the $u$ variables in terms of ``$y$'' variables,  not only at tree level but to all orders in the genus expansion~\cite{BinGeom,CountProblem,Arkani-Hamed:2019mrd,Gluons,MultPartFact,Zeros,Splits}. This makes it possible to manifest all the singularities of the amplitudes without the need for a manual ``blow up'' of singular regions on the worldsheet. In this work, we use this framework to derive several new results for $n$-point tree-level open string scattering. 
After reviewing the representation of string amplitudes in terms of $u$ variables in Sec.~\ref{sec:uvariables}, in Sec.~\ref{sec:massive} we construct explicit expressions for the factorization of $n$-point string amplitudes at general {\it massive} levels, which remarkably can be quite elegantly written in terms of products of lower-point {\it massless} amplitudes evaluated at {\it shifted kinematics}. Simple properties of the $u$ variables also help us determine the asymptotic behavior of the amplitude when subsets of kinematic variables become large, determining regimes with exponential (generalized hard scattering) and power-law (generalized Regge) behavior. This will lead us to simple, precise expressions for the amplitudes in these asymptotic regions, which we give in Sec.~\ref{sec:asymptotic}.

The knowledge of $u$ variables, residues on poles, and asymptotics also allows us to find a number of new series representations of the amplitudes in Sec.~\ref{sec:recursive}. These new representations converge in a wider range of kinematics than the one in which the integral representation is convergent. In particular, these can be specialized to give a dual resonant representation of the amplitude expressed as an infinite sum over residues. 

Finally, for the case of five-point tree-level scattering, in Sec.~\ref{sec:hypergeometric}  we present a new closed-form expression for the amplitude that---for the first time---coverges in all kinematic regimes.  This representation exploits certain interesting identities satisfied for the $_3F_2$ hypergeometric function. We give a conceptual explanation for some of these identities in terms of the ``split'' factorizations of Ref.~\cite{Splits} and indicate their extension to higher multiplicity. 

In Sec.~\ref{sec:numerics}, we present some numerical checks of the formulas derived in the text for both the asymptotic behavior and the new series representations of string amplitudes. 

The arsenal of new tools developed in this paper for evaluating string amplitudes and their physical properties opens up many avenues for future work, ranging from their application beyond tree level~\cite{Eberhardt:2023xck,Eberhardt:2022zay,Gross:1987ar,Gross:1987kza,Mende:1989wt}, to statistical properties of $n$-point string scattering in the large-$n$ limit and its relation to black holes~\cite{Addazi:2016ksu,Dvali:2014ila}, to tests of stringy deformations and the quest to prove that string theory is unique~\cite{MultPartFact,Cheung:2023adk,Cheung:2023uwn,David,Coon,Haring:2023zwu,Caron-Huot:2016icg,Cheung:2024uhn,Cheung:2024obl}.
We summarize and discuss future directions in Sec.~\ref{sec:discussion}.

\section{String Amplitudes, $u$ Variables and ${\cal F}$ Polynomials}\label{sec:uvariables}
The central object of study in this paper is the $n$-point tree-level KN factor in \Eq{eq:KN}, which we will hereafter refer to as the ``string amplitude.'' Given the ordering of the external states, we will employ the planar Mandelstam invariants,
\be
X_{i,j}= (p_i + p_{i+1} + \cdots +p_{j-1})^2.
\ee

\begin{figure}[t]
    \centering
    \includegraphics[width=\linewidth]{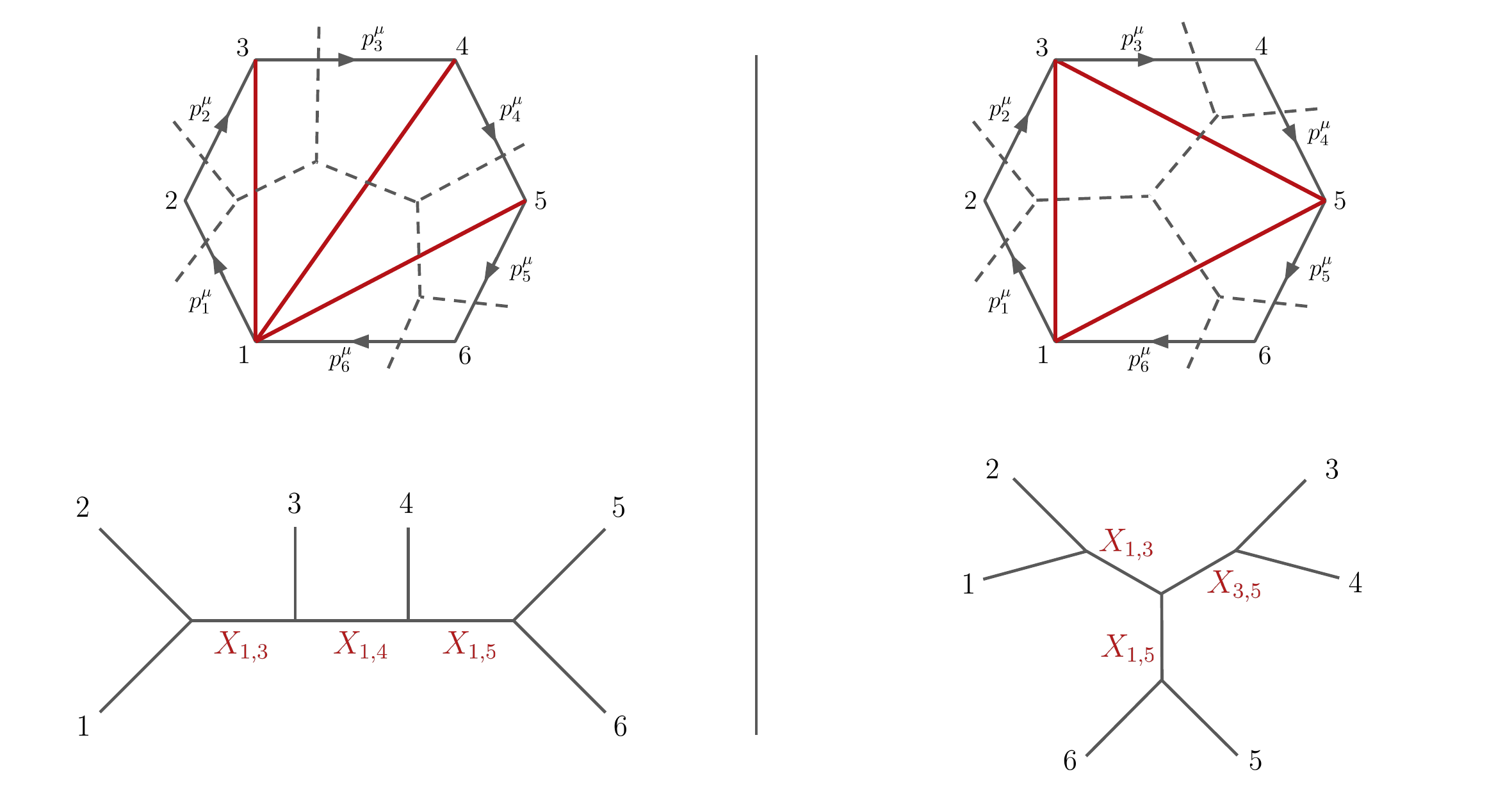}
    \caption{(Left)~Ray-like triangulation for the momentum polygon at $n=6$, with $i^\star =1$. The dual diagram is the half-ladder containing propagators $X_{1,j}$ with $j=3,4,5$. (Right)~ Mercedes-Benz triangulation and respective dual diagram.}
    \label{fig:RayTriang}
\end{figure}

Each factorization channel corresponds to the limit of vanishing $X_{i,j}\rightarrow 0$.  Drawing the momentum of each ordered external state head to tail, we obtain a closed \textit{momentum polygon}, where $X_{i,j}$ is the length squared of the chord $(i,j)$ extending  from $i$ to $j$ (see Fig.~\ref{fig:RayTriang}, top).  To manifest the dependence of $\mathcal{A}_n$ on the planar Mandelstam invariants, we rewrite \Eq{eq:KN} in terms of the  SL(2,$\mathbb{R}$) invariant cross-ratios $u_{i,j} = (z_{i-1,j} z_{i,j-1})/(z_{i,j}z_{i-1,j-1}) $, yielding
\be 
\mathcal{A}_n = \int \frac{d^{n} z}{{\rm SL}(2,\mathbb{R})} \frac{\prod_{i<j} u_{i,j}^{X_{i,j}}}{z_{1,2}\cdots z_{n,1}} 
. \label{eq:Au}
\ee

The  $u$ variables satisfy the set of nonlinear equations known as the \textit{$u$ equations} \cite{ABHY,BinGeom,AllLoop,CountProblem},
\eq{
u_{i,j} + \prod_{(i^\prime,j^\prime) } u_{i^\prime,j^\prime}^{\#\text{int}[(i,j),(i^\prime,j^\prime)]} = 1,
}{u_eq}
where $\#\text{int}[(i,j),(i^\prime,j^\prime)]=1$ or $0$ depending on whether or not the chord $(i^\prime,j^\prime)$ crosses the chord $(i,j)$. Remarkably, the solution space of these equations is $(n-3)$-dimensional and can be parameterized by the moduli $z_i$ of the gauge-fixed worldsheet.  

For the present analysis, we utilize the \textit{positive parameterization} of the $u$ variables in terms of so-called $y$ variables~\cite{BinGeom,CountProblem,Arkani-Hamed:2019mrd,Gluons,MultPartFact,Zeros,Splits},
\eq{
\mathcal{A}_n= \int_0^\infty \prod_{\mathcal{C}\in\mathcal{T}} \frac{dy_{\mathcal{C}}}{y_{\mathcal{C}}} \prod_{i<j} u_{i,j}^{X_{i,j}}[y_\mathcal{C}],
}{string_can_form}

This ``stringy canonical form'' has been crucial for illuminating many hidden properties of string amplitudes.  To start, we choose some base triangulation $\mathcal{T}$ of the momentum polygon.\footnote{Of course, the integral we obtain does not depend on the choice of underlying triangulation; however, the location of the singularities in the boundary of the integration domain will depend on the choice of $\mathcal{T}$ (see Ref.~\cite{AllLoop}).}  Then, for each chord $\mathcal{C}$ that appears in $\mathcal{T}$, we associate a variable $y_{\mathcal{C}}$. The explicit parameterization $u_{i,j}[y_{\mathcal{C}}]$ can be computed mechanically in the simple way described in Refs.~\cite{CountProblem,Gluons,Splits}.  Here we review some of the important features of this parameterization that will be relevant for our analysis. The $u_{i,j}[y_{\mathcal{C}}]$ are given by ratios of polynomials---the $\mathcal{F}$ polynomials, $\mathcal{F}_{i,j}(y_{\mathcal{C}})$---in the $y_{\mathcal{C}}$ variables, such that the integral can be written as 
\be 
\mathcal{A}_n = \int_0^\infty \prod_{\mathcal{C}\in\mathcal{T}} \frac{dy_{\mathcal{C}}}{y_{\mathcal{C}}}  y_{\mathcal{C}}^{X_{\mathcal{C}}} \prod_{i<j} \mathcal{F}_{i,j}[y_\mathcal{C}]^{-c_{i,j}},
\ee
where $X_\mathcal{C}$ is the Mandelstam variable associated to chord $\mathcal{C}$ entering the base triangulation, and $c_{i,j} = -2 p_i \cdot p_j$ with $i$ and $j$ non-adjacent. In the next subsection, we provide the explicit form of $\mathcal{F}_{i,j}[y_\mathcal{C}]$, and from it one can extract $u_{i,j}[y_\mathcal{C}]$ by collecting all the terms raised to the power $X_{i,j}$. On a given factorization channel $X_{i,j}\to0$, the integral develops a logarithmic singularity in the region where $u_{i,j}\to 0$, which in terms of the $y$ variables corresponds to a given region in the boundary of the integration domain. In particular, for the chords $\mathcal{C}\in\mathcal{T}$, the singularities $X_{\mathcal{C}} \to 0$ are located in the region $y_{\mathcal{C}} \to 0$, corresponding to the region where $u_{\mathcal{C}} \to 0$. 

An important implication of the $u$ equations in \Eq{u_eq} is that sending $u_{i,j}\to 0$ necessarily sends $u_{k,m}\to 1$ for all chords $(k,m)$ that cross $(i,j)$.  This is referred to as the \textit{binary} behavior of the $u$ variables. This means that in the amplitude in \Eq{string_can_form}, when we set $X_{i,j} \to 0$, the integral develops a singularity in the region $u_{i,j}\to 0$, and by the $u$ equations the contribution to the string amplitude from curves crossing $(i,j)$ goes to $1$.   We are thus left with the product of the chords that reside in the two lower-point polygons obtained by cutting the original polygon along the chord $(i,j)$.  Consequently, the amplitude factorizes into the product of two lower-point amplitudes.
For most of the text we will choose the base triangulation to be \textit{ray-like}, which is a triangulation containing chords $(i^\star,j)$ with $j\in \{i^\star+1,\cdots,i^\star-1\}$ (see Fig.~\ref{fig:RayTriang} (left) for the $n=6$ example with $i^\star=1$). This triangulation corresponds to a Feynman diagram in the {\it half-ladder} topology. Let us now present concrete examples what of what the $y$ parameterization looks like in these cases. 

\subsection{$y$ representation for ray-like base triangulations}

The ray-like triangulations are particularly nice because the ${\cal F}$ polynomials that appear have a very predictable structure. To understand how this works, let us consider some explicit examples. At four point, taking the base triangulation containing chord $(1,3)$---therefore propagator $X_{1,3}$---we obtain
\begin{equation}
	\mathcal{A}_4
	 = \int_0^\infty \frac{\diff y_{1,3}}{y_{1,3}} \, \, y_{1,3}^{X_{1,3}}(1+y_{1,3})^{-c_{1,3}} .
	\label{eq:4pt}
\end{equation}

For five-point scattering, for the base triangulation with propagators $\{ X_{1,3},X_{1,4}\}$, we find
\begin{equation}
\begin{aligned}
	\mathcal{A}_5
	= \int_0^\infty \frac{\diff y_{1,3} \diff y_{1,4}}{y_{1,3}y_{1,4}} \, \, y_{1,3}^{X_{1,3}} y_{1,4}^{X_{1,4}} \, \times (1+&y_{1,3})^{-c_{1,3}} (1+y_{1,4})^{-c_{2,4}} (1+y_{1,4}(1+y_{1,3}))^{-c_{1,4}}.
\end{aligned}
\label{eq:5pt}
\end{equation}

Similarly, for six-point scattering with propagators $\{ X_{1,3}, X_{1,4},X_{1,5}\}$, we have
\begin{equation}
\begin{aligned}
\mathcal{A}_6 &= \int_0^\infty \frac{\diff y_{1,3} \diff y_{1,4} \diff y_{1,5}}{y_{1,3}y_{1,4}y_{1,5}} \, \, y_{1,3}^{X_{1,3}} y_{1,4}^{X_{1,4}}  y_{1,5}^{X_{1,5}}\, \times  (1+y_{1,3})^{-c_{1,3}} (1+y_{1,4})^{-c_{2,4}} (1+y_{1,5})^{-c_{3,5}}\\
	&\;\;\;   \times (1+y_{1,4}(1+y_{1,3}))^{-c_{1,4}} (1+y_{1,5}(1+y_{1,4}))^{-c_{2,5}}   (1+y_{1,5}(1+y_{1,4}(1+y_{1,3})))^{-c_{1,5}}.
\end{aligned}
\label{eq:6pt}
\end{equation}
\begin{figure}[t]
    \centering
\includegraphics[width=\textwidth]{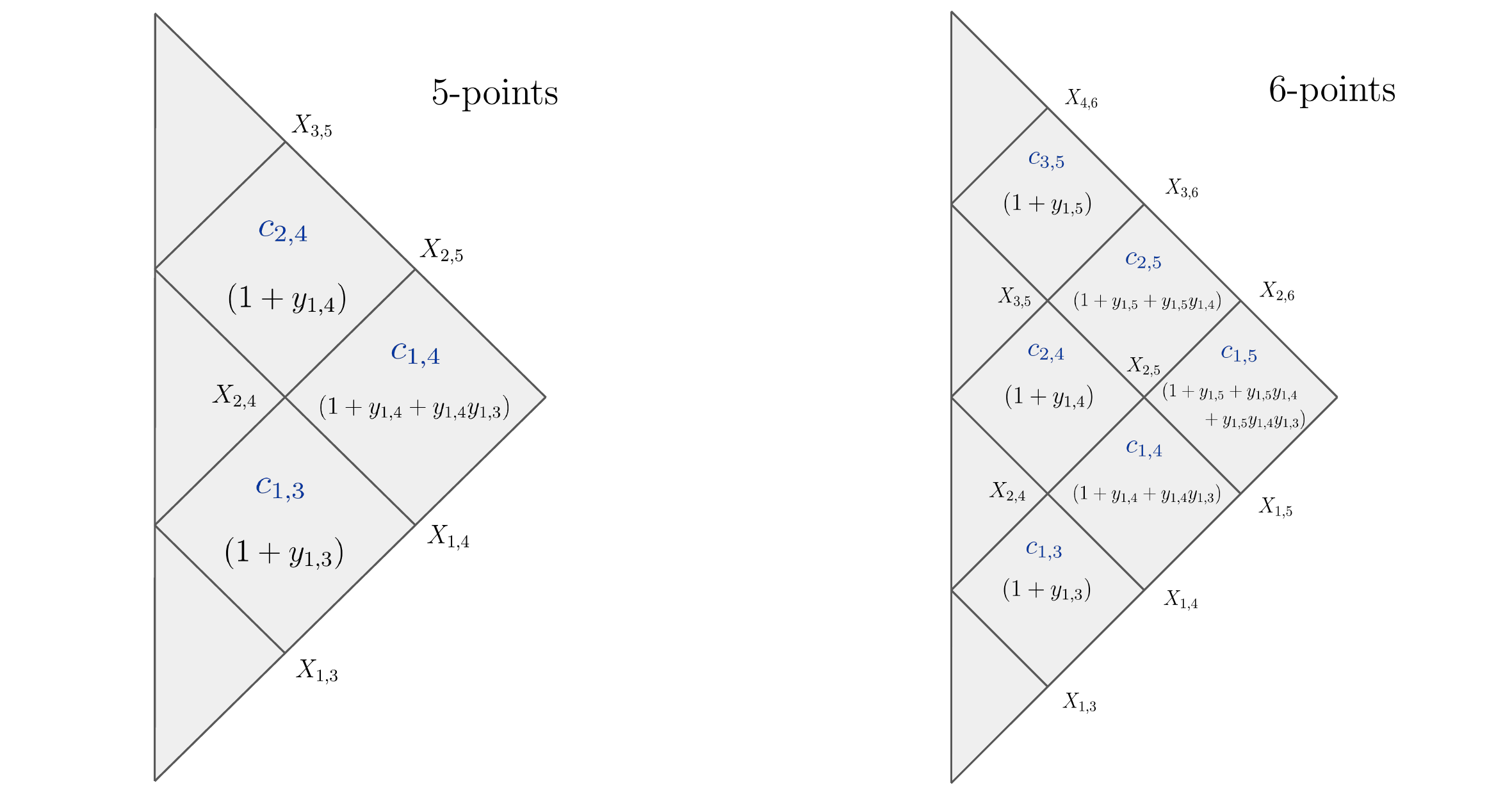}
    \caption{The kinematic mesh for the ray-like triangulations at five and six points. The points in the mesh are associated with the Mandelstam invariants $X_{i,j}$, and the diamonds to the $c_{i,j} = X_{i,j} + X_{i+1,j+1}-X_{i,j+1}-X_{i+1,j}$. Inside each diamond, we represent the respective ${\cal F}$ polynomial $\mathcal{F}_{i,j}$.}
    \label{fig:6ptFPol}
\end{figure}

We can see by eye that the $\mathcal{F}$ polynomials that appear have a nested structure.  There is a particularly nice way of organizing this structure of the polynomials appearing in the integrand using the {\it kinematic mesh}; see App~\ref{app:mesh} for details.  For example, let us consider the five and six-point amplitudes with triangulations $\{X_{1,3},X_{1,4}\}$ and $\{X_{1,3},X_{1,4},X_{1,5}\}$, respectively, which lead to integrands in Eqs.~\eqref{eq:5pt} and \eqref{eq:6pt}.  The corresponding kinematic meshes are presented in \Fig{fig:6ptFPol}. We now write each polynomial inside the mesh corresponding to the power $-c_{i,j}$ appearing in the string amplitude. We can observe that the polynomials corresponding to the meshes on the left are the simpler ones $(1+y_P)$---where $P$ denotes the chords entering the ray-like triangulation---and as we move towards the right they follow a predictable nested structure, 
\be 
{\cal F}_{i,j} = 1+\sum_{l=i+2}^j \prod_{k=l}^j y_{1,k},\label{eq:FF}
\ee 
for $i\leq j$. This particular organization will be very helpful in providing a systematic way of solving for $u$ variables in terms of those in the ray-like triangulations.

There is of course a  simple way of getting ${\cal F}$ polynomials for an {\it arbitrary} choice of ``base triangulation,'' given by the ``surfaceology'' formalism of Ref.~\cite{CountProblem} (also summarized in Refs.~\cite{Gluons,Splits,TropLag}), that we quickly review here.  We first draw our base triangulation via the dual fat graph, and we associate variables $y_e$ with all the edges of the fat graph. Any chord $(i,j)$ is associated with a curve on the fat graph that begins on the boundary $(i, i+1)$ and ends on the boundary $(j, j+1)$. This curve is associated with a word that begins with the starting boundary and simply records the left/right turns and roads encountered until the curve exits at the ending boundary. In order to compute the ${\cal F}$ polynomials, we ``trim'' this word, first by deleting any string that starts from the beginning boundary and turns right (continuously), or that turns left (continuously) into the ending boundary. Then we delete the boundary roads that may be left after this operation. This leaves us with a word of the form 
\begin{equation}
y_1 
\, \xrightarrow{{\rm turn}_1}\, y_2 \,\xrightarrow{{\rm turn}_2} \,\cdots \, \xrightarrow{{\rm turn}_{r-1}}\,y_r,
\label{eq:word}
\end{equation}
where ${\rm turn}_i$ can be either $L$ or $R$.

There is a simple counting problem associated with this word described in Ref.~\cite{CountProblem} that gives us the ${\cal F}$ polynomial, but here we simply summarize the expression. We first define $2 \times 2$ matrices $M_L(y), M_R(y)$ via
\begin{equation}
M_L(y)=\left(\begin{array}{c c} y & y \\ 0 & 1 \end{array} \right), \quad  M_R(y)=\left(\begin{array}{c c} y & 0 \\ 1 & 1 \end{array} \right).
\end{equation}

Then, for any word $W$ like the one in Eq.~\eqref{eq:word}, we define a polynomial 
\begin{equation}
F_W = (1, 1) \cdot M_{{\rm turn}_1}(y_1) M_{{\rm turn}_2}(y_2) \cdots M_{{\rm turn}_{r-1}}(y_r) \cdot \left( \begin{array}{c} 1 \\
 0 \end{array} \right).
\end{equation}
Finally, the ${\cal F}$-polynomial associated a given $c_{i,j}$ is given by the $F_{W}$ for the word associated to chord $(i+1,j+1)$, this is 
\begin{equation}
{\cal F}_{i,j} = F_{W_{i+1,j+1}}.
\end{equation}

We note that the ${\cal F}_{i,j}$ polynomials for those $(i,j)$ such that $(i+1,j+1)$ is a chord in the base triangulation are empty and hence those ${\cal F}$ polynomials are just equal to $1$. These are exactly the $c_{i,j}$ that are not included for this triangulation. 

Let us illustrate these rules for the ray-like triangulation at $n=5$. Consider ${\cal F}_{1,4}$. For this case, we have to compute the ${\cal F}$ polynomial for the curve $(2,5)$. The word for this curve is $$(23) \xrightarrow{L} (13) \xrightarrow{R} (14) \xrightarrow{R} (15).$$ 
In this case there is no ``trimming'', since nothing turns right out of (23) or turns left into $(15)$. We still delete the boundary roads (that is, roads $(23)$ and $(15)$) and are left with the word $$(13) \xrightarrow{R} (14).$$ 

If we look instead at ${\cal F}_{1,3}$, we look at the word for $(2,4)$, which is $$(23)\xrightarrow{L}(13)\xrightarrow{R}(14)\xrightarrow{L}(45).$$
This time, in trimming we remove everything going left into the boundary $(45)$, which deletes $(14)$. We also delete the boundary $(23)$, so we are left just with the word $(13)$. 
Then we compute the ${\cal F}$ polynomials as 
\begin{equation}
\begin{aligned}
&{\cal F}^{{\rm ray}}_{1,4} = (1,1) \cdot M_R(y_{1,3}) M_R(y_{1,4}) \cdot \left(\begin{array}{c} 1 \\ 0 \end{array} \right) = 1 + y_{1,4} + y_{1,4} y_{1,3}, \nonumber \\ 
&{\cal F}^{{\rm ray}}_{1,3} = (1,1) \cdot M_R(y_{1,3}) \cdot \left(\begin{array}{c} 1 \\ 0 \end{array} \right) = 1 + y_{1,3}.
\end{aligned}
\end{equation}

As another example, consider the ``Mercedes-Benz'' triangulation (see Fig.~\ref{fig:RayTriang}, right). For ${\cal F}_{1,3}$ we again look at the word for curve $(2,4)$, which is $(23)\xrightarrow{L}(13)\xrightarrow{L}(35)\xrightarrow{R}(45)$. There is no trimming, but we remove the boundaries to get the word $(13)\xrightarrow{L}(35)$. For ${\cal F}_{1,4}$ we look at the word for $(2,5)$ which is $(23)\xrightarrow{L}(13)\xrightarrow{R}(15)\xrightarrow{L}(56)$. Now again we trim from the part that turns left continuously until $(56)$ and delete boundary $(23)$, which leaves with the word $(13)$. Thus we find 
\begin{equation}
\begin{aligned}
&{\cal F}^{{\rm benz}}_{1,3} = (1,1) \cdot M_R(y_{1,3}) M_R(y_{3,5}) \cdot \left(\begin{array}{c} 1 \\ 0 \end{array} \right) = 1 + y_{1,3} + y_{3,5} y_{1,3}, \nonumber \\ 
&{\cal F}^{{\rm benz}}_{1,4} = (1,1) \cdot M_R(y_{1,3}) \cdot \left(\begin{array}{c} 1 \\ 0 \end{array} \right) = 1 + y_{1,3}.
\end{aligned}
\end{equation}
 
\section{Factorization on Massive Poles} \label{sec:massive}
Factorization is a fundamental feature of tree-level amplitudes. From the $u$ representation of the string amplitude, the factorization on massless poles, $X_{i,j}\rightarrow 0$, is automatic due to the binary property of the $u$ variables.  In this case the integral just factorizes into the product of two lower-point string amplitudes, as expected. 

However, an explicit determination of amplitudes on massive factorization channels, $X_{i,j}\rightarrow -n_{i,j}$, has not been available. 
Among other things, massive factorization beyond four point has proven to be extremely constraining in considering deformations of the string~\cite{MultPartFact}, so it would be useful to have explicit expressions for the residue on the poles. 

\subsection{Factorization on single poles}

It turns out that the $u$ representation together with the positive parameterization in terms of $y_\mathcal{C}$ allows us to write a general answer for the residue at level $n_{i,j}$ as a sum of \textit{massless} amplitudes evaluated on shifted kinematics. 

Let us begin by studying this phenomenon at five point.  To extract the residue at $X_{1,4}\rightarrow-n_{1,4}$, choosing the $y$ representation for base triangulation $\{X_{1,3},X_{1,4}\}$ we wish to compute
\begin{equation}
\underset{X_{1,4} \to -n_{1,4}}{\rm Res}
    \oint \frac{\diff y_{1,4}}{y_{1,4}}  y_{1,4}^{-n_{1,4}} \frac{\diff y_{1,3}}{y_{1,3}} y_{1,3}^{X_{1,3}}  (1 + y_{1,3})^{-c_{1,3}} (1 + y_{14}(1 + y_{1,3}))^{-c_{1,4}} (1 + y_{1,4})^{-c_{1,4}}.
\end{equation}

Expanding both $\mathcal F$ polynomials, $(1 + y_{1,4})^{-c_{2,4}}$ and $(1 + y_{1,4}(1 + y_{1,3}))^{-c_{1,4}}$, in powers of $y_{1,4}$ in order to extract the power of $y_{1,4}^{n_{1,4}}$ and calculate the residue, we obtain
\be
 \sum_{k_{2,4}, k_{1,4}=0}^{\infty} \delta_{k_{1,4}+k_{2,4},n_{1,4}} \binom{-c_{1,4}}{k_{1,4}}  \binom{-c_{2,4}}{k_{2,4}} \underbrace{\times \int \frac{\diff y_{1,3}}{y_{1,3}} y_{1,3}^{X_{1,3}} (1 + y_{1,3})^{-(c_{1,3} - k_{1,4})}}_{{\cal A}_4(X_{13},X_{24} + n_{1,4}- k_{1,4})},
\ee
where we use $c_{1,3} =X_{1,3} + X_{2,4} - X_{1,4} = X_{1,3} + X_{2,4} + n_{1,4}$ to recognize the second factor as the four-point massless string amplitude, evaluated at shifted kinematics. Using also that $n_{1,4} = k_{1,4} + k_{2,4}$, we find that 
\begin{equation}
\underset{X_{1,4} \to -n_{1,4}}{\rm Res} {\cal A}_5 = \sum_{k_{1,4} + k_{2,4} = n_{1,4}}\binom{n_{1,4} + X_{2,4} - X_{2,5}}{k_{1,4}}  \binom{-c_{2,4}}{k_{2,4}}  {\cal A}_4(X_{1,3},X_{2,4} +  k_{2,4}).
\label{eq:resX14}
\end{equation}

\begin{figure}[t]
    \centering
    \includegraphics[width=0.9\linewidth]{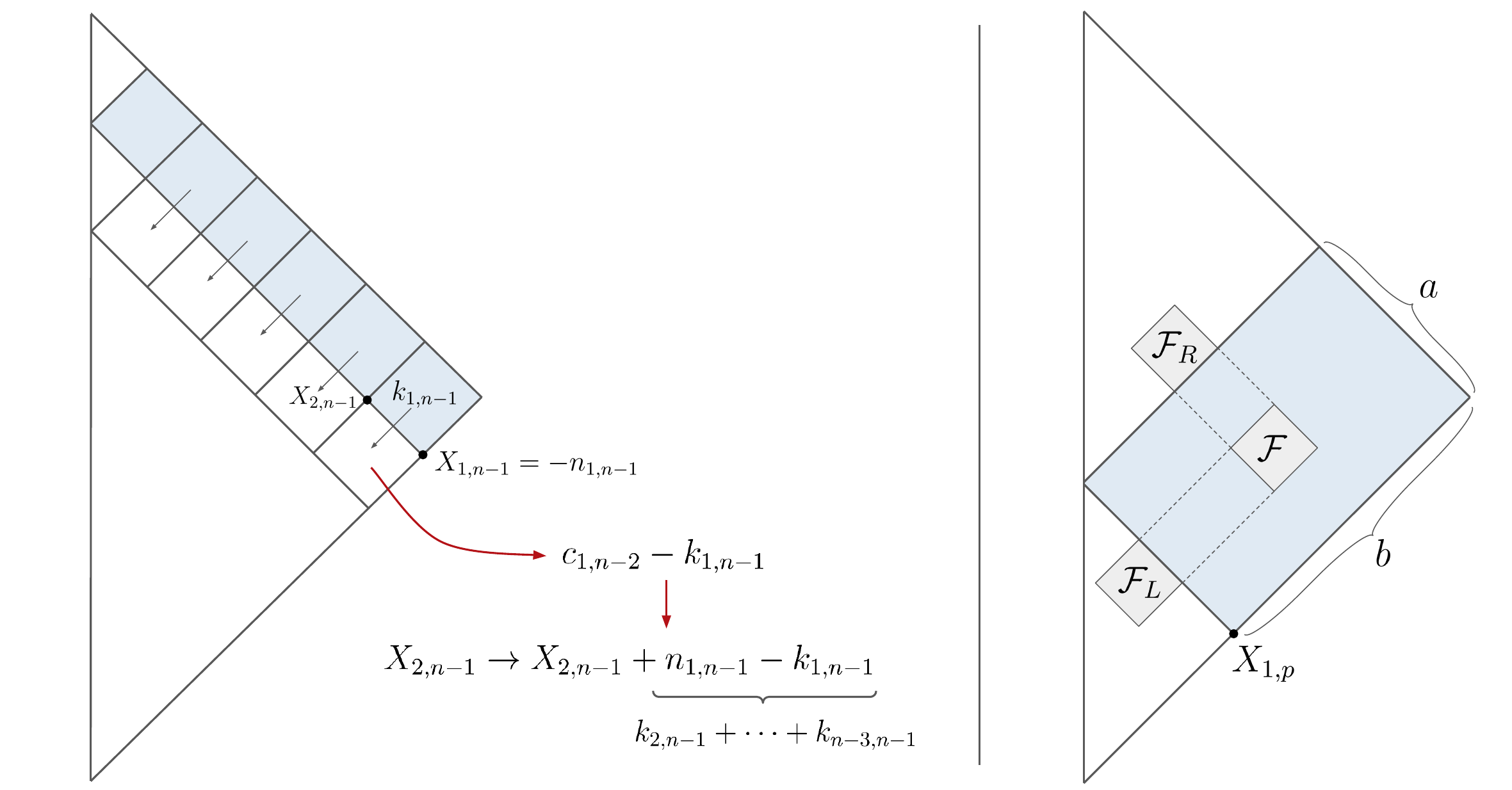}
    \caption{(Left) Kinematic mapping for factorization on the pole at $X_{1,n-1}=-n_{1,n-1}$. (Right) Factorization of $\mathcal{F}$ polynomials on the general massive level $X_{1,p}=-n_{1,p}$.}
    \label{fig:FFact}
\end{figure}

We can proceed in the same way for working out the factorization on a general massive pole. The simplest case corresponds to the ``collinear'' factorization where $X_{i,i+2}$ is set to a negative integer. Without loss of generality we can assume this is $X_{1,n-1} \to -n_{1,n-1}$. The dependence of the ${\cal F}$ polynomials on $y_{1,n-1}$ is especially simple (see Fig.~\ref{fig:FFact}, left). We have that 

\begin{equation}
{\cal F}_{i,n-1} = 1 + y_{1,n-1} f_{i,n-2},
\end{equation}
where $f_{i,j}$ are the ${\cal F}$ polynomials for the lower $(n-1)$-point problem. Binomial expanding in $y_{1,n-1}$ gives us a sum over the lower $(n-1)$ amplitude with shifted kinematics. 
Computing the residue on $y_{1,n-1}$ then gives
\begin{equation}
\underset{X_{1,n-1} \to -n_{1,n-1}}{\rm Res} {\cal A}_n = \sum_{k_{1,n-1} + \cdots k_{n-3,n-1} = n_{1,n-1}} \prod_{i=1}^{n-3} \binom{-c_{i,n-1}}{k_{i,n-1}} \times {\cal A}_{n-1}(\hat{X}),\label{eq:colR}
\end{equation}
where the lower amplitudes have shifted kinematics given by 
\begin{equation}
\hat{X}_{j,n-1} = X_{j,n-1} + (k_{j,n-1} + \cdots k_{n-3,n-1}).
\end{equation}
It is interesting that the residue is determined by sums of lower amplitudes with kinematics shifted by {\it positive} integers, a fact that will be useful in deriving general ``dual resonance'' expressions for the amplitude at all multiplicity.  

We can easily extend this result to consider factorization on a completely general pole $X_{i,j} \to -n_{i,j}$. By cyclic roation we can take this to be $X_{1,p} \to -n_{1,p}$ for some $p$. Focusing on any $\mathcal F$ polynomial that depends on $y_{1,p}$, 
we note again the crucial fact that the dependence on $y_{1,p}$ is extremely simple in a way that reveals the cut amplitudes. In particular, we see that
\begin{equation}
{\cal F} = {\cal F}_{R} + y_{1,p} {\cal F}_{L},
\end{equation}
where ${\cal F}_{L,R}$ are the ${\cal F}$ polynomials for the left and right factors on the cut (see Fig.~\ref{fig:FFact}, right). Expanding in $y_{1,p}$ just as we did above, we can write the massive residue of $\mathcal{A}_n$ as follows,
\be
\begin{aligned}
\mathop{\mathrm{Res}}_{X_{1,p} \to -n_{1,p}}{\cal A}_n&= \sum_{\substack{k_{a,b},\\
1 \leq a \leq p-2, \\ p\leq b \leq n-1}} \prod_{a,b} \binom{-c_{a,b}}{k_{a,b}} \delta_{\sum_{a,b} k_{a,b},n_{1,p}} 
{\cal A}_L\left(\hat X_{l,p}\right) {\cal A}_R \left( \hat X_{1,m} \right) \\
\hat X_{l,p} &= X_{l,p} + \sum_{l \leq a , b} k_{a,b}, \quad \text{with } l\in \{2,3, \cdots,p-2\}, \\
\hat X_{1,m} &=X_{1,m} + \sum_{m \leq b \leq n-1, a} k_{a,b}, \quad \text{with }m\in\{p+1,\cdots,n-1\}, 
\end{aligned}
\ee
where the lower-point amplitudes are evaluated on the shifted kinematics defined by the $\hat X$ variables.

We have derived all of these results by making a particular choice---the simplest ``ray-like triangulation''---for the base triangulation used to define the $y$ variables and ${\cal F}$ polynomials. It is easy to see that for {\it any} triangulation, the ${\cal F}$ polynomials have the same kind of expansion in the $y$ variables for any internal chord, and that this leads to a formula for the residue in terms of products of lower-point amplitudes. Indeed, interestingly, the expressions for the residues end up being exactly the same as we have seen here in the simplest case of the ray-like triangulation. 

There is another interesting choice that has been tacitly made, which we wish to highlight. Consider again the expression for the residue when sending $X_{1,n-1} \to -n_{1,n-1}$. Our expression is given in terms of a shifted lower-point amplitude, where $X_{j,n-1}$ is shifted. But this may appear peculiar: after all, the string amplitudes are not only cyclically symmetric, they are also dihedrally invariant in reversing the order $(1,2,3,\cdots,n) \to (n,n-1,\cdots,1)$. This asymmetry is built into the surface formalism, in the definition of ``laminations,'' which involve taking chords $(i,j)$ and associating them with curves where the endpoints are rotated cyclically in one or the other direction. The choice we have been using is cyclic rotation to the right, but we can also make the opposite choice, and that would give rise to a similar formula for the residue where $X_{1,j}$ is shifted instead of $X_{j,n-1}$.

\subsection{Massive factorization on complete diagrams}\label{sec:fact}

Having determined factorization on a single massive pole in terms of products of shifted lower-point amplitudes, we can of course recursively continue and compute the residue on a maximal collection of compatible poles evaluated at general negative integers. We can also use the ${\cal F}$ polynomials and $y$ variables to compute the residues on massive poles ``in one shot''. For instance, returning to our five-point example,
we can extract the residue $R_{n_{1,3},n_{1,4}}$ when $X_{1,3} \to -n_{1,3}, X_{1,4} \to -n_{1,4}$. Indeed, the integrand written in ${\cal F}$ polynomial form can be thought of as a generating function for these residues as 
\begin{equation}
(1 + y_{1,3})^{-c_{1,3}} (1 + y_{1,4} + y_{1,4} y_{1,3})^{-c_{1,4}} (1 + y_{1,4})^{-c_{2,4}} = \sum_{n_{1,3},n_{1,4}} y_{1,3}^{n_{1,3}} y_{1,4}^{n_{1,4}} R_{n_{1,3}, n_{1,4}}.
\end{equation}

To obtain an explicit form of the coefficients we can trivially perform a multinomial expansion in both $y_{1,3},y_{1,4}$ as
\be 
\begin{aligned}
(1 + y_{1,3})^{-c_{1,3}} &= \sum_{k_{1,3}} \binom{-c_{1,3}}{k_{1,3}} y_{1,3}^{k_{1,3}}\\ (1 + y_{1,4})^{-c_{2,4}} &= \sum_{k_{2,4}} \binom{-c_{2,4}}{k_{2,4}} y_{1,4}^{k_{2,4}}  \\ (1 + y_{1,4} + y_{1,4} y_{1,3})^{-c_{1,4}} &= \sum_{k^{(1)}_{1,4},k^{(2)}_{1,4}} \binom{-c_{1,4}}{k^{(1)}_{1,4},k^{(2)}_{1,4}} y_{1,4}^{k^{(1)}_{1,4} + k^{(2)}_{1,4}} y_{1,3}^{k^{(2)}_{1,4}}
\end{aligned}
\ee
to obtain 
\begin{equation}
R_{n_{1,3},n_{1,4}} = \sum_{k_{1,3}, k_{2,4}, k^{(1,2)}_{1,4}} \delta_{k_{1,3} + k^{(2)}_{1,4}, n_{1,3}} \delta_{k_{2,4} + k^{(1)}_{1,4} + k^{(2)}_{1,4}, n_{1,4}} \times \binom{-c_{1,3}}{k_{1,3}} \binom{-c_{2,4}}{k_{2,4}} \binom{-c_{1,4}}{k^{(1)}_{1,4},k^{(2)}_{1,4}} .
\end{equation}
We note that in this expansion we think of $R_{n_{1,3}, n_{1,4}}$ as a function of the $c$ variables; when evaluating the $c_{i,j}$ in terms of the remaining kinematic invariants we must of course put $X_{1,3} \to -n_{1,3}, \, X_{1,4} \to -n_{1,4}$. 

Of course, this equation is not deeply different than the one we have obtained by recursively applying our single-pole factorization. The two are related by expressing the multinomial coefficients above as a sum over binomial coefficients. But this multinomial form is more ``global'' and compact, and reflects our ability to give a non-recursive, direct construction of the ${\cal F}$ polynomials. 

In this way we can similarly define the residue on the massive poles when $X_{1,j} \to -n_{1,j}$ with $j \in \mathcal{T}^R =\{3, \cdots, n-1\}$, for any ray-like triangulation. Again the ${\cal F}$ polynomial form gives us the generating function for $R_{\{n_{1,j} \}}$ as 
\begin{equation}
\prod_{i<j} (1 + y_{1,j} + y_{1,j} y_{1,j-1} + \cdots + y_{1,j} \cdots y_{1,i})^{-c_{i,j}} = \sum_{\{n_{1,j}\}} \prod_{k\in \mathcal{T}^R} y_{1,k}^{n_{1,k}} R_{\{n_{1,j}\}},
\end{equation}
and we can give an explicit expression for $ R_{\{n_{1,j}\}}$ by performing a multinomial expansion of the powers on the left-hand side of the equation. Again this gives the coefficients $R$ in terms of the $c$ variables, and when we express these in terms of the remaining kinematics we must remember to put $X_{1,j} \to -n_{1,j}$, for all $j\in \mathcal{T}^R$.
For example at $n=6$, for the triangulation $\{(1,3),(1,4),(1,5)\}$, we have
\begin{equation}
\begin{aligned}
&R_{n_{1,3},n_{1,4},n_{1,5}} =\\&\qquad \sum_{k_{i,j}} \delta_{n_{1,3},k_{1,3}+k_{1,4}^{(2)}+k_{1,5}^{(3)}} \delta_{n_{1,4},k_{2,4}+k_{1,4}^{(1)}+k_{1,4}^{(2)}+k_{2,5}^{(2)}+k_{1,5}^{(2)}+k_{1,5}^{(3)}}\delta_{n_{1,5},k_{3,5}+k_{2,5}^{(1)}+k_{2,5}^{(2)}+k_{1,5}^{(1)}+k_{1,5}^{(2)}+k_{1,5}^{(3)}} \times \\
&\qquad\qquad \times \binom{-c_{1,3}}{k_{1,3}}  \binom{-c_{2,4}}{k_{2,4}} \binom{-c_{3,5}}{k_{3,5}}
\binom{-c_{1,4}}{k_{1,4}^{(1)},k_{1,4}^{(2)}} \binom{-c_{2,5}}{k_{2,5}^{(1)},k_{2,5}^{(2)}}  \binom{-c_{1,5}}{k_{1,5}^{(1)},k_{1,5}^{(2)},k_{1,5}^{(3)}} .
\end{aligned}
\label{eq:ResHalfLadd}
\end{equation}
As another interesting example, we can consider massive factorization on the ``Mercedes-Benz'' configuration at $n\,{=}\,6$ points, corresponding to the triangulation of the hexagon containing chords $\{X_{1,3},X_{3,5},X_{1,5}\}$ (see Fig.~\ref{fig:RayTriang}, right).  Here we are interested in computing the residue $R_{n_{1,3},n_{3,5},n_{1,5}}$ when $X_{1,3} \to -n_{1,3},X_{3,5} \to -n_{3,5}, X_{1,5} \to - n_{1,5}$, with generating function given by 
\begin{equation}
\begin{aligned}
&(1 + y_{1,3})^{-c_{1,4}} (1 + y_{3,5})^{-c_{3,6}} (1 + y_{1,5})^{-c_{2,5}}  (1 + y_{1,5} + y_{1,5} y_{1,3})^{-c_{1,5}} \\
& \times (1 + y_{3,5} + y_{3,5} y_{1,5})^{-c_{3,5}} (1 + y_{1,3} + y_{1,3} y_{3,5})^{-c_{1,3}}  = \sum y_{1,3}^{n_{1,3}} y_{3,5}^{n_{3,5}} y_{1,5}^{n_{1,5}} R_{n_{1,3},n_{3,5},n_{1,5}}.
\end{aligned}
\end{equation}

\section{Asymptotic Limits} \label{sec:asymptotic}
String amplitudes are known for their remarkable asymptotic features: already at four point they are exponentially suppressed in the hard scattering limit, while growing only polynomially in the \textit{Regge limit} of fixed momentum transfer and large center-of-mass energy. While remarkable,
on their own these features do not fix the amplitude uniquely to the beta function, and there is now a panoply of other objects being explored that exhibit some or all of these properties~\cite{Cheung:2023adk,Cheung:2023uwn,David,Coon,Haring:2023zwu}; however, invoking the additional feature of ``level truncation,'' string amplitudes have recently been shown to be the unique superpolynomially soft, dual resonant amplitudes at tree level~\cite{Cheung:2024uhn,Cheung:2024obl}.
In any case, the remarkable high-energy behavior of the four-point string amplitude is certainly among its most salient and consequential physical attributes.

We will now investigate how these limits work for general $n$-point scattering. When is the amplitude exponentially soft, and what subsets of kinematic invariants need to be made large for this to happen? What subsets instead give power-law behavior when made large, which we might call ``generalized Regge'' behavior? 

We will explore these questions in this section. The familiar case of the four-particle amplitude is so special as to not immediately suggest even the qualitatively correct statements. But as we will see, in the simple regime where all $X>0$ and the integral is convergent, the $u$ variables not only yield the answers to these questions, but in many cases they lead to precise asymptotic statements, where the asymptotic amplitudes factor into lower-point amplitudes. This gives us yet another example of what is now a ubiquitous phenomenon, where particle and string amplitudes are seen to factorize even away from poles~\cite{Zeros}. 

Note that having all $X_{i,j}$ positive cannot always be realized in Lorentzian kinematics. Indeed, it is easy to see that for an odd number of particles, it is impossible to make all $X_{i,j}>0$ with real Lorentzian kinematics, while for even points there are simple configurations where this is possible, where, e.g., the momenta $p_i$ are incoming for even $i$ and outgoing for odd $i$, with some further nonlinear restrictions on the $X$ variables. 
Our analysis for $X>0$ using $u$ variables can be thought of as giving us a powerful way of  understanding the saddle points that dominate the integral in this case. Moreover, we know that the exponential suppression for $X>0$ continues to holds when some $X$ variables are negative, and it will be interesting to extend our analysis to delineate the regions in $X$ where exponential suppression continues to hold. For the case of Regge behavior, the saddle points end up localizing close to the boundaries of the moduli space, where some $u$ variables go to $0$ or $1$, as expected on physical grounds. Thus, it is natural to expect these saddle points to continue to dominate even away from the all $X>0$ region where the analysis strictly holds. In Sec.~\ref{sec:numerics}, we will give evidence for this expectation by checking our ``precision Regge'' predictions against new representations for the amplitude that we will give in later sections, finding perfect agreement in all kinematic regions these representations allow us to access, even where we have many $X<0$.

As a further consequence of our generalized Regge limits, we will derive dual resonance of the string at $n$ point, for the first time recovered strictly using the amplitude itself.

\subsection{Exponential suppression}
The initial expectation about the $n$-point amplitude in string theory is that hard scattering at high energies should be exponentially suppressed, just as at four points. 
In the context of KN, the integral representation is convergent when $X_{i,j}>0$, and the ``hard scattering'' regime is intuitively the one in which all the $X_{i,j}$ are made large and positive.

The $u$ representation of the integral makes this exponential suppression manifest: since all the $u_{i,j}$ are bounded between $0$ and $1$, it follows that the $\prod_{i,j} u_{i,j}^{X_{i,j}}$ factors are exponentially small everywhere except in the regions where $u_{i,j} \to 1$. If we make all the $X_{i,j}$ large and positive, this would require all the $u_{i,j} \to 1$. But this is clearly impossible from the $u$ equations, as $u + \prod u = 1$ clearly cannot be satisfied if all the $u_{i,j} \to 1$.
As a result, exponential suppression is guaranteed when all the $X_{i,j}$ are large and positive.

\subsubsection{All $X_{i,j}$ large}\label{sec:AllXlarge}
This expectation---of exponential suppression for all $X>0$ large---can be quantified. For simplicity, suppose we make all the $X_{i,j}$ equal to some quantity $X$; then the KN factor becomes $U^X$ where $U = \prod_{i,j} u_{i,j}$. It is easy to see that $U$ is maximized for the cyclically symmetric configuration of the $z_j$, i.e., where under a conformal map to the unit circle, $w_j = \frac{1 + i z_j}{1 -i z_j}$, the $w_j$ are uniformly spaced. Maximizing $U$ in the positive region where all $0<u_{i,j}<1$ is equivalent to looking for a solution of the saddle point/scattering equations.  Now if the $X_{i,j}$ are $\mathbb{Z}_n$ cyclically invariant, then either the solutions of the scattering equations are themselves cyclically invariant, or if they break this ${\mathbb{Z}_n}$ discrete symmetry, there must be $n$ of them. But it was shown in Ref.~\cite{Arkani-Hamed:2019mrd} that for nonnegative $X$ and $c$ variables, there is a {\it unique} real, positive solution of the scattering equations. This unique solution must therefore be the cyclically invariant one we identified. This argument applies not just to the case where all the $X_{i,j}$ are equal, but more generally to a cyclically symmetric configuration where $X_{i,j}$ = $X_k$ with $k=|i-j|$.  
 
The $u_{i,j}=\frac{(w_{i-1} - w_j)(w_i - w_{j-1})}{(w_{i-1} - w_{j-1})(w_i - w_j)}$ are readily computed from the cyclically symmetric configuration where $w_j = {\rm exp}(2 \pi i j /n)$, and we find 
\begin{equation}
u_{i,j} = \frac{\sin(\frac{\pi(i-j - 1)}{n})\sin(\frac{\pi (i-j + 1)}{n})}{\sin^2(\frac{\pi (i-j)}{n})}.
\end{equation}

Hence when the $X_{i,j} = X_{k=|i-j|}$ are large, we find the exponential suppression of the amplitude is given by 
\begin{equation}
{\cal A}_n \xrightarrow{X_{i,j} = X_{k=|i-j|} \gg 1} \prod_{k=2}^{\lfloor n/2\rfloor} \left[\frac{\sin\left(\frac{\pi(k - 1)}{n}\right) \sin \left(\frac{\pi (k + 1)}{n}\right)}{\sin^2( \frac{\pi k}{n})}\right]^{n_k X_k},
\end{equation}
where $n_k = n$ unless $k=n/2$ for $n$ even, in which case $n_{k=n/2} = n/2$.

In the special case where all the $X_{i,j}$ are equal, the product $U$ of all the $u_{i,j}$ simplifies to 
$U = 2^{-n} \sec^n (\pi/n)$, and so for large equal $X$ we have an especially elegant expression for the exponential suppression of the amplitude,
\begin{equation}
{\cal A}_n \xrightarrow{X_{i,j}=X \gg 0} \frac{1}{\left[2 {\rm cos}(\pi/n)\right]^{n X}}.\label{eq:expsup1}
\end{equation}

These expressions capture the exponential suppression at large $X$, but we can of course do better by computing the Gaussian integral around the saddle point. When all the $X$ variables are equal, this tells us more precisely that 
\begin{equation}
{\cal A}_n \xrightarrow{X_{i,j}=X \gg 0} C_n \times \left(\frac{2 \pi}{X} \right)^{(n-3)/2} \times \frac{1}{\left[2 {\rm cos}(\pi/n)\right]^{n X}} \times \left[ 1 + O(1/X) \right],\label{eq:expsup2}
\end{equation}
where $C_n$ is an $X$-independent constant given by the determinant of the double-derivative matrix evaluated on our cyclically symmetric saddle point, which we can compute for small $n$ as $C_4 = \sqrt{2},\, C_5 = \sqrt{2/(25 - 11 \sqrt{5})},\, C_6 = \sqrt{27/2}$, etc.

In particular, when $n$ is large, we asymptote to a simple universal behavior ${\cal A}_n \to 2^{-n X}$, up to power-law corrections. It is interesting to compare this with the behavior of the amplitude at large $n$ in the field theory limit, which for all $X_{i,j}$ equal is just the Catalan number counting of diagrams and scales as $4^n/X^{n-3}$. Thus at large $n$, we can write the amplitude as 
\begin{equation}
{\cal A}_n(X) \to e^{- n a(X)}
\end{equation}
where 
\begin{equation}
a(X) \to \left\{\begin{array}{ll} \log (X/4), & X \ll 1 \\ X \log 2, & X \gg 1. \end{array} \right.
\end{equation}

The fact that the $n$ depedence factors out as an exponential in the prefactor strongly suggests that, even when then $X_{i,j}$ are not all equal, if they are chosen to be sufficiently smooth for large $n$, there will be a well defined ``large-$n$
limit'' of string amplitudes at both low and high energies; we leave this fascinating topic for  future work.

\subsubsection{Subsets of  $X_{i,j}$ large}
\label{sec:subsetLarge}

So far we have seen that we have exponential suppression when all $X$ are large simply because we cannot make all the respective $u$ variables go to $1$ on the support of the $u$ equations. It is then clear that exponential suppression can also arise even if not all $X_{i,j}$ are made large. We simply have to identify a set of chords for which the $u$ equations make it impossible for all the associated $u$ variables to go to $1$. If instead it is possible to have all $u_{i,j} \to 1$, then we have instead polynomial growth, which we will come back to momentarily. This observation reduces the determination of whether the amplitude exhibits exponential or polynomial behavior to an essentially combinatorial question about patterns of crossing chords. This problem is an interesting cousin of the more familiar question of when collections of $u$ variables can go to zero together, which relates to the compatibility of the poles of the amplitude. 

A very simple example leading to exponential suppression is if we send a single $X_{i,j} \to \infty$, in which case the integral will be dominated by the region where the corresponding $u_{i,j} \to 1$. In this region, \textit{at least one} of the $u_{k,m}$ corresponding to curves that cross $X_{i,j}$ has to go to zero, so that the equation $u_{i,j} + \prod_{k,m} u_{k,m}=1$ is satisfied. Therefore if we consider the limit where $X_{i,j}$ as well as $X_{k,m}$---for all $(k,m)$ that cross $(i,j)$---get large, we have that the amplitude will be exponentially suppressed, as we cannot set all the $u$ variables to $1$ simultaneously.

\begin{figure}[t]
    \centering
    \includegraphics[width=0.5\linewidth]{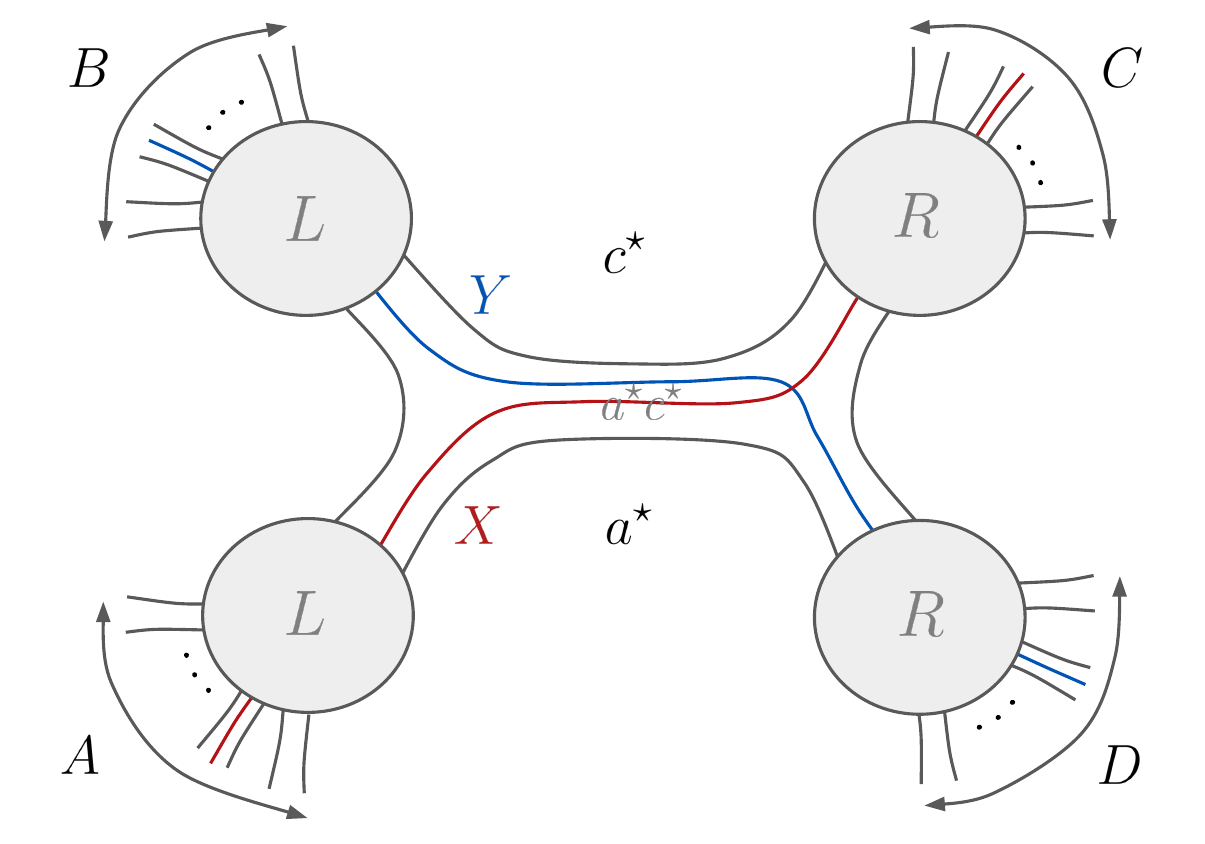}
    \caption{Effective four-point problem containing particles in subsets $A$, $B$, $C$, and $D$.}
    \label{fig:ABCD}
\end{figure}

This picture has a simple generalization, associated with the ``generalized'' $u$ equations: given a division of the particle labels into four subsets of adjacent indices, $A,B,C,D$, we have
\begin{equation}
\prod_{a\in A, c\in C} u_{a,c} + \prod_{b\in B, d\in D} u_{b,d} =1,  
\end{equation}
so if we send $X_{a,c}$ as well as $X_{b,d}$ for all $a\in A, b\in B, c\in C$ and $d\in D$ to infinity, the amplitude will be exponentially suppressed.
This is a multiparticle generalization of the exponential suppression at four point, where each of particles 1 through 4 has now been replaced by a collection of multiple particles.
In particular, suppose we set $X_{a,c} = X$ and $X_{b,d} = Y$, for all $a \in A, c\in C, b\in B$ and $d\in D$ (see Fig.~\ref{fig:ABCD}), and define $U_X = \prod_{a\in A, c\in C} u_{a,c}$ and $U_Y = \prod_{b\in B, d\in D} u_{b,d}$, in which case we can write the amplitude as 
\begin{equation}
\begin{aligned}
&{\cal A} = \int \frac{\diff y_{a^\star,c^\star}}{y_{a^\star,c^\star}}U_X^X U_Y^Y \int \prod_{P\neq (a^\star,c^\star)} \frac{\diff y_P}{y_P} \prod_{X^\prime \neq X,Y} u_{X^\prime}^{X^\prime} =\int \frac{\diff y_{a^\star,c^\star}}{y_{a^\star,c^\star}}U_X^X (1-U_X)^Y  (\cdots) \\
&\to \frac{X^X Y^Y}{(X+Y)^{X+Y}} \times \int \prod_{P\neq (a^\star,c^\star)} \frac{\diff y_P}{y_P} \prod_{X^\prime \neq X,Y} u_{X^\prime}^{X^\prime} 
\end{aligned}
\end{equation}
where in the last expression, we solved for the saddle, $U_X = X/(X+Y)$. Here, $(a^\star,c^\star)$ is the chord associated to the propagator in the underlying base triangulation entering the effective four-point problem mentioned above (where particle 1 is replaced by set $A$, 2 by $B$, and so on).
That is, we have $\log{\cal A} = X \log X + Y\log Y - (X+Y)\log(X+Y)+\cdots$, directly akin to the familiar form of the hard scattering of the four-point string.

\subsection{Exponential suppression and ``split'' factorization}

A remarkable feature of string amplitudes that is made obvious by thinking in terms of $u$ variables is a pattern of ``split'' factorization away from poles. Following the discussion of Ref.~\cite{Splits}, we can choose two sub-polygons $S_1,S_2$ that overlap on a triangle, and then {\it define} kinematics for the big surface by setting $X^{\rm split}= X_1 + X_2$, where $X_{1,2}$ is restriction of $X$ to $S_{1,2}$ (see Ref.~\cite{Splits} for more details). On this locus in kinematic space, we have that 
\begin{equation}
{\cal A}(X^{\rm split}) = {\cal A}_{S_1}(X_1) \times {\cal A}_{S_2}(X_2),
\end{equation}
which defines a lower-dimensional locus in kinematic space, away from poles, where the amplitude factors. This follows from writing the $u$ variables of a subsurface as a monomial in the $u$ variables of the full surface \cite{Splits}. 

An obvious extension of this argument tells us that even if we are away from this locus in kinematic space, if we scale to large $X$ values so that we obtain split kinematics, the amplitude will also factorize. In other words, if we set 
\begin{equation}
X = z X^{\rm split} + x
\end{equation}
and send $z \to \infty$, then for $z X^{\rm split}$ positive we are localizing to a unique solution of the saddle point equations, and so the exponentially suppressed amplitudes factorize just as before:
\begin{equation}
{\cal A}(z X^{\rm split} + x) \xrightarrow{z \to \infty} {\cal A}_1(z X_1) \times {\cal A}_2(z X_2) .
\end{equation}

\subsubsection{Minimal kinematics with exponential suppression}
The examples above are perhaps the simplest infinite class of kinematics where the amplitude is exponentially suppressed, but other classes can also be straightforwardly identified. Consider a chord $X_1$, and look at the set of chords $\{X^{{\rm cross}}_1\}$ that cross it; suppose that a subset $\{Y^{{\rm cross}}_1\}$ of these chords are {\it not} made large. Thus at least one of the $u$ variables in the set $\{Y^{{\rm cross}}_1\}$ must be sent to zero. Let us do the same for another chord $X_2$. Then if the chords in $\{Y^{{\rm cross}}_1\}$ intersect all the chords in $\{Y^{{\rm cross}}_2\}$, there are no pairs of $u$ variables in $\{Y^{{\rm cross}}_1\},\, \{Y^{{\rm cross}}_2\}$ that can be set to zero, and hence we have exponential suppression.

The simplest example of this phenomenon occurs at $n=7$ points. Suppose that we take $X_{1,3},X_{1,6},X_{2,4},X_{2,7},X_{5,7}$ to all be large, so all the corresponding $u$ variables must be set to $1$. From $u_{1,6} \to 1$, we conclude that either $u_{3,7}$ or $u_{4,7}$ must be set to zero. Similarly from $u_{1,3} \to 1$ we conclude that either $u_{2,5}$ or $u_{2,6}$ must be set to zero. But any of the chords in $\{(3,7),(4,7)\}$ intersect any of those in $\{(2,5),(2,6)\}$, and hence it is impossible to set one to zero in each set. We can again get a quantitative understanding of the exponential suppression in this example, if we set all the $X_{1,3},X_{1,6},X_{2,4},X_{2,7},X_{5,7}=X$ equal and large. Then the exponential suppression is given by $U^X$ where $U=u_{1,3} u_{1,6} u_{2,4} u_{2,7} u_{5,7}$. We have concluded that it is impossible to have $U \to 1$, and in fact one can show that $U$ has an upper bound of $1/4$. Thus, in the limit where $X$ is large, we find that the seven-particle amplitude scales as 
\begin{equation}
{\cal A}_7 \xrightarrow{X_{1,3},X_{1,6},X_{2,4},X_{2,7},X_{5,7}=X\gg 0} C \times \frac{1}{X^2} \times \frac{1}{4^X}.
\end{equation}
where as above the $X$-independent constant $C$ multiplying the power-law scaling with $X$ follows from doing the Gaussian integral around the saddle. 

\begin{figure}[t]
    \centering
    \includegraphics[width=0.8\linewidth]{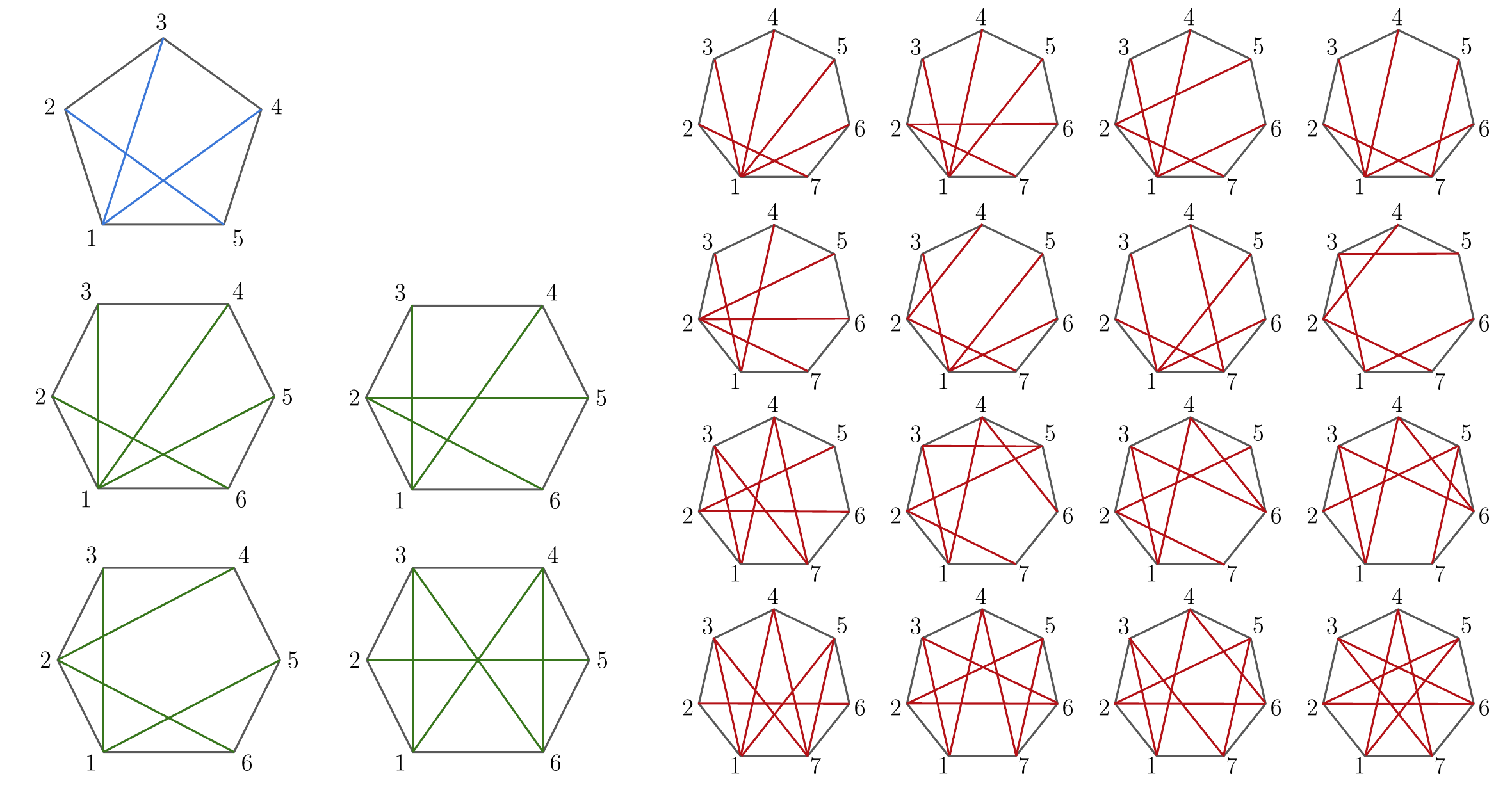}
    \caption{``Minimal'' set of kinematics that give exponential suppression for $n=5$ (blue), $n=6$ (green), and $n=7$ (red).}
    \label{fig:expsup}
\end{figure}

It is an interesting combinatorial problem to classify all the ``minimal'' sets of kinematics that give exponential suppression. By ``minimal,'' we simply mean finding a set $\{X_{{\rm large}}\}$ of chords for which all the $u$ variables cannot be set to $1$, but such that if any one of the chords is removed from $\{X_{{\rm large}}\}$, the corresponding $u$ variables {\it can} be set to $1$. Up to $n=6$, all of these choices of minimal exponentially soft kinematics are simply associated with subsurfaces. For $n=7$, all the cyclic classes of minimal exponentially soft kinematics are shown in Fig.~\ref{fig:expsup} (as well as those for $n=5,6$). We leave the solution of the combinatorial problem of determining these minimal kinematics for all $n$ to future explorations.

\subsection{Regge limits} Let us now move to the case where the amplitude exhibits polynomial behavior, which we will call Regge behavior from this point onward. We will explore different regimes of Regge scaling and lay out the strategy to derive the asymptotic behavior of the amplitude.
As in the case of exponential softness, our study of Regge limits will not be exhaustive. Rather, we will study the general principles governing which choices of kinematics can lead to Regge growth and illustrate these observations with some salient examples.
As a particularly important example, we will show that $n$-point scattering obeys sufficient Regge scaling to allow the string to be written in dual resonant form, for any number of external particles.

Before proceeding to explicit examples, we will first quickly highlight the important asymptotic approximation that ultimately lets us derive the results that follow. This is the observation that, considering the limit where a ``stringy integral'' is dominated by the region where a given $u \to 0$, in this regime we can write it as
\begin{equation}
I = \int_0^1 \frac{\diff u}{u} u^X (1-u a_1)^{A_1} (1-u a_2)^{A_2} \cdots  (1-u a_k)^{A_k} \times F(\tilde{u}),
\label{eq:ReggeApprox}
\end{equation}
where $X$ is the kinematic variable associated $u$, and $a_i$ are some polynomials of the remaining $u$ variables, $A_i$ are some kinematics $X_{i,j}$ corresponding to curves that cross $u$, and $F(\tilde{u})$ is some function/integral of the remaining $u$ variables, namely part of some lower-point stringy integral. Since the integral is dominated near $u \to 0$ (because the $A_i$ are large), we can approximate $(1-a_i u)^{A_i} \approx \exp{\left(-a_i u A_i\right)}$ and do the integral up to $u \to \infty$. This gives a good approximation,
\begin{equation}
I \to \Gamma(X) \times \left(\frac{1}{a_1 A_1 + a_2 A_2 + \cdots a_k A_k}\right)^X \times F(\tilde{u}) \vert_{u=0}.
\end{equation}

\subsubsection{Regge scaling with one chord large}\label{sec:singlechord}
Let us start by considering a simple example where we are one step away from exponential suppression, the limit where all $X_{j,n}$ for $j \in \{2,3, \cdots, n-2\}$ are large but $X_{1,n-1}$ is \textit{not} large.\footnote{If $X_{1,n-1}$ were also large, we would be in a case like those described in Sec.~\ref{sec:subsetLarge}, and so would have exponential suppression, but in the present case we are still in a regime where we can set set all $u_{j,n}\to 1$ provided $u_{1,n-1}\to 0$.} The $u$ equations for $u_{j,n}$ read
\begin{equation}
u_{j,n}+ u V_j = 1,
\end{equation}
defining $u = u_{1,n-1}$ and writing $V_j$ for the product of the $u$ variables of the remaining curves that cross $(j,n)$. Then we can write the KN factor as follows,
\begin{equation}
u^{X_{1,n-1}} \prod_{j}(1-u V_j)^{X_{j,n}}\prod_{X\in \mathcal{P}_{n-1}} u_X^X,
\label{eq:KN_ray}
\end{equation}
where the last term corresponds to all the curves $X$ living inside the analogous problem at $n-1$ point involving particles $(1,2,\cdots,n-1) = \mathcal{P}_{n-1}$. For more control over what happens when the $X_{j,n}$ become large, let us further consider the case where $X_{j,n}=X$ for all $j$, so that we can rewrite the amplitude in Eq.~\eqref{eq:KN_ray} as  
\begin{equation}
\int \frac{\diff y_{1,n-1}}{y_{1,n-1}} u^{X_{1,n-1}} (1-u)^X   \prod_{p\neq (1,n-1)} \frac{\diff y_{p}}{y_{p}} \prod_{X_{i,j}\in \mathcal{P}_{n-1}} u_{i,j}^{X_{i,j}}.
\end{equation}

To avoid exponential suppression we need to have $u\to 0$, which corresponds to the region where $y_{1,n-1} \to 0$. In this limit, the dependence of $u_X$ on $y_{1,n-1}$ drops out, the amplitude manifestly factorizes, and the second factor becomes precisely the lower-point amplitude:
\begin{equation}
\frac{\Gamma(X_{1,n-1})\Gamma(X)}{\Gamma(X_{1,n-1}+X)}  \mathcal{A}_{n-1}(1,2,\cdots,n-1)\xrightarrow[]{X\to \infty} \Gamma(X_{1,n-1}) X^{-X_{1,n-1}} \mathcal{A}_{n-1}(1,2,\cdots,n-1).
\end{equation}

This argument is completely general for the case in which we pick a given curve $X_{i,j}$ and send \textit{all} the curves that cross it $X_{k,m} \equiv X \to \infty$, for which we obtain
\begin{equation}
\mathcal{A}_n \to \Gamma(X_{i,j}) X^{-X_{i,j}} \times \mathcal{A}_L\times \mathcal{A}_R,
\end{equation}
where $\mathcal{A}_L= \mathcal{A}_L(i,i+1, \cdots,j-1)$ and $ \mathcal{A}_R= \mathcal{A}_R(j,j+1, \cdots,i-1)$ are the two lower-point amplitudes we obtain when we go on the $X_{i,j}$ cut.
This is a beautiful generalization of the familiar Regge behavior of the four-point amplitude ${\cal A}_4 \rightarrow \Gamma(X_{2,4})X_{1,3}^{-X_{2,4}}$ when $X_{1,3}$ is large.

\subsubsection{Regge scaling with two or more chords}
\begin{figure}[t]
    \centering
    \includegraphics[width=\linewidth]{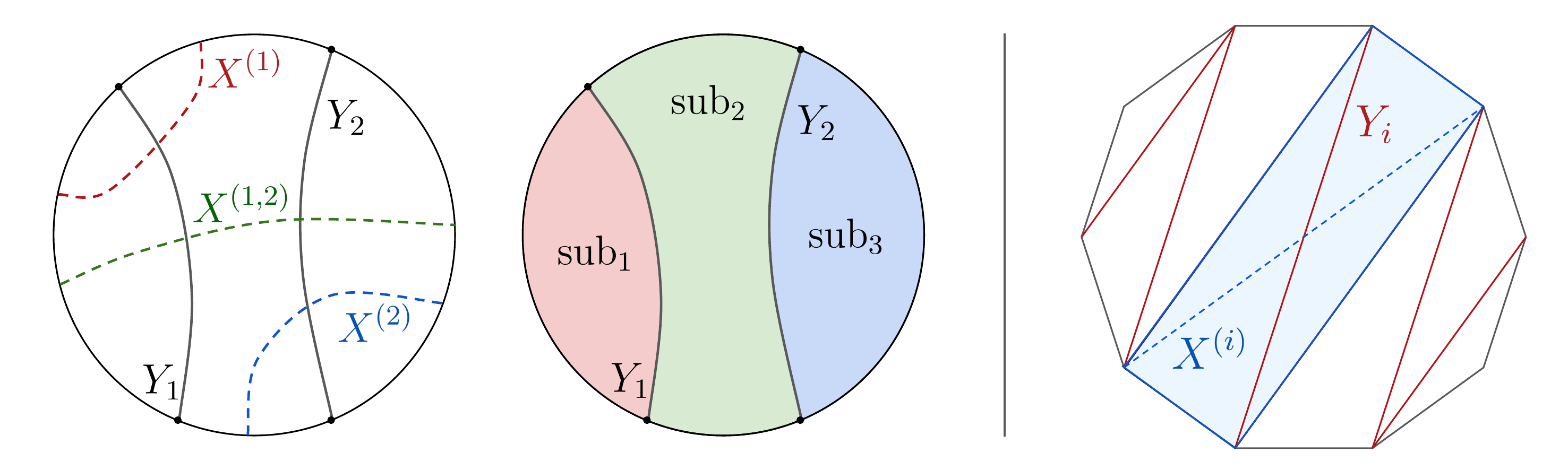}
    \caption{(Left) Surface and two chords $Y_1$ and $Y_2$ as well as the respective chords that cross them, $X^{(1)},X^{(2)}$, and $X^{(1,2)}$; highlighted in red, green, and blue, we have the three subsurfaces we obtain when we cut through $Y_1$ and $Y_2$. (Right) Given a chord $Y_i$ in a triangulation, it is always inside a square determined by the chords in the triangulation at the edges of the polygon (highlighted in blue). Then we define the mutation of a chord $Y_i$ inside the triangulation,  $X^{(i)}$, to be the  other diagonal of the square that $Y_i$ lives in.}
    \label{fig:Mut}
\end{figure}

Let us now proceed to a slight generalization of the case above, where we pick two different non-crossing chords $Y_1$ and $Y_2$ and send all the chords $X_{k,m}$ that cross {\it either} $Y_1$ or $Y_2$ to infinity. In particular, let us consider the set of $X_{i,j}$ that cross $Y_{1}/Y_{2}$ by $S^{(1)}/S^{(2)}$, and of those that cross both by $S^{(1,2)}$ (see Fig.~\ref{fig:Mut}, left). For simplicity, suppose we set all the $X_{i,j}\in S^{(1)}$ to $X^{(1)}$, those in $ S^{(2)}$ to $X^{(2)}$ and $X_{i,j} \in S^{(1,2)}$ to $X^{(1,2)}$. Once more, when taking $X^{(1)},X^{(2)},X^{(1,2)}\rightarrow\infty$,  we will be localizing the integral in the region where $u_{Y_1},u_{Y_2} \to 0$, factoring out the piece going like $u_{Y_1}u_{Y_2}$ in the KN factor, so that we have 
\begin{equation}
\begin{aligned}
&\prod_{X\in S^{(1)}} u_X^{X^{(1)}}\prod_{X\in S^{(2)}} u_X^{X^{(2)}}\prod_{X\in S^{(1,2)}} u_X^{X^{(1,2)}-X^{(1)}-X^{(2)}} \left( \prod_{X\notin S^{(1)},S^{(2)},S^{(1,2)}} u_X^X\right)\\
&=(1\,{-}\,u_{Y_1})^{X^{(1)}}(1\,{-}\,u_{Y_2})^{X^{(2)}}   \!\prod_{X\in S^{(1,2)}}(1\,{-}\,u_{Y_1}u_{Y_2}V_X)^{X^{(1,2)}-X^{(1)}-X^{(2)}} \left( \prod_{X\notin S^{(1)},S^{(2)},S^{(1,2)}} u_X^X\right),
\end{aligned}
\label{eq:X1X2}
\end{equation}
where we write $V_X$ for the remaining $u$ variables $ \neq u_{Y_1},u_{Y_2}$ that cross $u_X$ with $X\in S^{(1,2)}$. Note that the chords on the last factor, $X\notin S^{(1)},S^{(2)},S^{(1,2)}$,  are precisely those that live inside the three lower sub-polygons we obtain by cutting along curves $Y_1$ and $Y_2$. From Eq.~\eqref{eq:X1X2}, we see that in the limit $u_{Y_1},u_{Y_2} \to 0$, if $X^{(1,2)}\sim X^{(1)},X^{(2)}$ then the piece coming from $X\in X^{(1,2)}$ is negligible compared to those from $X^{(1)}$ and $X^{(2)}$. Thus, just as in the asymptotic limit, when $X^{(1)},X^{(2)},X^{(1,2)}\to \infty$ the integral becomes localized to the region $u_{Y_1},u_{Y_2} \to 0$ and factorizes as follows,
\begin{equation}
\mathcal{A}_n \to \Gamma(Y_1) \left(X^{(1)}\right)^{-Y_1}  \Gamma(Y_2) \left(X^{(2)}\right)^{-Y_2}  \times {\cal A}_{{\rm sub},1}\times {\cal A}_{{\rm sub},2}\times {\cal A}_{{\rm sub},3},
\end{equation}
where we now have two Regge factors, and the $\mathcal{A}_{{\rm sub},i}$ stand for the three lower-point amplitudes we obtain when we cut along $Y_1,Y_2$ (see Fig.~\ref{fig:Mut}, left).

It is now trivial to extend this result to the case where we keep fixed a full triangulation, $Y_i \in T$, and take every other chord large. In this case, all $\mathcal{A}_{{\rm sub},i}$ correspond to three-point amplitudes, and we simply obtain
\begin{equation}
\mathcal{A}_n \to \prod_{Y_i \in T} \Gamma(Y_i)\left(X^{(i)}\right)^{-Y_i}  ,\label{eq:Reggetri}
\end{equation}
where $X^{(i)}$ is the mutation of $Y_i$ inside the triangulation $T$ (see Fig.~\ref{fig:Mut}, right).

\subsubsection{Factorization and precision Regge scaling}
 We have already described the power-law scaling when one chord is taken to be large. But we can do much better than predicting the scaling in this limit; we can determine the precise behavior of the amplitude, which is determined by products of lower-point amplitudes with shifted kinematics, giving another example of the striking phenomenon of ``factorization away from poles'' seen for particle/string amplitudes. 

Consider again taking a single Mandelstam variable $X_{i,j}$ large. We start by looking at the first example we gave in Sec.~\ref{sec:singlechord}, but where instead we send $X_{1,n-1}\to \infty$. In this case, the integral will localize in the region where $u_{1,n-1}\to 1$ and thus at least one of the $u_{j,n}$ has to go to zero, so that the $u$ equation for $u_{1,n-1}$ is satisfied. Let us consider the region where $u_{J,n} \to 0$ and the remaining $u_{j,n}$ are generic. For this region, we find
\begin{equation}
\begin{aligned}
&\int (\text{rest})\vert_{y_{J,n}=0} \times \underbrace{\int \frac{\diff y_{J,n}}{y_{J,n}} u_{J,n}^{X_{J,n}}(1 - u_{J,n} \prod_{j\neq J} u_{j,n})^{X_{1,n-1}}}_{\to \Gamma(X_{J,n}) \left(\frac{1}{X_{1,n-1}\prod_{j\neq J} u_{j,n}}\right)^{X_{J,n}}}\\
 & \to \Gamma(X_{J,n}) \left(X_{1,n-1}\right)^{-X_{J,n}} \times \underbrace{\int \prod_{P\neq (J,n)} \frac{\diff y_P}{y_P} \prod_{(k,m)\in \mathcal{P}_{n-1}}u_{k,m}^{X_{k,m}-\delta_{(k,m),(j,n)}X_{J,n}}}_{\mathcal{A}^{(J)}_{L}(X_{j,n}\to X_{j,n}-X_{J,n})  \times \mathcal{A}^{(J)}_{R}(X_{j,n}\to X_{j,n}-X_{J,n}) },
\end{aligned}
\end{equation}
where we obtain precisely the lower-point amplitudes appearing in the factorization along $X_{J,n}$,  $\mathcal{A}^{(J)}_{L}(1,2,\cdots,J,n)$ and $\mathcal{A}^{(J)}_{R}(J,J+1,\cdots,n)$,  but evaluated for {\it shifted kinematics}. The full asymptotic form of the amplitude is then given by a sum over all such regions,
\begin{equation}
\mathcal{A}_n \to \sum_J \Gamma(X_{J,n}) \left(X_{1,n-1}\right)^{-X_{J,n}} \mathcal{A}^{(J)}_{L}( X_{j,n}-X_{J,n})\times  \mathcal{A}^{(J)}_{R}( X_{j,n}-X_{J,n}),
\end{equation}
where the sum comprises regions in which different $u_{J,n} \to 0$.

This picture generalizes if we send any  single $X_{i,j}$ large. Let us denote the set of chords crossing $(i,j)$ by ${\cal C}$. Then we have that ${\cal A}_n$ goes to  
\begin{equation}
\mathcal{A}_n \to \sum_{X_c\in \mathcal{C}} \!\Gamma(X_c) X_{i,j}^{-X_c} \mathcal{A}^{X_c}_{L}( \hat{X}^L_{k,m} )  
\times \mathcal{A}^{X_c}_{R}( \hat{X}^R_{k,m}), 
\label{eq:singleRegge}
\end{equation}
where
\begin{equation}\hat{X}^{L/R}_{k,m} = \begin{cases} X^{L/R}_{k,m} {-}  X_{c}, & \text{if } (k,m)\in \mathcal{C}\\
X^{L/R}_{k,m}, & \text{otherwise,}
\end{cases}
\end{equation}
and ${\cal A}_{L,R}^{X_{c}}$ denote the left and right amplitudes when cutting on $X_{c}$, on the kinematics $\hat{X}^{L,R}$.
For generic values of the constant $X_c$ variables, one term in Eq.~\eqref{eq:singleRegge} will dominate in the single Regge limit. However, near the spurious poles that occur in ${\cal A}_L$ or ${\cal A}_R$ at unphysical kinematics, the sum over the different sets of chords in Eq.~\eqref{eq:singleRegge} ensures that the Regge limit remains regular. We will numerically test Eq.~\eqref{eq:singleRegge} in Sec.~\ref{sec:numerics}.

\subsubsection{Dual Regge limit} 
Let us now consider a Regge limit that is in a sense dual to the one we derived in Eq.~\eqref{eq:Reggetri}.
That is, given some triangulation $T$, rather than keeping all Mandelstams corresponding to the chords of $T$ fixed, let us instead take them all {\it large}.
A particularly simple case is when we consider the triangulation to be \textit{ray-like}. If we make $X_{i^\star,j}$ large for all $j$, then all the respective $u_{X} \to 1$, and we force a single $u$ variable to go to zero, namely, that of the only curve that crosses all these chords, $X_{i^\star-1,i^\star+1}$. For a more general triangulation, $u_{X\in T} \to 1$ will localize on different regions where different $u$ variables go to zero, so just as before, to correctly predict the asymptotic behavior we must sum over all such regions. 
 
Let us look at a particular example at six points for the case of the triangulation $T=\{X_{1,3},X_{3,6},X_{4,6}\}$. When $u_{1,3},u_{3,6},u_{4,6} \to 1$, we can have the $u$ equations be satisfied either by sending $u_{2,5} \to 0$, or by sending both $u_{2,4}$ and $u_{1,5} \to 0$. We consider each of these regions in turn.

We start by looking at the case where $u_{2,5}\to 0$. From the $u$ equations, we have
\be
u_{1,3}= 1-u_{2,5}(u_{2,4}u_{2,6}), \quad u_{3,6}= 1-u_{2,5}(u_{2,4}u_{1,5}u_{1,4}), \quad u_{4,6}= 1-u_{2,5}(u_{3,5}u_{1,5}),
\ee
so at leading order when $u_{2,5}\to0$, the string integrand gives
\be
\Gamma(X_{2,5}) \left(\frac{1}{X_{1,3}u_{2,4}u_{2,6}+ X_{3,6}u_{2,4}u_{1,5}+ X_{4,6}u_{3,5}u_{1,5} }\right)^{X_{2,5}}  \times (\text{KN})^{(X_{2,5})}_{L} \times (\text{KN})^{(X_{2,5})}_{R},
\ee
where $(\text{KN})^{(X_{2,5})}_{L}$ and $ (\text{KN})^{(X_{2,5})}_{R}$ stand for the lower-point KN factors we obtain on the $X_{2,5}$ cut. Now for simplicity, let us take $X_{1,3} = X_{3,6}=X_{4,6}=X$, and note that on the support of $u_{2,5}=0$ we have that $u_{2,4}+u_{3,5}=1$ and $u_{1,5}+u_{2,6}=1$, which allow us to write $u_{2,4}u_{2,6}+ u_{2,4}u_{1,5}+ u_{3,5}u_{1,5}=1-u_{3,5}u_{2,6}$. Using 
\be
(1-u_{3,5}u_{2,6})^{-X_{2,5}} = \sum_{k=0}^\infty (-1)^k \binom{-X_{2,5}}{k}u_{3,5}^k u_{2,6}^k,
\ee
we find that the contribution to the Regge limit from $u_{2,5}\to 0$ is
\be
\Gamma(X_{2,5}) X^{-X_{2,5}} \sum_{k=0}^\infty (-1)^k \binom{-X_{2,5}}{k} \times \mathcal{A}_4(X_{2,4},X_{3,5}+k) \mathcal{A}_4(X_{1,5},X_{2,6}+k).\label{eq:Reggepre1}
\ee

Performing a similar analysis for the regime $u_{2,4},u_{1,5}\to 0$, for the string integral we find
\be
\Gamma(X_{2,4}) \left(\frac{1}{X_{1,3}u_{2,5}}\right)^{X_{2,4}} \Gamma(X_{1,5}) \left(\frac{1}{X_{4,6}u_{2,5}}\right)^{X_{1,5}} \tilde{(\text{KN})},
\ee
where $\tilde{(\text{KN})}$ is the product of the lower-point KN factors we obtain by going on the $X_{2,4},X_{1,5}$ cuts, which is simply a four-point factor in this case. Therefore we obtain the following Regge limit contribution in this regime:
\be
\Gamma(X_{2,4})X_{1,3}^{-X_{2,4}} \Gamma(X_{1,5}) X_{4,6}^{-X_{1,5}} \times \mathcal{A}_4(X_{1,4},X_{2,5}-X_{2,4}-X_{1,5}).\label{eq:Reggepre2}
\ee

The full Regge limit is then given by the sum of the two contributions in Eqs.~\eqref{eq:Reggepre1} and \eqref{eq:Reggepre2}, that is,
\be
\begin{aligned}
{\cal A}_6 &\rightarrow  \Gamma(X_{1,5})\Gamma(X_{2,4}) \mathcal{A}_4(X_{1,4},X_{2,5}-X_{2,4}-X_{1,5}) X^{-X_{1,5}-X_{2,4}} \\&\qquad + \Gamma(X_{2,5}) X^{-X_{2,5}} \sum_{k=0}^\infty (-1)^k \binom{-X_{2,5}}{k} \times \mathcal{A}_4(X_{2,4},X_{3,5}+k) \mathcal{A}_4(X_{1,5},X_{2,6}+k)
\end{aligned}
\ee
at large $X=X_{1,3}=X_{3,6}=X_{4,6}$.

\section{New Series Representations} \label{sec:recursive}

As we have stressed repeatedly, the integral representations for string amplitudes are rather formal, converging only for a narrow range of kinematics where all $X_{i,j}>0$, and even there, their numerical evaluation even for modestly large $n$ is not straightforward. It is clearly desirable, most importantly for conceptual but also for practical reasons, to find representations of string amplitudes that can be efficiently computed for all kinematics, or failing that, at least for wider kinematics than the one afforded by the integral representation. 

One way to do this is to  define the contour of integration to make the integral well defined by incorportating the $i\epsilon$ prescription \cite{WittenIEps,SL1,SLP2,ManschotIEps}. But while this works in principle, in practice it is not easy to evaluate the integrals when the kinematic invariants with negative real parts become even modestly large. And even for all positive kinematics we have found that in practice we cannot use these numerical methods to, e.g., explicitly check our predictions for exponentially small or Regge limits of the amplitudes. 

In this section we will instead use a variety of ideas to give new infinite series representations of string amplitudes. These do not converge for all kinematics, but do extend well beyond the $X>0$ region. Among other things, they will allow us to explicitly check the asymptotic behavior we established analytically in the previous sections. 

\subsection{Solving for $u$'s in terms of $u$'s} Instead of parameterizing the space of solutions of the $u$ equations in terms of the positive $y$ coordinates, we can also simply solve for $u$ variables in terms of some $(n-3)$-dimensional subset of $u$ variables. A particularly simple choice is to solve for all the $u$ variables in terms of those on a \textit{ray-like} triangulation. For example, at five points, we can solve for all $u$ variables in terms of $u_{1,3},u_{1,4}$. In addition, one can easily check that the measure $\diff y_{1,3}/y_{1,3} \diff y_{1,4}/y_{1,4}$ becomes $\diff u_{1,3}/[u_{1,3}(1-u_{1,3})] \diff u_{1,4}/[u_{1,4}(1-u_{1,4})]$, allowing us to write the amplitude as follows,
\begin{equation}
 \mathcal{A}_5 = \int_{0}^1 \frac{\text{d} u_{1,3}}{u_{1,3}(1{-}u_{1,3})}   \frac{\text{d} u_{1,4}}{u_{1,4}(1{-}u_{1,4})} u_{1,3}^{ X_{1,3}} u_{1,4}^{X_{1,4}} (1{-}u_{1,3})^{X_{2,4}} (1{-}u_{1,4})^{X_{3,5}}(1{-}u_{1,3}u_{1,4})^{-c_{2,4}}.
\end{equation}
Now expanding $(1-u_{1,3}u_{1,4})^{-c_{2,4}}$, we find
\begin{equation}
\begin{aligned}
 \mathcal{A}_5 &= \sum_{n=0}^{\infty} (-1)^n \begin{pmatrix} -c_{2,4} \\ n \end{pmatrix}  \int_{0}^1 \text{d} u_{1,3} u_{1,3}^{ X_{1,3}+n-1} (1{-}u_{1,3})^{X_{2,4}-1} \int_0^1 \text{d} u_{1,4} u_{1,4}^{X_{1,4}+n-1}  (1{-}u_{1,4})^{X_{3,5}-1} \\
 &= \sum_{n=0}^{\infty} (-1)^n \begin{pmatrix} -c_{2,4} \\ n \end{pmatrix} \mathcal{A}_4(X_{1,3}+n,X_{2,4})  \times \mathcal{A}_4(X_{1,4}+n,X_{3,5})\\
 &= \sum_{n=0}^{\infty} (-1)^n \begin{pmatrix} -c_{2,4} \\ n \end{pmatrix} \frac{\Gamma(X_{1,3}+n)\Gamma(X_{2,4})}{\Gamma(X_{1,3}+n+X_{2,4})} \times \frac{\Gamma(X_{1,4}+n)\Gamma(X_{3,5})}{\Gamma(X_{1,4}+n+X_{3,5})}.
\end{aligned}
\label{eq:5ptExpansionU}
\end{equation}
Thus, we have managed to write the five-point amplitude as a sum of shifted four-point amplitudes. 

Let us postpone the discussion of the domains of convergence of this expansion for a moment and look at a six-point example. Choosing ray-like triangulation $\{(1,3),(1,4),(1,5)\}$, we can solve for six-point $u$ variables and write the amplitude as follows,
\begin{equation}
\begin{aligned}
    \mathcal{A}_6 = \int_0^1 \prod_{i=3,4,5}\frac{\text{d} u_{1,i}}{u_{1,i}(1-u_{1,i})} &u_{1,3}^{X_{1,3}}u_{1,4}^{X_{1,4}}u_{1,5}^{X_{1,5
    }} (1-u_{1,3})^{X_{2,4}}  (1-u_{1,4})^{X_{3,5}} (1-u_{1,5})^{X_{4,6}}\\
    & (1-u_{1,3}u_{1,4})^{-c_{2,4}}(1-u_{1,4}u_{1,5})^{-c_{3,5}} (1-u_{1,3}u_{1,4}u_{1,5})^{-c_{2,5}}.
\end{aligned}
\end{equation}

Expanding the polynomials with exponents $c_{3,5}$ and $c_{2,5}$,  we obtain
\begin{equation}\hspace{-6mm}
\begin{aligned}
    \mathcal{A}_6 &=\!\!\sum_{k_{2,5},k_{3,5}=0}^{\infty} \!\!(-1)^{k_{2,5}+k_{3,5}} \begin{pmatrix} -c_{3,5} \\ k_{3,5} \end{pmatrix} \begin{pmatrix} -c_{2,5} \\ k_{2,5} \end{pmatrix} \int_0^1 \!\prod_{i=3,4,5}\!\!\frac{\text{d} u_{1,i}\,u_{1,3}^{X_{1,3}{+}k_{2,5}}u_{1,4}^{X_{1,4}{+}k_{2,5}{+}k_{3,5}}}{u_{1,i}(1{-}u_{1,i})}\times\\ 
    &  \quad \quad \quad \times u_{1,5}^{X_{1,5}+k_{2,5}+k_{3,5}} (1{-}u_{1,3})^{X_{2,4}}  (1{-}u_{1,4})^{X_{3,5}} (1{-}u_{1,5})^{X_{4,6}}
     (1{-}u_{1,3}u_{1,4})^{-c_{2,4}}\\     &=\sum_{k_{2,5},k_{3,5}=0}^{\infty} (-1)^{k_{2,5}+k_{3,5}} \left(\begin{array}{c} -c_{3,5} \\ k_{3,5} \end{array} \right) \left(\begin{array}{c} -c_{2,5} \\ k_{2,5} \end{array} \right) \mathcal{A}_4(X_{1,5}+k_{2,5}+k_{3,5},X_{4,6}) \times \\
     &  \quad \quad \quad \times \mathcal{A}_5(X_{1,3}+k_{2,5},X_{1,4}+k_{2,5}+k_{3,5},X_{2,4},X_{2,5},X_{3,5}),
\end{aligned}\hspace{-10mm}
\end{equation}
which then gives us the six-point amplitude in terms of shifted lower-point amplitudes. Of course, we could further write the five-point amplitude in terms of shifted four-point, in which case we express the six-point ampllitude purely in terms of four-point. This is equivalent to binomial expanding the powers of $(1 - u_{1,3} u_{1,4})$, $(1 - u_{1,4} u_{1,5})$, and $(1 - u_{1,3} u_{1,4} u_{1,5})$, yielding 
\be 
\begin{aligned}
{\cal A}_6 &= \sum_{k_{2,4},k_{3,5},k_{2,5}} (-1)^{k_{2,4}} \begin{pmatrix} -c_{2,4} \\ k_{2,4} \end{pmatrix} (-1)^{k_{3,5}} \begin{pmatrix} -c_{3,5} \\ k_{3,5} \end{pmatrix} (-1)^{k_{2,5}} \begin{pmatrix} -c_{2,5} \\ k_{2,5} \end{pmatrix}\times \\ &\times  {\cal A}_4(X_{1,3} {+} k_{2,4} {+} k_{2,5}, X_{2,4}) {\cal A}_4(X_{1,4} {+} k_{2,4} {+} k_{3,5} {+} k_{2,5}, X_{3,5}) {\cal A}_4(X_{1,5} {+} k_{3,5} {+} k_{2,5},X_{4,6}).
\end{aligned}\label{eq:6pointbigsum}
\ee

\subsection{Subsurface $u$ parametrization and amplitude recursion}

We can continue in this way and directly solve for $u$ variables in terms of a maximal set of compatible $u_{i,j}$, but we can do this in a conceptually cleaner  way that naturally leads to a recursive computation of the amplitude. Consider any chord $(i,j)$ that cuts the polygon into left and right pieces. We recall the fundamental fact that the $u$ variables for a curve living on a   subsurface is given by the ``extension formula,'' as the product over the $u$ variables for all the ways the curve can be extended into the full surface \cite{Splits}. It is thus natural to try and parametrize the $u$ variables of the full surface in terms of the $u$ variables $U_L,U_R$ for the left and right pieces together with $u_{i,j}$ itself. 

Let us see how this can be done in the simplest case where we cut on the chord $(1,n-1)$, so we just have the lower $(n-1)$-gon surface. Note that the $u$ variables for the lower-point problem are given by the extension formulae as 
\begin{equation}
U_{i,n-1} = u_{i,n-1} u_{i,n}; \quad  {\rm all \, other}\; U_{i,j}=u_{i,j} \, \, {\rm for} \,\, i,j<n-1.
\end{equation}

Our goal is to express all the $u_{i,j}$ in terms of the $U$ variables and $u_{1,n-1}$. To begin with, the $u$ equation for the chord $(2,n)$ tells us that 
\begin{equation}
u_{2,n} = 1 - u_{1,3} u_{1,4} \cdots u_{1,n-1}.
\end{equation}

Next, we consider the extended $u$ equation for $u_{2,n} u_{3,n}$,
\begin{equation}
u_{2,n} u_{3,n} = 1 - u_{1,4} u_{1,5} \cdots u_{1,n-1}.
\end{equation}

Since we have already solved for $u_{2,n}$ this permits us to solve for $u_{3,n}$ as 
\begin{equation}
u_{3,n} = \frac{1 - u_{1,4} \cdots u_{1,n-1}}{1 - u_{1,3} \cdots u_{1,n-1}},
\end{equation}
and continuing in this way, the extended $u$ equation for $u_{2,n} \cdots u_{i,n}$ lets us solve for 
\begin{equation}
u_{i,n} = \frac{1 - u_{1,i+1} \cdots u_{1,n-1}}{1 - u_{1,i} \cdots u_{1,n-1}}.
\end{equation}

This now allows us to solve for $u_{i,n-1}$ via 
\begin{equation}
u_{i,n-1} = \frac{U_{i,n-1}}{u_{i,n}} = \frac{U_{i,n-1} (1 - u_{1,i} \cdots u_{1,n-1})}{1 - u_{1,i+1} \cdots u_{1,n-1}}.
\end{equation}

We can now express the KN factor for the $n$-point amplitude in a  nicely factorized form, as 
\begin{equation}
\begin{aligned}
\prod_{p=1}^{n-3} u_{p,n-1}^{X_{p,n-1}} \prod_{q=2}^{n-2} u_{q,n}^{X_{q,n}} = &\,u_{1,n-1}^{X_{1,n-1}} \times 
\prod_{r = 2}^{n-2} U_{r,n-1}^{X_{r,n-1}} \times(1 - u_{1,3} \cdots u_{1,n-1})^{-c_{2,n-1}} \\
&\quad (1 - u_{1,4} \cdots u_{1,n-1})^{-c_{3,n-1}} \cdots \times (1 - u_{1,n-1})^{X_{n-2,n}}.
\end{aligned}
\end{equation}

Furthermore, it is easy to see that the Parke-Taylor form also factorizes nicely, as 
\begin{equation}
\omega^{(n)} = \omega^{(n-1)} \times \frac{ d u_{1,n-1}}{u_{1,n-1} (1 - u_{1,n-1})}.
\end{equation}

Binomial expanding all the factors containing $u_{1,n-1}$, other than $u_{1,n-1}^{X_{1,n-1}}$ and $(1 - u_{1,n-1})^{X_{n-2,n}}$, gives us a recursive expression in terms of products of shifted $(n-1)$-point amplitudes and four-point amplitudes: 
\begin{equation}
\begin{aligned}
{\cal A}_n = \sum_{k_{2,n-1},\cdots, k_{n-3,n-1}} \prod_j (-1)^{k_{j,n-1}} \binom{-c_{j,n-1}}{k_{j,n-1}}& \times {\cal A}_4(X_{1,n-1} + \sum_{j=2}^{n-3} k_{j,n-1},X_{n-2,n}) \\
&\times {\cal A}_{n-1}(\hat{X}_{1,p} = X_{1,p} + \sum_{j<p} k_{p,n-1}).
\end{aligned}
\end{equation}

We can recurse this equation until we are left with a sum of products of four-point amplitudes weighted by binomial coefficients and evaluated at shifted kinematics.  

\subsection{Domain of convergence for infinite series}

Let us consider the six-point case of the sum in Eq.~\eqref{eq:6pointbigsum} as an example. We note that each ${\cal A}_4(X,Y)=\Gamma(X) \Gamma(Y)/\Gamma(X+Y)$ appearing in the sum manifests some poles; indeed all the poles in $X_{1,3},X_{2,4},X_{1,4},X_{3,5},X_{1,5},X_{4,6}$ are manifest in this representation. But the poles in $X_{2,5},X_{2,6},X_{3,6}$ are not manifest, and must arise from the infinite sums. This observation already tells us to expect that our infinite series representation cannot be well defined for all kinematics, and must minimally require $X_{2,5},X_{2,6},X_{3,6} > 0$ for convergence. In fact, something even stronger is needed. Using the large-$k$ asymptotics for the binomial coefficients, 
\begin{equation}(-1)^k \left(\begin{smallmatrix} a\\ k \end{smallmatrix}\right) \to \frac{k^{-a - 1}}{\Gamma(-a)} \quad  \text{and} \quad  {\cal A}_4(X + k, Y) \to \Gamma(Y) k^{-Y},
\end{equation}
at large $k_{2,4},k_{2,5},k_{3,5}$---which we will write as $k_{1,2,3}$ here for brevity---the infinite series in Eq.~\eqref{eq:6pointbigsum} is well approximated by 
\begin{equation}
\begin{aligned}
&{\cal A}_6 \xrightarrow{{\rm large}\, k_{1,2,3}} \\& \frac{\Gamma(X_{2,4}) \Gamma(X_{3,5}) \Gamma(X_{4,6})}{\Gamma(c_{2,4}) \Gamma(c_{3,5}) \Gamma(c_{2,5})}   \sum_{k_1,k_2,k_3}  \frac{k_1^{c_{2,4}} k_2^{c_{3,5}} k_3^{c_{2,5}}}{k_1 k_2 k_3} (k_1 {+} k_3)^{-X_{2,4}} (k_1 {+} k_2 {+} k_3)^{-X_{3,5}}(k_2 {+} k_3)^{-X_{4,6}} .
\end{aligned}\label{eq:6pointbigsumapprox}
\end{equation}

There are some obvious domains where this sum converges. For instance, clearly if $c_{2,4}
,c_{2,5},c_{3,5}<0$ as well as $X_{2,4},X_{3,5},X_{4,6}>0$, every term is damped for large $k$ and the sum is highly convergent. But of course this is overkill. In order to study the true domain of convergence of this series, we can set $k_i = e^{t_i}$, convert the sums to integrals, and study the asymptotics in $t_{1,2,3}$ space for all $3! = 6$ orderings of the $t_i$. The result is the following interesting region of convergence: 
\be 
\begin{aligned}
X_{2,5},X_{2,6},X_{3,6}&>0\\
-X_{3,5}+X_{3,6} + X_{2,5}&>0\\
-X_{2,5} + X_{2,4} + X_{2,6} + X_{3,5}&>0  \\ -X_{3,6} + X_{2,6} + X_{3,5} + X_{4,6}&>0\\ X_{2,4} - X_{2,5} + X_{2,6} + 2 X_{3,5} - X_{3,6} + X_{4,6}&>0 .
\end{aligned}
\ee
These inequalities guarantee that the sum over the $k_i$ is convergent in all directions. It is of course also possible that there are cancellations between different regions of the sum that could lead to a wider domain of convergence, but this conservative region is already interesting and allows us to evaluate the amplitude far away from the all $X>0$ region where the integral representation is convergent. 

For general $n$, this representation of the amplitude certainly converges in a much wider region of $X$ space than the integral representation that demands all $X>0$. But it is true that at large $n$, we must still demand that most of the $X$ variables are positive. For instance, recursing all the way down to the product of four-point amplitudes, the $\Gamma$ functions can capture all the poles in $X_{1,3},\ldots,X_{1,n-1}$ and $X_{2,n}, \ldots, X_{n-2,n}$, but none of the other poles are manifestly present; the series converges only when all the rest of the $X_{i,j}$ are positive. Of course we are winning over the integral representation by having a rapidly converging, analytic expression, but conceptually, the extension away from $X>0$ is being driven largely by the analytic continuation of the four-point amplitude given by the beta function. We will later give an exact analytic expression for the five-point amplitude, and this can be used to further extend the range of validity of our series representations. But at large $n$, there are $O(n^2)$  $X$ variables, and $O(n^2)$ of them must be positive; only $O(n)$ of them can have either sign. Some new ideas are needed to find series representations where all, or at least $O(n^2)$, of the $X$ variables can have either sign. 

\subsection{Dual resonance} 

A powerful tool for computing tree-level  $n$-point string amplitudes is the miraculous property of dual resonance, which says that the amplitude can be expressed as a sum of terms {\it purely in a single channel}.  Hence, the residues in that single channel are sufficient to define the entire amplitude.  Dual resonance is a hallmark of string theory, evoking the iconic picture of a multi-legged string diagram whose exchanges can all be sculpted into a single channels by deforming the worldsheet.

The dual resonant representations follow from the statement that string amplitudes vanish in certain high-energy limits.  At four point, for example, the open string amplitude vanishes in the Regge limit, $\lim_{|s|\rightarrow\infty} {\cal A}_4^{\rm tree} =0$ at fixed $t>0$.  We can then use an unsubtracted dispersion relation in the $s$ channel~\cite{Cheung:2023adk} to recast the amplitude in the dual resonant form,
\be 
{\cal A}_4^{\rm tree} = \sum_{n=0}^\infty \frac{R_n(t)}{s+n},\label{eq:DR}
\ee 
where the residues are $R_n(t) = \prod_{k=1}^n \left(1-\frac{t}{k}\right)$.

The same logic applies at five point, where we have $\lim_{|X_{1,3}|,|X_{1,4}|\rightarrow\infty} {\cal A}_5^{\rm tree} =0$ at fixed $X_{2,4},X_{2,5},X_{3,5}>0$.  In this case the dual resonant form is
\be
{\cal A}_5^{\rm tree} = \sum_{m,n=0}^\infty \frac{R_{m,n}(X_{2,4},X_{3,5},X_{2,5})}{(X_{1,3}+m)(X_{1,4}+n)},\label{eq:DR5}
\ee
where the residue is
\begin{equation}
\begin{aligned}R_{m,n}(X_{2,4},X_{3,5},X_{2,5}) &=\frac{(1-X_{2,4})_{m}(1-X_{3,5})_{n}}{m!n!}\,{}_{3}F_{2}\!\left[\!\begin{array}{c}
{-}m,{-}n,X_{2,4}{+}X_{3,5}{-}X_{2,5}\\
{-}m{+}X_{2,4},{-}n{+}X_{3,5}
\end{array};1\right]\!,\!
\end{aligned}
\label{eq:res}\hspace{-1mm}
\end{equation}
defining the Pochammer symbol $(a)_{n}=\Gamma(a+n)/\Gamma(a)$.

There is an easy argument establishing the dual resonant representation at all $n$, as an expansion in poles associated with any triangulation. Let us start at $n=5$ and think of the amplitude as a function of $X_{1,4}$.
If the crossing chords $X_{2,5}, X_{3,5}$ are positive, then by single-chord Regge (Sec.~\ref{sec:singlechord}), when $X_{1,4}$ gets large the amplitude vanishes at infinity. Therefore as usual from Cauchy we can 
write it as a sum over $X_{1,4}$ poles, where the residues are given in terms of lower (four-point) amplitudes where $X_{j,4}$ are shifted by {\it positive} integers (as given explicitly in Eq.~\eqref{eq:resX14}). This allows us to trivially continue and recursively express the four-point amplitudes as well, in any channel we like. The reason is simply that the vanishing at infinity in any $X$ variable only depends on the positivity of the $X$ variables crossing it. In this way, given any triangulation $T$, we simply ask for all the $X$ variables of the chords {\it not} in the triangulation to be positive, and we can successively compute the amplitude recursively, at all steps only shifting kinematics further positive, until we arrive at the dual resonant representation as a sum over residues on all massive poles of $T$. 

Given that we have learned to explicitly compute the residues on any massive poles, we can explicitly write down dual resonant representations involving sum over poles of arbitrary partial triangulations. 

For instance there is a recursive form where we sum over poles only in a single channel, say for simplicity over the collinear channel where $X_{1,n-1}=-n_{1,n-1}$,
\begin{equation}
{\cal A}_n = \sum_{n_{1,n-1}} \frac{R_{n_{1,n-1}}}{X_{1,n-1} + n_{1,n-1}}
\end{equation}
with $R_{n_{1,n-1}}$ given in Eq.~\eqref{eq:colR}. 
We can proceed to the opposite extreme, summing over all poles in a given triangulation $T$ as 
\begin{equation}
{\cal A}_n = \sum_{n_i} R_{n_1, \cdots, n_{n-3}} \times \prod_{i \in T} \frac{1}{X_i + n_i}\label{eq:DRfull}
\end{equation}
where the residues can be read off from the multinomial expansion of the ${\cal F}$ polynomials as explicitly shown in Sec.~\ref{sec:fact}. 

\subsection{Comparing the subsurface and dual resonant representations}

It is interesting to compare the two methods we have given for recursively determining the amplitude, using factorization to give the dual resonant forms, and the subsurface parametrization of $u$ variables. Both ideas involve a factorizing/recursive structure, for the $u$ variables directly or for the expansion of ${\cal F}$ polynomials, and the expressions look very similar. But they are not quite identical, and the equality between them is an interesting identity for the amplitude. We can see this already in the simplest example of $n=5$. Using the subsurface $u$ parametrization, we can extract the residue from \eqref{eq:5ptExpansionU} on the pole where $X_{1,4} \to -n_{1,4}$, which gives 
\begin{equation}
R^{(u)}_{n_{1,4}} = \sum_{k_{2,4}=0}^{n_{1,4}} (-1)^{k_{2,4}} \binom{-c_{2,4}}{k_{2,4}} {\cal A}_4(X_{1,3} + k_{2,4},X_{2,4}) \binom{n_{1,4} - k_{2,4} - X_{3,5}}{n_{1,4} - k_{2,4}}.
\end{equation}

We can compare this with the same residue computed using the factorization of the ${\cal F}$ polynomials, which determines our dual resonant expansion, \eqref{eq:resX14}
\begin{equation}
R^{({\cal F})}_{n_{1,4}} = \sum_{k_{2,4}=0}^{n_{1,4}} \binom{-c_{2,4}}{k_{2,4}} {\cal A}_4(X_{1,3}, X_{2,4} + k_{2,4}) \binom{n_{1,4} + X_{2,4} - X_{2,5}}{n_{1,4} - k_{2,4}}.
\end{equation}

Despite their similarities, the expressions are not manifestly equal; nonetheless they are of course equal. This shows that the ``subsurface'' and ``factorization'' recursions give in general different series expansions, which are non-trivially equal to the same final amplitude. 

We note that in comparison with the ``$u$'s in terms of $u$'s'' series representation above, the dual resonant form has a more restricted domain of convergence. We have to assume that {\it all} the $X$ variables not in the triangulation are positive. For general $n$, the $u$-recursive form allows $2(n-3)$ of the $X$ variables to have any sign, while the dual resonant form only allows the $(n-3)$ $X$ variables in the triangulation to have any sign.

\section{Five Point in Full}\label{sec:hypergeometric}

In this section, we will make use of some special identities satisfied by various generalized hypergeometric functions---specifically the Thomae and Whipple identities---to extend the domain of convergence of string amplitudes, allowing us to evaluate them in broader regions of kinematic space than are evaluable using the standard KN formula.

We first tackle this problem for the case of five-point scattering, where we will find complete success: a new form of the amplitude that is evaluable {\it everywhere}, for arbitrary kinematics, in direct analogy with what the Euler beta function~\eqref{eq:Euler} accomplishes at four point. 

In the five-point KN formula of Eq.~\eqref{eq:KN}, we recognize the integral definition of the ${}_3 F_2$ generalized hypergeometric function,\footnote{This double-integral identity can be obtained from one application of Euler's integral transform to relate ${}_3 F_2$ to an integral of $_{2} F_1$, Gauss's summation theorem to relate the latter to a ratio of gamma functions, and finally the integral form of the Euler beta function to recast the gamma functions as a second integral.}
\be 
{\cal A}_5=\frac{\Gamma(X_{1,3})\Gamma(X_{2,4})\Gamma(X_{3,5})\Gamma(X_{1,4})}{\Gamma(X_{1,3}+X_{2,4})\Gamma(X_{3,5}+X_{1,4})} {}\,_{3}F_{2}\left[\begin{array}{c}
X_{1,3},\,X_{1,4},\,X_{2,4}+X_{3,5}-X_{2,5}\\
X_{1,3}+X_{2,4},\,X_{3,5}+X_{1,4}
\end{array};1\right],\label{eq:5ptstandard}
\ee 
as has been recognized since dawn of string theory~\cite{Bialas:1969jz}.
The form of the amplitude in Eq.~\eqref{eq:5ptstandard} converges (modulo physical poles) if and only if $X_{2,5} > 0$. Note that the expression we found via the series representation of $u$'s in terms of $u$'s in Eq.~\eqref{eq:5ptExpansionU} also allow us to derive Eq.~\eqref{eq:5ptstandard} via the series representation of the ${}_3 F_2$.

Physically, this amplitude should be invariant under cyclic permutations of the external particles, i.e., ${\cal A}_5 (X_{1,3},X_{1,4},X_{2,4},X_{2,5},X_{3,5})$ should equal ${\cal A}_5 (X_{2,4},X_{2,5},X_{3,5},X_{1,3},X_{1,4})$ in the domain where they both converge.
At the level of the KN integral, we can implement this cyclic transformation via a simple change of integration variables~\cite{Nakanishi:1971ve,Fairlie:1970di}. Iterating, we find the full orbit of cyclic transformations of Eq.~\eqref{eq:5ptstandard}, with domains of convergence $X_{1,3}>0$, $X_{2,4}>0$, $X_{3,5}>0$, $X_{1,4}>0$, or $X_{2,5}>0$,
which give expressions for the amplitude that agree where domains overlap.
There remains an infinite wedge unaccounted for in the space of Mandelstams, where all the planar invariants are $<0$, representing $1/32$ of the domain.
One can show that it is impossible to choose {\it real} momenta in a {\it Lorentzian} spacetime for which this occurs, so Eq.~\eqref{eq:5ptstandard} and its cyclic cousins are sufficient to describe all real Lorentzian $2\rightarrow 3$ processes at tree level in open string theory.
However, in characterizing scattering amplitudes, we often have occasion to go to unphysical kinematics and inquire into the behavior of amplitudes outside of these narrow physical requirements; for that reason, it would be useful to have an explicit form of the five-point string amplitude that can be evaluated at {\it arbitrary} values of the Mandelstams, regardless of sign. It is precisely such a representation that we discover in this section.

A virtue of the hypergeometric representation of the amplitude in Eq.~\eqref{eq:5ptstandard} is that it makes manifest properties that are hard to see at the level of the integral, even accounting for cyclic invariance and the $\mathbb{Z}_2$ flip symmetry of the factorization channel.
For example, the generalized hypergeometric amplitude is symmetric on separate permutations of its upper and lower parameters, so 
\be
\begin{aligned}
&\Gamma(X_{1,3}+X_{2,4}-X_{1,4})\Gamma(X_{3,5}+X_{1,4}-X_{1,3})\times {\cal A}_5 (X_{1,3},X_{1,4},X_{2,4},X_{2,5},X_{3,5})
\\&= \Gamma(X_{2,4})\Gamma(X_{3,5}) \times {\cal A}_5(X_{1,3},X_{1,4},X_{3,5}{+}X_{1,4}{-}X_{1,3},X_{2,5},X_{1,3}{+}X_{2,4}{-}X_{1,4}).
\end{aligned}
\ee
Note that for notational convenience, we are representing the five-point amplitude as a function ${\cal A}_5(a_1,a_2,a_3,a_4,a_5)$, where the $a_i$ are related to the kinematics as most obviously read off from the mesh picture moving from the bottom to the top, i.e., $a_1 = X_{1,3}$, $a_2=X_{1,4}$, $a_3=X_{2,4}$, $a_4=X_{2,5}$, and $a_5=X_{3,5}$. Thus, in the right-hand side of the equation above, we mean that we are evaluating the five-point amplitude at kinematics $x_{i,j}$ with $x_{1,3}=X_{1,3}, x_{1,4} = X_{1,4}, x_{2,4} = X_{3,5} + X_{1,4} - X_{1,3}, x_{2,5} = X_{2,5}, x_{3,5} = X_{1,3} + X_{2,4} - X_{1,4}$.\footnote{We will follow this convention for ordering the $X_{i,j}$ variables for all $n$. For instance, at six points we will write ${\cal A}_6(a_1, a_2, \ldots, a_9) = {\cal A}_6(X_{1,3},X_{1,4},X_{1,5},X_{2,4},X_{2,5},X_{2,6},X_{3,5},X_{3,6},X_{4,6})$.} 

Acting repeatedly with this transformation in combination with cyclic invariance, we arrive at a striking identity, which is best expressed in terms of the $X_{i,j}$ and $c_{i,j}$:
\be 
\frac{{\cal A}_5(X_{1,3},X_{1,4},X_{2,4},X_{2,5},X_{3,5})}{\Gamma(X_{1,3})\Gamma(X_{2,4})\Gamma(X_{3,5})\Gamma(X_{1,4})\Gamma(X_{2,5})} = \frac{{\cal A}_5(c_{1,3},c_{2,5},c_{1,4},c_{3,5},c_{2,4})}{\Gamma(c_{1,3})\Gamma(c_{1,4})\Gamma(c_{2,4})\Gamma(c_{2,5})\Gamma(c_{3,5})}.
\ee
This identity makes the hidden zeros found in Ref.~\cite{Zeros} manifest, as when we pick $i^\star$ and set $c_{i^\star,j}=-n_{i^\star,j}$, for $n_{i^\star,j}$ positive integers, on the right-hand side the amplitude does not have a pole, since we are going simultaneously on two incompatible poles, but the denominator blows up. Therefore the right-hand side gives us zero and automatically implies that ${\cal A}_5(X_{1,3},X_{1,4},X_{2,4},X_{2,5},X_{3,5})$ must vanish in this locus.

\subsection{First kinematic extension from Thomae relations}

The generalized hypergeometric functions, and the so-called Clausenian hypergeometric function ${}_3 F_2(1)$ in particular, have been of interest to mathematicians for centuries, and possess a rich web of identities and transformations that is still being explored~\cite{Milgram,Bailey}. 
Notably, the Thomae transformation~\cite{Thomae,Bailey} allows for a reshuffling of the upper and lower parameters of ${}_3 F_2$. Concretely, the Thomae transformation can be written as
\be
\begin{aligned}
{}_{3}F_{2}\left[\begin{matrix}
a_{1},\,a_{2},\,a_{3}\\
b_{1},\,b_{2}
\end{matrix};1\right]&=\frac{\Gamma(b_{1})\Gamma(b_{1}{+}b_{2}{-}a_{1}{-}a_{2}{-}a_{3})}{\Gamma(b_{1}{-}a_{1})\Gamma(b_{1}{+}b_{2}{-}a_{2}{-}a_{3})} {}_{3}F_{2}\left[\begin{matrix}
a_{1},\,b_{2}-a_{2},\,b_{2}-a_{3}\\
b_{2},\,b_{1}+b_{2}-a_{2}-a_{3}
\end{matrix};1\right],
\end{aligned}
\ee
which holds when $b_{1}+b_{2}-a_{1}-a_{2}-a_{3}>0$ and $b_{1}-a_{1}>0$.
Repeated application of the Thomae transformation, along with  the symmetries on the $a_i$ and $b_i$ parameters, allows for the derivation of yet more identities, e.g., 
\be 
\begin{aligned}
{}_{3}F_{2}\left[\begin{matrix}
a_{1},\,a_{2},\,a_{3}\\
b_{1},\,b_{2}
\end{matrix};1\right]&= \frac{\Gamma(b_{1})\Gamma(b_{2})\Gamma(b_{1}{+}b_{2}{-}a_{1}{-}a_{2}{-}a_{3})}{\Gamma(a_{1})\Gamma(b_{1}{+}b_{2}{-}a_{1}{-}a_{2})\Gamma(b_{1}{+}b_{2}{-}a_{1}{-}a_{3})}\times \\&\qquad\times  {}_{3}F_{2}\left[\begin{matrix}
b_{1}{-}a_{1},\,b_{2}{-}a_{1},\,b_{1}{+}b_{2}{-}a_{1}{-}a_{2}{-}a_{3}\\
b_{1}{+}b_{2}{-}a_{1}{-}a_{2},\,b_{1}{+}b_{2}{-}a_{1}{-}a_{3}
\end{matrix};1\right].
\end{aligned}
\ee

Accounting for the symmetry of the orderings of the sets of upper $a_i$ and lower $b_i$ parameters in the ${}_3 F_2$, there are a priori 12 different ways of applying the two Thomae transformations above to our standard ${}_3 F_2$ form of the amplitude.
Doing so, and after modding out by the ordering redundancy of hypergeometric parameters in the output, we find ten distinct representations of the amplitude, five of which comprise our standard ${}_3 F_2$ form and its cyclic relatives. Thus, one of the Thomae transformations instantiates cyclic permutation of the amplitude.
The other transformation gives us a new representation of the five-point amplitude,
\be 
\begin{aligned}
{\cal A}_5 &=\frac{\Gamma(X_{1,3})\Gamma(X_{2,4})\Gamma(X_{3,5})\Gamma(X_{1,4})\Gamma(X_{2,5})}{\Gamma(X_{2,5}+X_{1,3})\Gamma(X_{1,4}+X_{2,5})\Gamma(X_{2,4}+X_{3,5}-X_{2,5})}\times \\& \qquad \times {}_{3}F_{2}\left[\begin{array}{c}
X_{2,5}+X_{1,3}-X_{3,5},\,X_{1,4}+X_{2,5}-X_{2,4},\,X_{2,5}\\
X_{2,5}+X_{1,3},\,X_{1,4}+X_{2,5}
\end{array};1\right],\label{eq:5ptThomae}
\end{aligned}
\ee
along with its four cyclic permutations. These converge on a different domain than Eq.~\eqref{eq:5ptstandard}, namely $X_{2,4}+X_{3,5}-X_{2,5}>0$---that is, $c_{2,4}>0$---and its cyclic relatives; as before, on overlapping regions of convergence, these new expressions agree with each other and with the cyclic orbit of Eq.~\eqref{eq:5ptstandard}.
A beautiful feature of this representation of the amplitude obtained from the Thomae transformation is that, unlike the standard form \eqref{eq:5ptstandard}, {\it all} of the physical poles are manifest in Eq.~\eqref{eq:5ptThomae}; that is, we have a factor of $\Gamma(X_{1,3})\Gamma(X_{2,4})\Gamma(X_{3,5})\Gamma(X_{1,4})\Gamma(X_{2,5})$. It is an interesting question whether generalizations of these relations to higher-point string amplitudes might give us access to representations that make all of the physical poles manifest. As we explain in Sec.~\ref{sec:Thomae_Splits}, understanding the generalization of Thomae transformations from a different perspective lets us conclude that they are not sufficient to achieve this at higher points for general kinematics.

Together, the above representations of the amplitude cover all of the kinematic space except for a (still infinite) sliver in which all of the inequalities are violated, representing $1/1024$ of the space.
This remaining wedge of the kinematic space---in which $X_{1,3}$, $X_{2,4}$, $X_{3,5}$, $X_{1,4}$, $X_{2,5}$, $c_{1,3}$, $c_{2,4}$, $c_{3,5}$, $c_{1,4}$, and $c_{2,5}$ are all $<0$---must still be understood. 

In our analysis here, we have committed the cardinal physics sin of using an obscure mathematical identity as a black box, without understanding where it comes from conceptually. Indeed the early proofs of Thomae in the literature involve a large sequence of manipulations on infinite sums, with little insight and no clue for how these might be generalized to higher points. The situation changed in the work of Brown \cite{FBrown}, who did provide a deeper explanation of Thomae, which made use of the representation of the string integrals using $u$ variables. In a moment, we will show that this understanding is in fact a special case of the very general phenomenon of  ``split'' factorization of string amplitudes away from poles recently seen in Refs.~\cite{Zeros,Splits}. This observation will immediately allow us to find generalizations of the Thomae identity at all $n$, though as we we will see, for $n\,{>}\,5$ this will give us identities on restricted kinematics. But before constructing this generalization, we will sin again, by using a {\it different} obscure hypergeometric identity to give us full control of the five-point amplitude.

\subsection{Full kinematic extension: the Whipple identity}

\begin{figure}[h]
	\centering
	\includegraphics[width=0.84\linewidth]{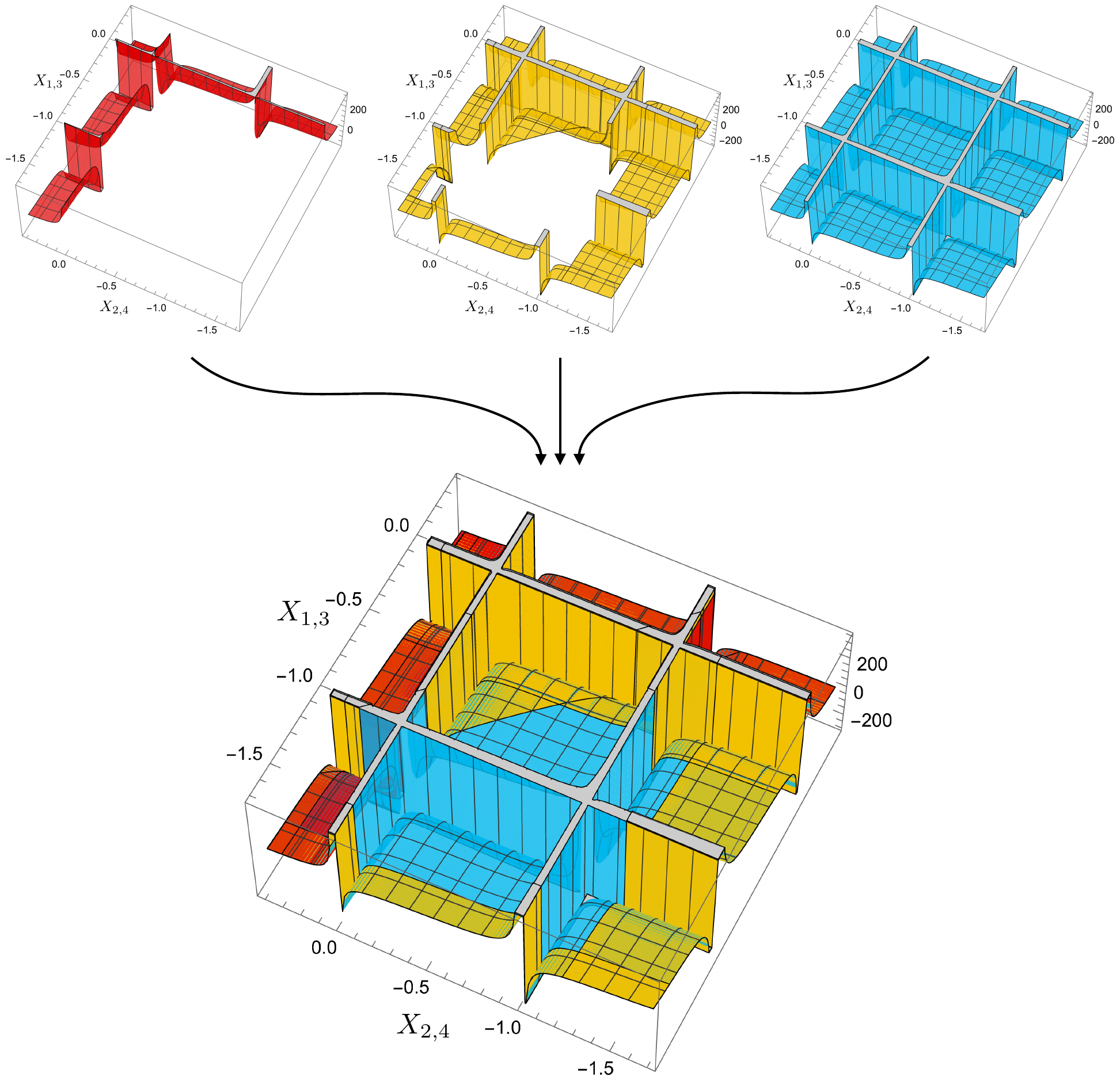}
	\caption{Illustration of our extension of the five-point string amplitude to arbitrary kinematics. We fix $(X_{3,5},X_{1,4},X_{2,5})=(-0.7,-0.8,-0.4)$. The standard form of the amplitude (red)---Eq.~\eqref{eq:5ptstandard} and its cyclically permuted cousins---covers the region where $\max\{X_{1,3},X_{1,4},X_{2,4},X_{2,5},X_{3,5}\}>0$.
 The form obtained via Thomae transformations (yellow)---Eq.~\eqref{eq:5ptThomae} and its cyclic analogues---covers the space where $\max\{c_{1,3},c_{1,4},c_{2,4},c_{2,5},c_{3,5}\}>0$. Remarkably, the form we find via Whipple's identity (blue) in Eq.~\eqref{eq:5ptWhipple} covers the {\it entire} space, giving the five-point string amplitude for arbitrary kinematics for the first time.}
	\label{fig:5pt}
\end{figure}

To find the string amplitude in the forbidden region, we make use of Whipple's identity~\cite{Whipple,Bailey}, which relates a ${}_3 F_2$ hypergeometric function evaluated at $+1$ to a ${}_6 F_5$ evaluated at $-1$, specifically
\be 
\begin{aligned}
&{}_{6}F_{5}\left[\begin{matrix}
a,\,1+\frac{1}{2}a,\,b,\,c,\,d,\,e\\
\frac{1}{2}a,\,1+a-b,\,1+a-c,\,1+a-d,\,1+a-e
\end{matrix};-1\right]\\&=\frac{\Gamma(1+a-d)\Gamma(1+a-e)}{\Gamma(1+a)\Gamma(1+a-d-e)}{}\,_{3}F_{2}\left[\begin{matrix}
1+a-b-c,\,d,\,e\\
1+a-b,\,1+a-c
\end{matrix};1\right].
\end{aligned}
\ee
Applying this identity to our earlier expressions for the amplitude, we obtain a new representation of the five-point string amplitude,
\be  
\begin{aligned}
&{\cal A}_5 =\frac{\Gamma(X_{1,3})\Gamma(X_{2,4})\Gamma(X_{3,5})\Gamma(X_{1,4})\Gamma(X_{2,5})\Gamma(X_{2,4}+X_{3,5}+X_{1,4})}{\Gamma(X_{1,3}+X_{2,4})\Gamma(X_{2,4}+X_{3,5})\Gamma(X_{3,5}+X_{1,4})\Gamma(X_{1,4}+X_{2,5})}\times\\&\qquad\times{}_{6}F_{5}\!\left[\begin{smallmatrix}
X_{2,4},X_{1,4},X_{3,5}{+}X_{1,4}{-}X_{1,3},X_{2,4}{+}X_{3,5}{-}X_{2,5},X_{2,4}{+}X_{3,5}{+}X_{1,4}{-}1,\frac{X_{2,4}+X_{3,5}+X_{1,4}+1}{2}\\
X_{1,3}+X_{2,4},\,X_{2,4}+X_{3,5},\,X_{3,5}+X_{1,4},\,X_{1,4}+X_{2,5},\,\frac{X_{2,4}+X_{3,5}+X_{1,4}-1}{2}
\end{smallmatrix}\!\!;-1\right].
\end{aligned}\label{eq:5ptWhipple}
\ee
As one can numerically verify, the single expression in Eq.~\eqref{eq:5ptWhipple} converges {\it everywhere}. (The one caveat is where $X_{2,4}+X_{3,5}+X_{1,4}$ equals a negative integer, but the limit to this measure-zero subregion of kinematic space 
can be easily evaluated numerically.)
The result in Eq.~\eqref{eq:5ptWhipple} automatically agrees with the representations of the amplitude in Eqs.~\eqref{eq:5ptstandard}, \eqref{eq:5ptThomae}, and their cyclic permutations where it overlaps their domains of convergence, and it gives an explicit expression for the analytic continuation to the entire space for arbitrary $(X_{1,3},X_{1,4},X_{2,4},X_{2,5},X_{3,5})$; see Fig.~\ref{fig:5pt} for an illustration.
This is the first time that a complete representation of the five-point string amplitude has been found.

\clearpage
\subsection{Understanding of Thomae from ``split'' factorizations}
\label{sec:Thomae_Splits}

At five points, we were able to use Thomae transformations to extend the domain of convergence to almost all of kinematic space (all but an infinite wedge comprising 1/1024 of the space of Mandelstams), and we subsequently used the Whipple identity to obtain a formula that converged everywhere. 
The Thomae transformations themselves can be thought of as relabelings of the worldsheet variables of integration, and hence we should expect generalized versions of them to apply at higher than five points.
We will find that such generalizations give us relations between $n$-point amplitudes evaluated at different \textit{restricted} kinematics, so they are not as powerful as the general five-point result.

In particular, as proposed by Ref.~\cite{FBrown}, we can understand the Thomae transformation at five points as a factorizing  property of the six-point string amplitude into the product of a four-point amplitude times a five-point one. We now provide an interpretation of this splitting as the factorization near zeros pointed out in Refs.~\cite{Zeros,Splits}, for the case of the skinny rectangle, i.e., the factorization that we get in the kinematical locus where we pick $i^\star$ and set $c_{i^\star,j}=0 $ for all $j\neq j^\star$. This ultimately allows us to derive generalizations of the Thomae transformation for higher-point tree-level string amplitudes.

\begin{figure}[t]
    \centering
    \includegraphics[width=0.4\textwidth]{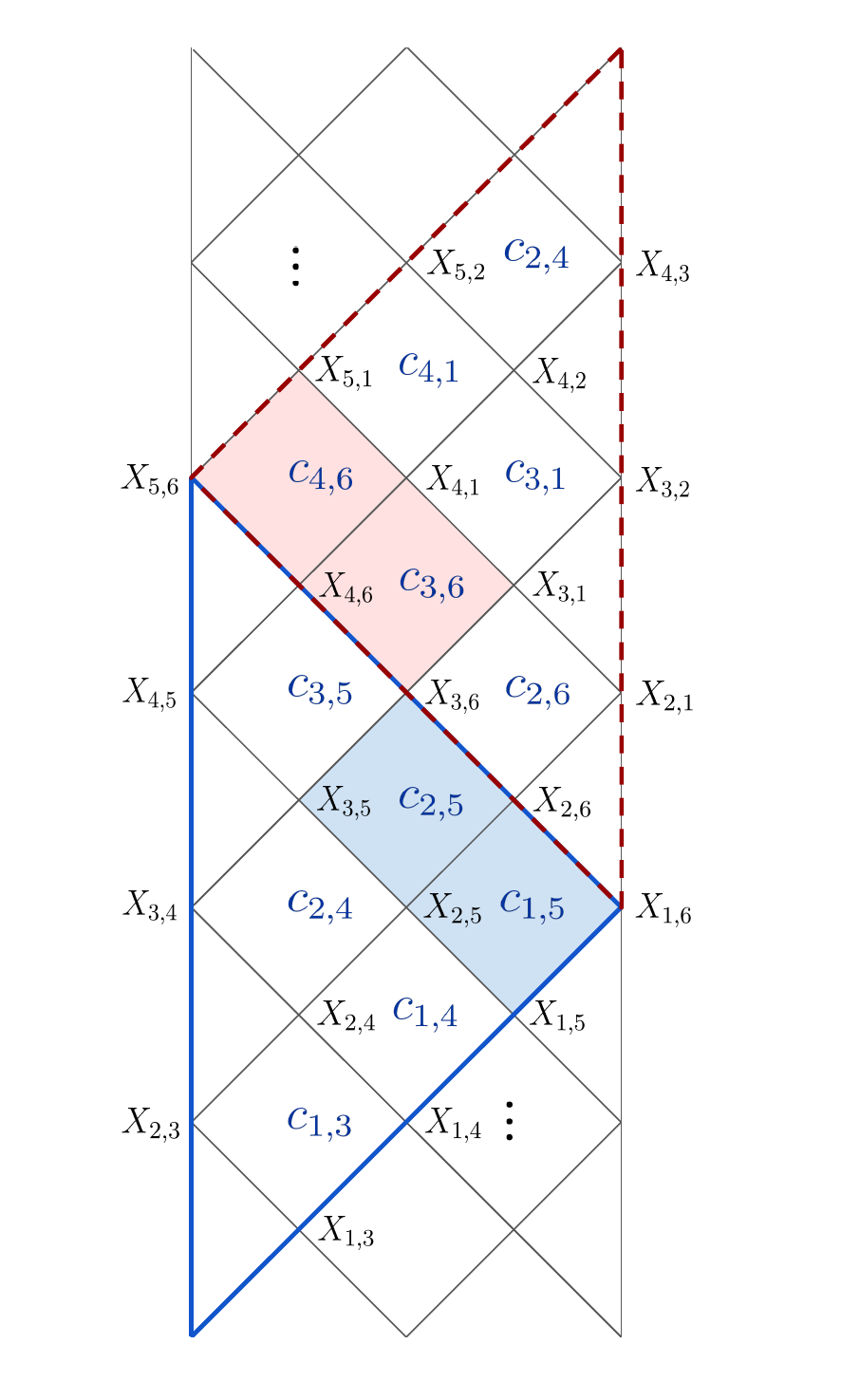}
    \caption{Factorization near zeros used to derive the Thomae transformations. First we consider the factorization associated to the skinny rectangle highlighted in blue, where we set both $c_{1,5}=c_{2,5}=0$ but keep $c_{3,5}\neq 0$. Then we further go on the locus associated to the factorization of the red skinny rectangle, where we set $c_{3,6}=c_{4,6}=0$ but keep $c_{2,6} \neq 0$.}
    \label{fig:Thomae}
\end{figure}

Let us look at Fig.~\ref{fig:Thomae} and consider the part of the kinematic mesh delimited in blue. Then setting $c_{1,5}=c_{2,5}=0$ in the stringy integral leads to the following factorization,
\begin{equation}
    \mathcal{A}_6(c_{1,5}=c_{2,5}=0)= \mathcal{A}_4(X_{1,5},X_{4,6}) \times \mathcal{A}_5(X_{1,3},X_{1,4},X_{2,4},X_{2,6},X_{3,6}),
    \label{eq:fact1}
\end{equation}
which is the splitting into a four-point times a five-point amplitude mentioned in Ref.~\cite{FBrown}. Alternatively, setting $c_{3,6}=c_{4,6}=0$, we get a different splitting that can be automatically read off by looking at the region of the mesh delimited in red, from which we find
\begin{equation}
    \mathcal{A}_6(c_{3,6}=c_{4,6}=0)= \mathcal{A}_4(X_{2,6},X_{1,5}) \times \mathcal{A}_5(X_{3,6},X_{4,6},X_{2,4},X_{2,5},X_{3,5}).
    \label{eq:fact2}
\end{equation}

Now notice that, precisely since each of the two skinny rectangles is contained in its own triangular region, if we start with the six-point amplitude and go on the locus where the four $c_{i,j}$ are set to zero, we can equate Eqs.~\eqref{eq:fact1} and \eqref{eq:fact2} to obtain
\begin{equation}
\begin{aligned}
\mathcal{A}_5(X_{1,3},X_{1,4},X_{2,4},X_{2,6},X_{3,6}) = \frac{\Gamma(X_{2,6})\Gamma(X_{1,5}{+}X_{4,6})}{\Gamma(X_{4,6})\Gamma(X_{1,5}{+}X_{2,6})} {\times} \mathcal{A}_5(X_{3,6},X_{4,6},X_{2,4},X_{2,5},X_{3,5})
\end{aligned}
\end{equation}
on the support $c_{1,5}=c_{2,5}=c_{3,6}=c_{4,6}=0$. Notice that, at six points, we initially have nine independent Mandelstams $X_{i,j}$, but after fixing the four $c_{i,j}$ above to zero, we only have five free variables remaining, exactly those of the five-point problem. Implementing these constraints, we can solve in terms of $X_{i,j}$,
\be
\begin{aligned}
  X_{2,6}&=X_{2,5}-X_{1,5}\\
    X_{3,5} &= X_{1,3}\\
    X_{3,6} &=X_{1,3} - X_{1,5}\\
    X_{4,6} &= X_{1,4} - X_{1,5} ,
\end{aligned}
\ee
and obtain a final equation depending only on $\{X_{1,3},X_{1,4},X_{1,5},X_{2,4},X_{2,5}\}$,
\begin{equation}
\begin{aligned}
&\mathcal{A}_5(X_{1,3},X_{1,4},X_{2,4},X_{2,5}-X_{1,5},X_{1,3} - X_{1,5}) \\&= \quad \frac{\Gamma(X_{2,5}-X_{1,5})\Gamma(X_{1,4})}{\Gamma(X_{1,4} - X_{1,5})\Gamma(X_{2,5})} \mathcal{A}_5(X_{1,3} - X_{1,5},X_{1,4} - X_{1,5},X_{2,4},X_{2,5},X_{1,3}).
\end{aligned}
\end{equation}
This is precisely the Thomae transformation introduced in the previous section.

Now looking at Fig.~\ref{fig:Thomae}, we know that we could have achieved the factorizations of the blue and red triangle by setting to zero a different subset of the $c_{i,j}$. In particular, we could have considered instead  setting $c_{3,5}=c_{1,5}=0$ in the blue skinny rectangle. Then we find the following relation,
\begin{equation}
\mathcal{A}_5(X_{1,3},X_{1,4},X_{2,4},X_{2,6},X_{3,5}) = \frac{\Gamma(X_{2,6})\Gamma(X_{1,5}{+}X_{4,6})}{\Gamma(X_{4,6})\Gamma(X_{1,5}{+}X_{2,6})} {\times} \mathcal{A}_5(X_{3,6},X_{4,6},X_{2,4},X_{2,5},X_{3,5}),
\end{equation}
on the support $c_{1,5}=c_{3,5}=c_{3,6}=c_{4,6}=0$. Similarly solving these constraints in terms of the same $X_{i,j}$,
\be 
\begin{aligned}
   X_{2,6}&=X_{2,5}-X_{1,5}\\
    X_{3,5}& = X_{1,3}-X_{1,4}\\
    X_{3,6}& =X_{1,3} - X_{1,5}\\
    X_{4,6}& = X_{1,4} - X_{1,5} ,
\end{aligned}
\ee
we can rewrite this relation as follows:
\begin{equation}
\begin{aligned}
&\mathcal{A}_5(X_{1,3},X_{1,4},X_{2,4},X_{2,5}-X_{1,5},X_{1,3}-X_{1,4}) = \\
&\quad \frac{\Gamma(X_{2,5}-X_{1,5})\Gamma(X_{1,4})}{\Gamma(X_{1,4} - X_{1,5})\Gamma(X_{2,5})} \times \underbrace{\mathcal{A}_5(X_{1,3} - X_{1,5},X_{1,4} - X_{1,5},X_{2,4},X_{2,5},X_{1,3}-X_{1,4})}_{\mathcal{A}_5(\tilde{X}_{1,3},\tilde{X}_{1,4},\tilde{X}_{2,4},\tilde{X}_{2,5},\tilde{X}_{3,5})}.
\end{aligned}
\end{equation}
On the right-hand side, we have written the kinematic arguments of the five-point amplitude as $\tilde X_{i,j}$. Note that these kinematics are not free, but instead satisfy some {\it linear relations}, such as $\tilde{X}_{1,3}-\tilde{X}_{3,5} = \tilde{X}_{1,4}$. Therefore, for this particular choice of $c_{i,j}$ that we set to zero, we obtain a relation between five-point amplitudes that involves restricted kinematics. It turns out that this will always be the case for any choice of $c_{i,j}$ other than the one we presented in the previous paragraph. Similarly, at higher points we are able to build generalizations of Thomae transformations that relate $n$-point amplitudes for restricted kinematics. 

As a final example at five point, we can take
\begin{equation}
    \mathcal{A}_5(X_{1,3},X_{1,4},X_{2,4},X_{2,5},X_{3,5}) = \frac{\Gamma(X_{2,6})\Gamma(X_{1,5}{+}X_{4,6})}{\Gamma(X_{4,6})\Gamma(X_{1,5}{+}X_{2,6})} {\times} \mathcal{A}_5(X_{3,6},X_{4,6},X_{2,4},X_{2,5},X_{3,5})
\end{equation}
with $c_{2,5}=c_{3,5}=c_{3,6}=c_{4,6}=0$.
Solving the constraints in terms of the $X_{i,j}$, we set
\be
\begin{aligned}
 X_{2,6}&=X_{1,4}+X_{2,5}-X_{1,5}\\
    X_{3,5} &= X_{1,3}-X_{1,4}\\
    X_{3,6} &=X_{1,3} - X_{1,5}\\
    X_{4,6} &= X_{1,4} - X_{1,5},
\end{aligned} 
\ee
from which we obtain the relation
\begin{equation}
\begin{aligned}
&\mathcal{A}_5(X_{1,3},X_{1,4},X_{2,4},X_{2,5},X_{1,3}-X_{1,4}) = \\
&\;\;\frac{\Gamma(X_{1,4}+X_{2,5}-X_{1,5})\Gamma(X_{1,4})}{\Gamma(X_{1,4} - X_{1,5})\Gamma(X_{1,4}+X_{2,5})} \times \mathcal{A}_5(X_{1,3} - X_{1,5},X_{1,4} - X_{1,5},X_{2,4},X_{2,5},X_{1,3}-X_{1,4}),
\end{aligned}
\end{equation}
where again the right-hand side has restricted kinematic arguments.

Proceeding in the analogous way for the red skinny rectangle, we find the additional six relations:
\begin{equation}
    \begin{aligned}
    &\mathcal{A}_5(X_{1,3},X_{1,4},X_{2,4},X_{2,6},X_{3,6})= \frac{\Gamma(X_{2,6})\Gamma(X_{1,5}+X_{4,6})}{\Gamma(X_{4,6})\Gamma(X_{1,5}+X_{2,6})}  \mathcal{A}_5(X_{1,3},X_{4,6},X_{1,5},X_{2,4},X_{2,5})\\
    & \quad \text{with } c_{1,5}=c_{2,5}=c_{2,6}=c_{4,6}=0,\\
    &\mathcal{A}_5(X_{1,3},X_{1,4},X_{2,4},X_{2,6},X_{3,5})= \frac{\Gamma(X_{2,6})\Gamma(X_{1,5}+X_{4,6})}{\Gamma(X_{4,6})\Gamma(X_{1,5}+X_{2,6})}  \mathcal{A}_5(X_{1,3},X_{4,6},X_{1,5},X_{2,4},X_{2,5})\\
    & \quad \text{with } c_{1,5}=c_{3,5}=c_{2,6}=c_{4,6}=0,\\
    &\mathcal{A}_5(X_{1,3},X_{1,4},X_{2,4},X_{2,5},X_{3,5})= \frac{\Gamma(X_{2,6})\Gamma(X_{1,5}+X_{4,6})}{\Gamma(X_{4,6})\Gamma(X_{1,5}+X_{2,6})}  \mathcal{A}_5(X_{1,3},X_{4,6},X_{1,5},X_{2,4},X_{2,5})\\
    & \quad \text{with } c_{2,5}=c_{3,5}=c_{2,6}=c_{4,6}=0,\\
    &\mathcal{A}_5(X_{1,3},X_{1,4},X_{2,4},X_{2,6},X_{3,6})= \frac{\Gamma(X_{2,6})\Gamma(X_{1,5}+X_{4,6})}{\Gamma(X_{4,6})\Gamma(X_{1,5}+X_{2,6})}  \mathcal{A}_5(X_{1,3},X_{1,4},X_{1,5},X_{2,4},X_{2,5})\\
    & \quad \text{with } c_{1,5}=c_{2,5}=c_{2,6}=c_{3,6}=0,  \\
    &\mathcal{A}_5(X_{1,3},X_{1,4},X_{2,4},X_{2,6},X_{3,5})= \frac{\Gamma(X_{2,6})\Gamma(X_{1,5}+X_{4,6})}{\Gamma(X_{4,6})\Gamma(X_{1,5}+X_{2,6})}  \mathcal{A}_5(X_{1,3},X_{1,4},X_{1,5},X_{2,4},X_{2,5})\\
    & \quad \text{with } c_{1,5}=c_{3,5}=c_{2,6}=c_{3,6}=0,  \\
    &\mathcal{A}_5(X_{1,3},X_{1,4},X_{2,4},X_{2,5},X_{3,5})= \frac{\Gamma(X_{2,6})\Gamma(X_{1,5}+X_{4,6})}{\Gamma(X_{4,6})\Gamma(X_{1,5}+X_{2,6})}  \mathcal{A}_5(X_{1,3},X_{1,4},X_{1,5},X_{2,4},X_{2,5})\\
    & \quad \text{with } c_{2,5}=c_{3,5}=c_{2,6}=c_{3,6}=0.  
    \end{aligned}
\end{equation}
One can check that in all the above cases the amplitudes are for restricted kinematics. 

We have seen that for five-point relations we begin with near-zero factorizations at six points. A very specific choice for ``skinny'' factorization gives a relation between five-point amplitudes at  generic kinematics, but more generally we find relations for restricted kinematics. This analysis can be trivially extended to any number of points, but it can easily be seen that for {\it all} choices of factorizations, we find identities similar to the ones above, with restricted kinematics. 

\section{Numerical Checks}\label{sec:numerics}

Having discovered new representations of tree-level string amplitudes that allow us to extend their domain of evaluability, as well as new expressions for their asymptotic high-energy behavior in various generalized Regge and hard scattering limits, we are now equipped to test our analytical results numerically.

\subsection{Five-point numerics}

\begin{figure}[t]
\begin{center}
\includegraphics[width=0.8\textwidth]{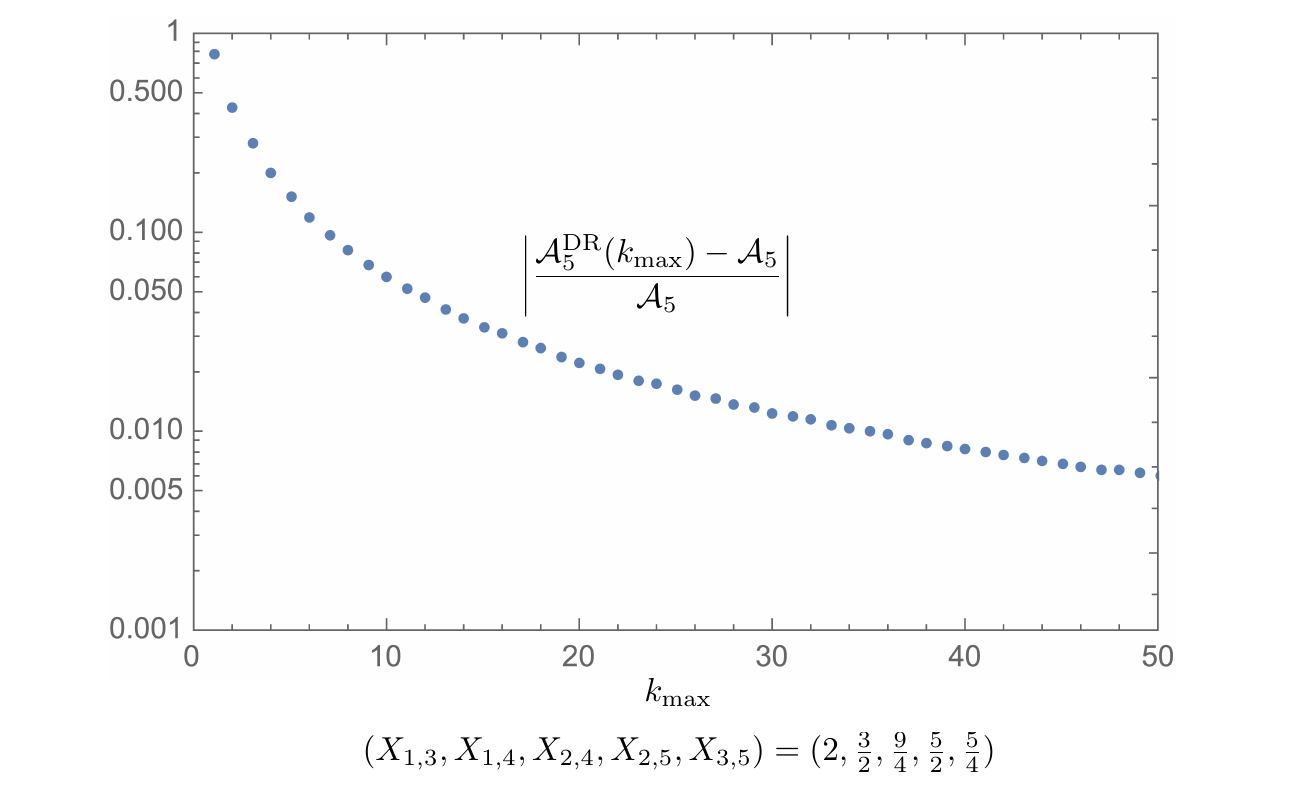}
\end{center}\vspace{-9mm}
\caption{Illustration of the convergence of the dual resonant form of the five-point amplitude in Eq.~\eqref{eq:DR5}, with both channels summed up to $k_{\max}$, compared to the analytic form of the amplitude in Eq.~\eqref{eq:5ptWhipple}.}
\label{fig:conv5}
\end{figure}

Let us first consider the five-point amplitude. We will use the form we found in Eq.~\eqref{eq:5ptWhipple}, which can be evaluated at arbitrary values of the Mandelstam invariants.
Comparing against the dual resonant form of the amplitude in Eq.~\eqref{eq:DR5}, where we cap the sums in both $m$ and $n$ at some integer $k_{\max}$, we evaluate at the following kinematics,
\be 
(X_{1,3},X_{1,4},X_{2,4},X_{2,5},X_{3,5})=(2,\tfrac{3}{2},\tfrac{9}{4},\tfrac{5}{2},\tfrac{5}{4}),
\ee
and find that the sum tends toward convergence, as shown in Fig.~\ref{fig:conv5}.

\begin{figure}[t]
\begin{center}
\includegraphics[width=0.8\textwidth]{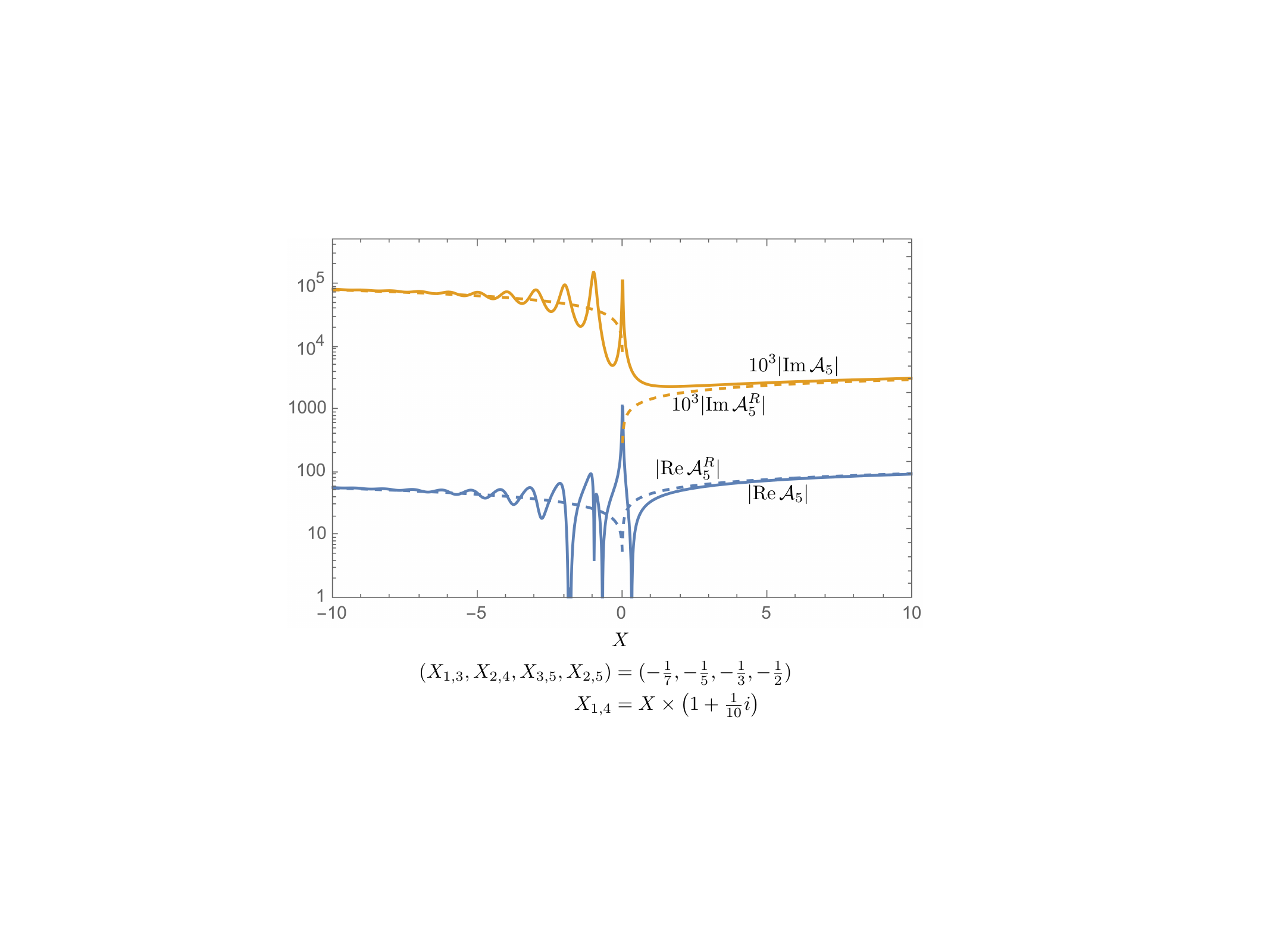}
\end{center}\vspace{-9mm}
\caption{Check of the single Regge limit for the five-point string amplitude, where all planar invariants except $X_{1,4}$ are fixed as above, and we scan in $X_{1,4}$ along the ray parallel to $1+\frac{i}{10}$. We see that for large $|X_{1,4}|$, the Regge limit provides an excellent approximation. In evaluating the five-point amplitude, we used the form in Eq.~\eqref{eq:5ptWhipple} that can be used for arbitrary Mandelstams.
The poles along the negative real $X_{1,4}$ axis are evident.}
\label{fig:Regge5}
\end{figure}

Let us now turn to the Regge limit. We fix $X_{1,3}$, $X_{2,4}$, $X_{3,5}$, and $X_{2,5}$ to constants and evaluate in the limit of large $X_{1,4}$, where from Eq.~\eqref{eq:singleRegge} we predict the single Regge limit ${\cal A}_5^R$ given by
\be
{\cal A}_5 \rightarrow {\cal A}_5^R = X_{1,4}^{-X_{2,5}}\Gamma(X_{2,5}){\cal A}_4(X_{2,4},X_{3,5}-X_{2,5})+X_{1,4}^{-X_{3,5}}\Gamma(X_{3,5}){\cal A}_4(X_{1,3},X_{2,5}-X_{3,5}),\label{eq:Reggedemo}
\ee
where ${\cal A}_4(s,t)=\Gamma(s)\Gamma(t)/\Gamma(s+t)$ is the four-point amplitude.
For generic $X_{i,j}$, one of the two terms in Eq.~\eqref{eq:Reggedemo} will dominate, except at the spurious pole where $X_{2,5}=X_{3,5}$, where the factors of $\Gamma(\pm X_{2,5}\mp X_{3,5})$ cancel.
Concretely, let us choose
\be 
(X_{1,3},X_{2,4},X_{3,5},X_{2,5}) = (-\tfrac{1}{7},-\tfrac{1}{5},-\tfrac{1}{3},-\tfrac{1}{2}),
\ee
and write 
\be 
X_{1,4} = X\times(1+\tfrac{1}{10}i).
\ee
Strictly speaking, we proved Eq.~\eqref{eq:singleRegge} for $X$ is large and positive, but as we expect that the only relevant saddles are on the boundary of the moduli space (i.e., those we have identified, where some $u\rightarrow 0$),  Eq.~\eqref{eq:Reggedemo} should hold more generally, for large negative $X$ as well.
We see in Fig.~\ref{fig:Regge5} that this is indeed the case.

\begin{figure}[t]
\begin{center}
\includegraphics[width=0.8\textwidth]{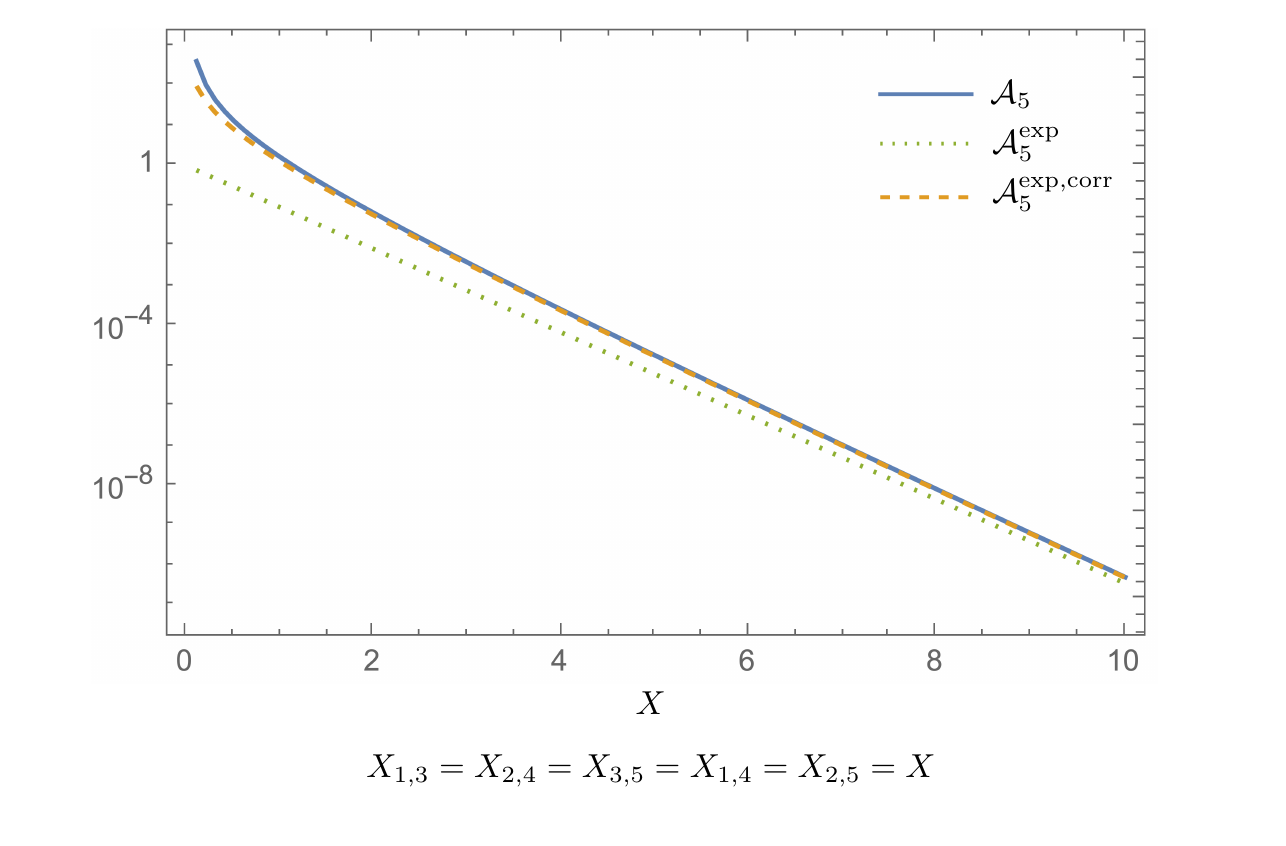}
\end{center}\vspace{-9mm}
\caption{Check of the hard scattering limit for the five-point string amplitude, where all planar invariants are set to $X$. We compare ${\cal A}_5$, the amplitude itself, against our leading exponential prediction~\eqref{eq:exponential5} and subleading correction~\eqref{eq:5subleading}, finding excellent agreement.}
\label{fig:exponential5}
\end{figure}

We can also check the hard scattering limit we derived in Eq.~\eqref{eq:expsup1}, where we have
\be
{\cal A}_5\rightarrow {\cal A}_5^{\rm exp} =  \frac{1}{[2\cos(\pi/5)]^{5X}}\label{eq:exponential5}
\ee
when $X_{1,3}=X_{2,4}=X_{3,5}=X_{1,4}=X_{2,5}=X$ is taken large. Note that Eq.~\eqref{eq:expsup1} holds up to power-law corrections, where the leading correction is a multiplicative factor of $\propto X^{(3-n)/2}$.
These were computed explicitly in Eq.~\eqref{eq:expsup2} by doing the Gaussian integral around the saddle point, in which case we find an  expression at five point that we repeat here for convenience,
\be
{\cal A}_5 \rightarrow {\cal A}_5^{\rm exp,corr}={\cal A}_5^{\rm exp}\times \frac{2\sqrt{2}\pi}{X\sqrt{25-11\sqrt{5}}}.\label{eq:5subleading}
\ee
Numerically evaluating ${\cal A}_5$ and scanning in $X$, we find that it indeed matches ${\cal A}_5^{\rm exp}$ for $X\gg 1$, and this matching is improved by the subleading correction in Eq.~\eqref{eq:5subleading}; see Fig.~\ref{fig:exponential5}.

\subsection{Six-point numerics}

\begin{figure}[t]
\begin{center}
\includegraphics[width=0.8\textwidth]{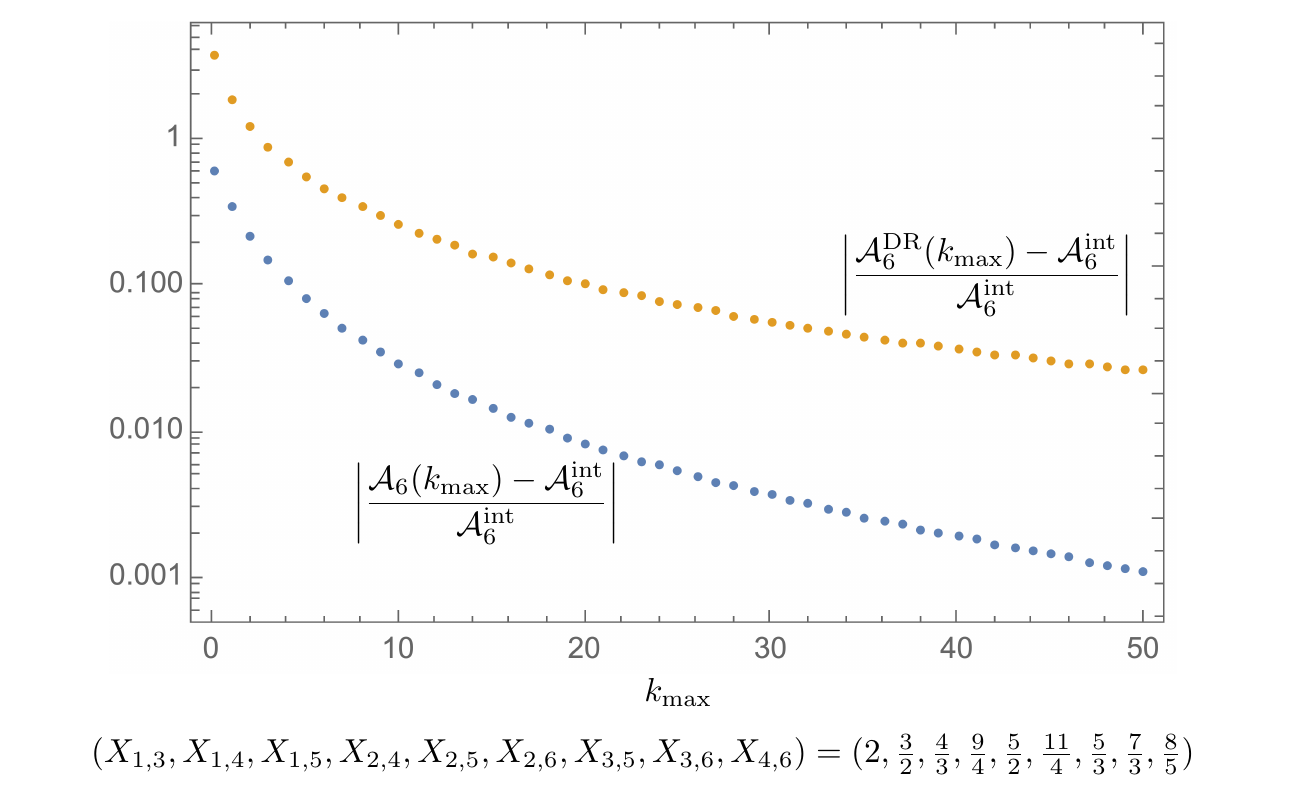}
\end{center}\vspace{-9mm}
\caption{Blue: Illustration of the convergence of the ``$u$'s in terms of $u$'s'' sum form of the six-point amplitude in Eq.~\eqref{eq:6pointbigsum}, for the given kinematics, where $k_{2,4},k_{2,5},k_{3,5}$ are capped at $k_{\max}$. Orange: The same, but for the dual resonant form of the amplitude in Eq.~\eqref{eq:DR6}.}
\label{fig:conv61}
\end{figure}

Next, we can numerically check our results at six point.
Here, we will use the new summation form of the six-point amplitude that we derived from solving for ``$u$'s in terms of $u$'s'' in Eq.~\eqref{eq:6pointbigsum}.
As a check of the convergence of this sum form of the amplitude, let us evaluate evaluate Eq.~\eqref{eq:6pointbigsum} by summing over $k_{2,4},k_{2,5},k_{3,5}\in[0,k_{\max}]$, defining ${\cal A}_6(k_{\rm max})$, where the true amplitude ${\cal A}_6$ is ${\cal A}_6(\infty)$.
We choose the following kinematics,
\be 
(X_{1,3},X_{1,4},X_{1,5},X_{2,4},X_{2,5},X_{2,6},X_{3,5},
X_{3,6},X_{4,6})=(2,\tfrac{3}{2},\tfrac{4}{3},\tfrac{9}{4},\tfrac{5}{2},\tfrac{11}{4},\tfrac{5}{3},\tfrac{7}{3},\tfrac{8}{5}),\label{eq:Xchoicepos6}
\ee
and will compare against the integral form of the amplitude in Eq.~\eqref{eq:6pt}, which converges for the choice of $X_{i,j}$ in Eq.~\eqref{eq:Xchoicepos6}, for which we find via numerical integration that ${\cal A}_6^{\rm int} \simeq 0.512$.
We will further compare against the dual resonant form of the amplitude from Eq.~\eqref{eq:DRfull}, which we write as
\be
{\cal A}_6^{\rm DR}(k_{\rm max}) = \sum_{n_1=0}^{k_{\max}} \sum_{n_2=0}^{k_{\max}} \sum_{n_3=0}^{k_{\max}} \frac{R_{n_1,n_2,n_3}}{(X_{1,3}+n_1)(X_{1,4}+n_2)(X_{1,5}+n_3)},\label{eq:DR6}
\ee
where the true amplitude is ${\cal A}_6^{\rm DR}(\infty)$, and we compute $R_{n_1,n_2,n_3}$ using Eq.~\eqref{eq:ResHalfLadd}.
In Fig.~\ref{fig:conv61}, we can see the trend toward convergence of all of these methods of computing the amplitude, thus numerically verifying our results.

\begin{figure}[t]
\begin{center}
\includegraphics[width=0.8\textwidth]{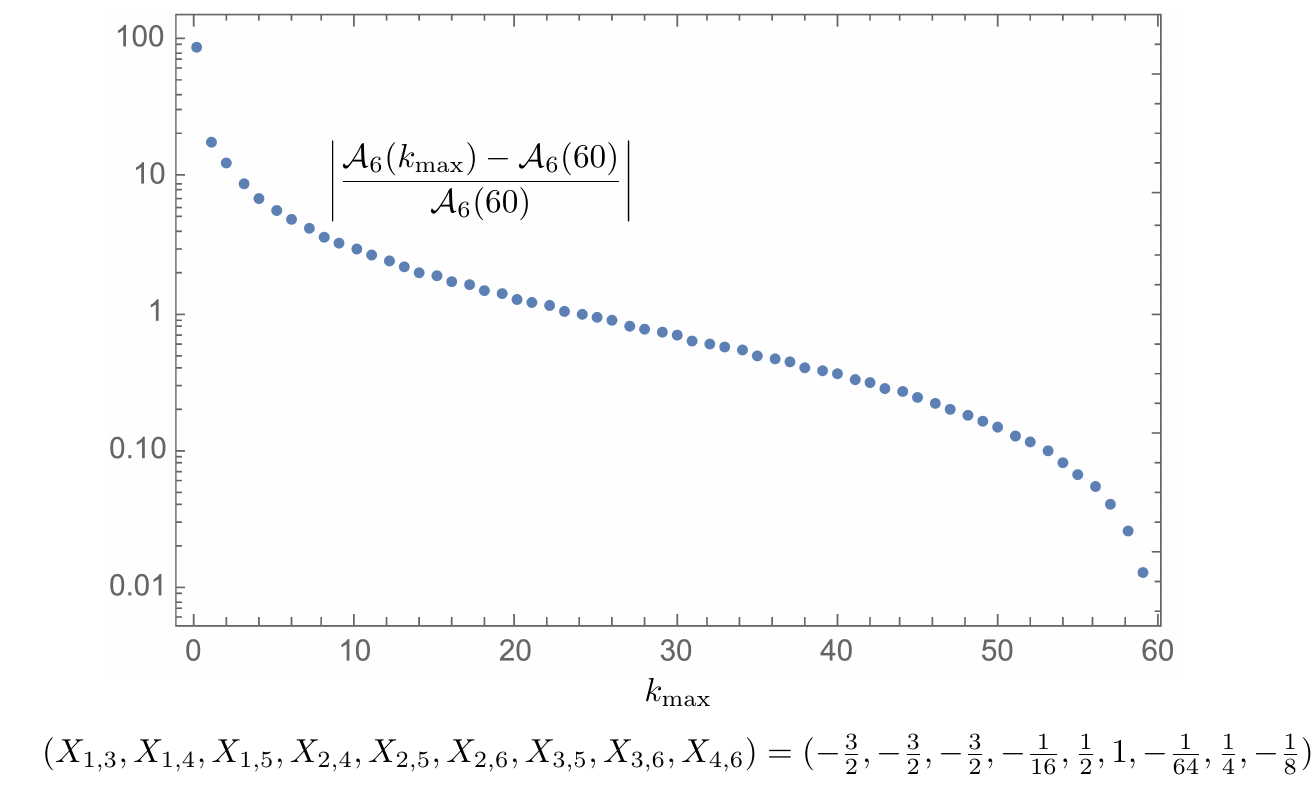}
\end{center}\vspace{-9mm}
\caption{Illustration of the convergence of the ``$u$'s in terms of $u$'s'' sum form of the six-point amplitude in Eq.~\eqref{eq:6pointbigsum}, for the given kinematics, where $k_{2,4},k_{2,5},k_{3,5}$ are capped at $k_{\max}$.}
\label{fig:conv62}
\end{figure}

As a further illustration of convergence, for the case of six $X$ variables taken negative, let us choose the following kinematics,
\be
\hspace{-1mm}(X_{1,3},X_{1,4},X_{1,5},X_{2,4},X_{2,5},X_{2,6},X_{3,5},
X_{3,6},X_{4,6})\,{=}\,(-\tfrac{3}{2},-\tfrac{3}{2},-\tfrac{3}{2},-\tfrac{1}{16},\tfrac{1}{2},1,-\tfrac{1}{64},\tfrac{1}{4},-\tfrac{1}{8}).
\ee
Scanning in $k_{\max}$, we can again see the trend toward convergence in Fig.~\ref{fig:conv62}.

\begin{figure}[t]
\begin{center}
\includegraphics[width=0.8\textwidth]{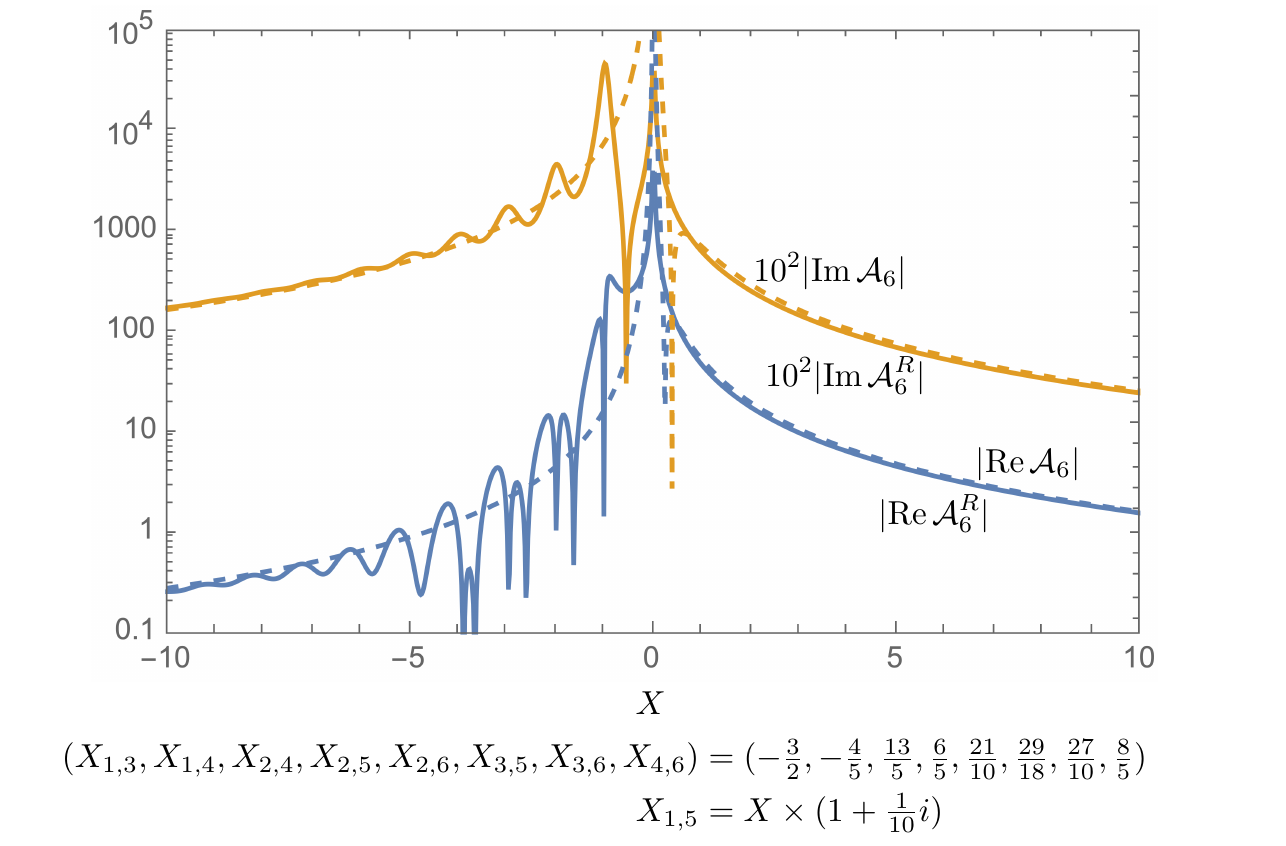}
\end{center}\vspace{-5mm}
\caption{Regge limit at six point, with $c_{2,4},c_{2,5},c_{3,5}
>0$ (i.e., outside the obvious convergence region), for the kinematics listed above. The amplitude (solid curves), as computed from the sum in Eq.~\eqref{eq:6pointbigsum} for $k_{2,4},k_{2,5},k_{3,5}\leq k_{\max}$ for $k_{\max}=30$, is compared against the Regge limit prediction given in Eq.~\eqref{eq:A6R} (dashed).}
\label{fig:Regge61}
\end{figure}

Turning now to the Regge limit, let us compute the amplitude at fixed $X_{1,3}$, $X_{1,4}$, $X_{2,4}$, $X_{2,5}$, $X_{2,6}$, $X_{3,5}$, $X_{3,6}$, and $X_{4,6}$, and scan in $X_{1,5}$.
We will choose kinematics such that this sum numerically converges with a simple cutoff for large $k_{2,4},k_{2,5},k_{3,5}$ in Eq.~\eqref{eq:6pointbigsum}.
With this numerical evaluation of the six-point amplitude in hand, we will compare the results to our Regge limit prediction in Eq.~\eqref{eq:singleRegge}, which in the limit of large $X_{1,5}$ gives
\be 
\begin{aligned}
{\cal A}_6 \rightarrow {\cal A}_6^R &= X_{1,5}^{-X_{2,6}}\Gamma(X_{2,6})  {\cal A}_5(X_{2,4},X_{2,5}, X_{3,5}, X_{3,6} - X_{2,6}, X_{4,6} - X_{2,6}) \\&\;\;\; + 
  X_{1,5}^{-X_{3,6}} \Gamma(X_{3,6}) {\cal A}_4(X_{1,3}, X_{2,6} - X_{3,6}) {\cal A}_4(X_{3,5}, X_{4,6} - X_{3,6}) \\&\;\;\; + 
X_{1,5}^{-X_{4,6}} \Gamma(X_{4,6}) {\cal A}_5(X_{1,3}, X_{1,4}, X_{2,4}, X_{2,6} - X_{4,6}, X_{3,6} - X_{4,6}),
\end{aligned}\label{eq:A6R}
\ee
where we write the arguments of the five-point amplitude given in Eq.~\eqref{eq:5ptWhipple} in the canonical ordering, that is, ${\cal A}_5(X_{1,3},X_{1,4},X_{2,4},X_{2,5},X_{3,5})$.
Writing $X_{1,5}=X\times(1+\tfrac{i}{10})$, let us first choose the following kinematics,
\be (X_{1,3},X_{1,4},X_{2,4},X_{2,5},X_{2,6},X_{3,5},X_{3,6},X_{4,6} )=(-\tfrac{3}{2},-\tfrac{4}{5},\tfrac{13}{5},\tfrac{6}{5},\tfrac{21}{10},\tfrac{29}{18},\tfrac{27}{10},\tfrac{8}{5}),
\ee
which corresponds to $c_{2,4},c_{2,5},c_{3,5}>0$, outside of the obvious convergence region of Eq.~\eqref{eq:6pointbigsum} noted just below Eq.~\eqref{eq:6pointbigsumapprox}.
We see in Fig.~\ref{fig:Regge61} that the amplitude is well approximated by the Regge limit~\eqref{eq:A6R} at large positive $X$, as required.

\begin{figure}[t]
\begin{center}
\includegraphics[width=0.8\textwidth]{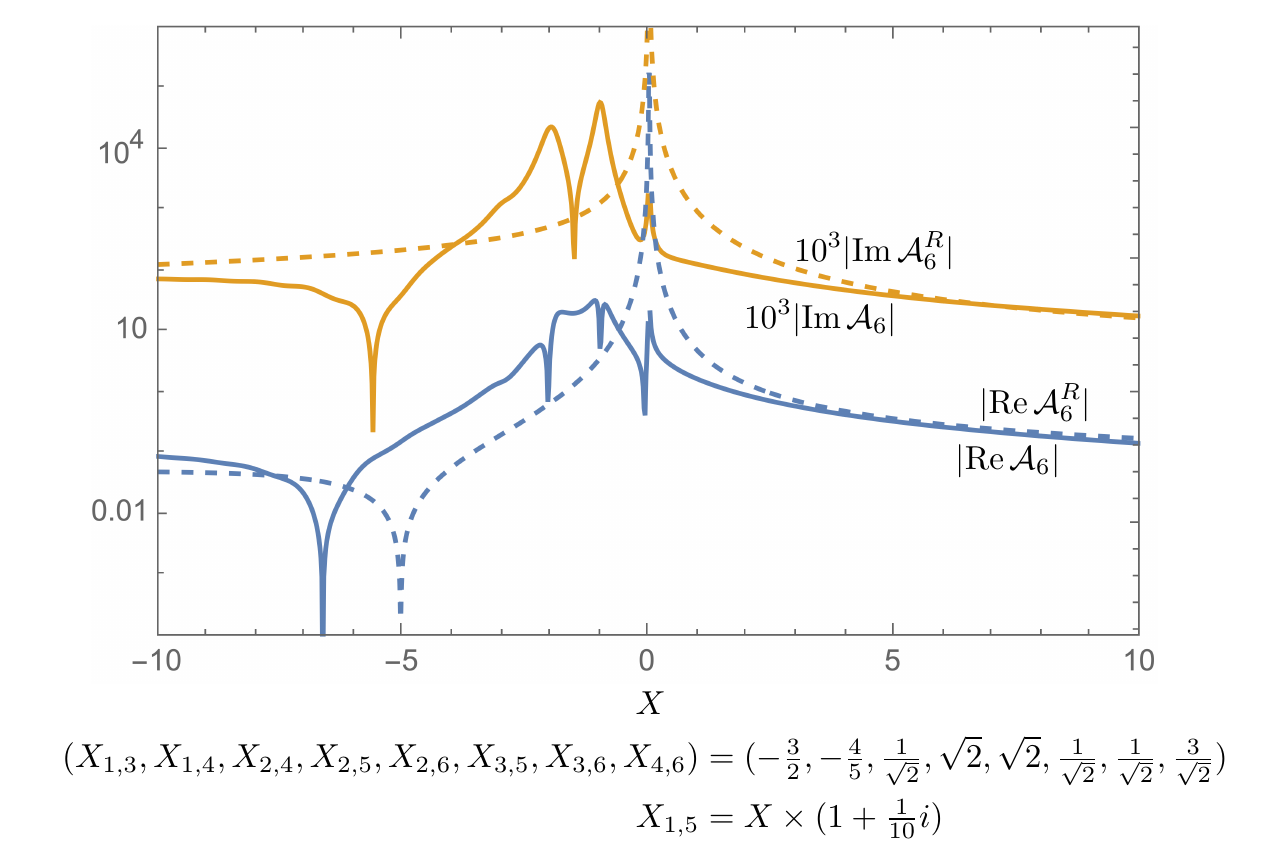}
\end{center}\vspace{-5mm}
\caption{A final example of the Regge limit at six point, with kinematics chosen such that the ${\cal A}_4 \times {\cal A}_4$ term in Eq.~\eqref{eq:A6R} dominates. The amplitude (solid curves), computed from the sum in Eq.~\eqref{eq:6pointbigsum} for $k_{2,4},k_{2,5},k_{3,5}\leq k_{\max}$, in this case for $k_{\max}=50$, is again compared against the Regge limit prediction given in Eq.~\eqref{eq:A6R} (dashed).}
\label{fig:Regge63}
\end{figure}

We can also make a choice of kinematics such that the ${\cal A}_4 \times {\cal A}_4$ term in the six-point Regge limit in Eq.~\eqref{eq:A6R} dominates over the five-point terms, which occurs when $X_{3,6}$ is smaller than $X_{2,6}$ and $X_{4,6}$.
An example such choice for which the sum in Eq.~\eqref{eq:6pointbigsum} converges is 
\be
(X_{1,3},X_{1,4},X_{2,4},X_{2,5},X_{2,6},X_{3,5},X_{3,6},X_{4,6} )=(-\tfrac{3}{2},-\tfrac{4}{5},\tfrac{1}{\sqrt{2}},\sqrt{2},\sqrt{2},\tfrac{1}{\sqrt{2}},\tfrac{1}{\sqrt{2}},\tfrac{3}{\sqrt{2}}), 
\ee
where we will scan in $X_{1,5}=X\times(1+\tfrac{i}{10})$ as before.
We again find agreement with the Regge limit, as shown in Fig.~\ref{fig:Regge63}.

\begin{figure}[t]
\begin{center}
\includegraphics[width=0.8\textwidth]{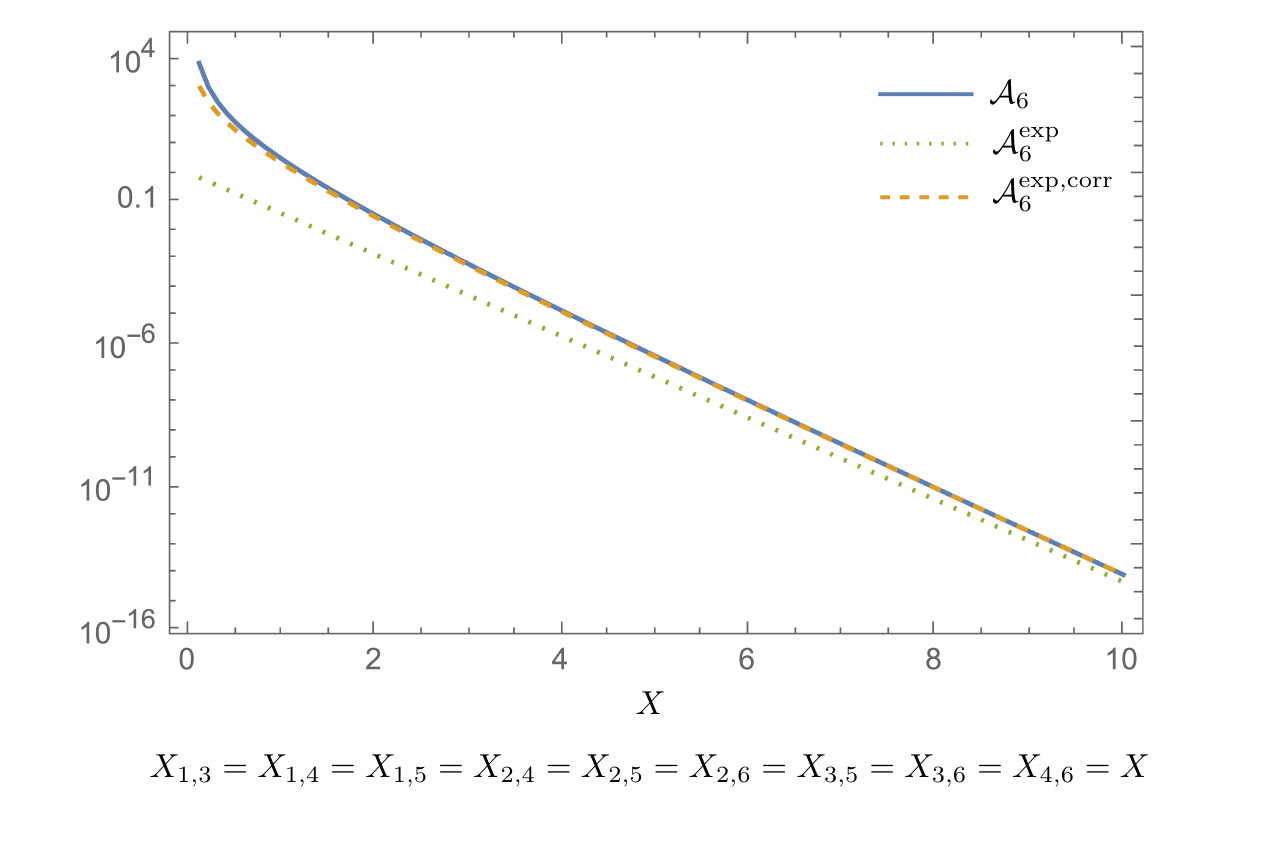}
\end{center}\vspace{-9mm}
\caption{Check of the hard scattering limit for the six-point string amplitude, where all planar invariants are set to $X$. We compare ${\cal A}_6$, the amplitude itself computed in Eq.~\eqref{eq:6pointbigsum}, against our leading exponential prediction~\eqref{eq:exponential6} and subleading correction~\eqref{eq:6subleading}, finding excellent agreement.}
\label{fig:exponential6}
\end{figure}

Let us also confirm the hard scattering limit in Eq.~\eqref{eq:expsup1}, which at six points gives
\be
{\cal A}_6\rightarrow {\cal A}_6^{\rm exp} =  \frac{1}{[2\cos(\pi/6)]^{6X}}\label{eq:exponential6}
\ee
when $X_{1,3}=X_{1,4}=X_{1,5}=X_{2,4}=X_{2,5}=X_{2,6}=X_{3,5}=X_{3,6}=X_{4,6}=X$ is taken large.
Including the power-law corrections we computed in Eq.~\eqref{eq:expsup2},  we have
\be
{\cal A}_6 \rightarrow {\cal A}_6^{\rm exp,corr}={\cal A}_6^{\rm exp}\times \frac{6\sqrt{3}\pi^{3/2}}{X^{3/2}}.\label{eq:6subleading}
\ee
As at five point, we evaluate ${\cal A}_6$ numerically---in this case using Eq.~\eqref{eq:6pointbigsum}---and scan in $X$, finding that it is well matched by ${\cal A}_6^{\rm exp}$ for $X\gg 1$ and even better with the subleading corrections in Eq.~\eqref{eq:6subleading}; see Fig.~\ref{fig:exponential6}.

\section{Discussion}\label{sec:discussion}

In this paper we have initiated a systematic exploration of many fundamental properties of all-$n$ tree-level open string amplitudes, exposing both new qualitative facts about the amplitudes and establishing various useful new representations. We close by mentioning a few of the many interesting open avenues for further exploration suggested immediately by our results. 

It would certainly be of great interest to establish the analogues of all of our results for closed string amplitudes at tree level. This can most directly be done using the KLT formalism to express closed string amplitudes in terms of open ones, but it would be more interesting to understand the basic features of factorization on massive poles, as well as the generalized fixed-angle and Regge asymptotics, more directly from the closed string integral representation as the square of the open string one. The positive parameteriaztion of the $u$ variables in terms of the $y$ variables is naively tailored to a given color ordering, but it has long been appreciated that all the orderings also have a nice understanding in the language of the $u$ variables, and it would be well motivated to leverage this fact to give a more intrinsic understanding of closed string amplitudes without using the double copy idea as a crutch. 

Perhaps the ultimate aim of these investigations would be a closed-form expression for the tree-level amplitudes that can be evaluated for all kinematics, as we have provided for the case of five-point scattering. Much of the magic there was connected to the ``Thomae'' identities relating the amplitude at different kinematic points. We have given a simple conceptual explanation of these identities and seen how they can be generalized to all $n$, but this generalization only works for constrained kinematics at general $n$. And we have yet to find a conceptual explanation of the ``Whipple'' identity crucially needed to find an expression valid at all kinematics, and so we do not yet know how to extend it to all $n$. It would be fascinating to understand these identities more deeply and use them to give analytic expressions for string amplitudes valid for the widest range of kinematics possible.

\vspace{\baselineskip}

\begin{center} 
{\bf Acknowledgments}
\end{center}
\noindent 
We thank Cliff Cheung for useful discussions.
The work of N.A.H. is supported by the DOE (Grant No. DE-SC0009988), by the Simons Collaboration on Celestial Holography, and the ERC UNIVERSE+ synergy grant. Further support was made possible by the Carl B. Feinberg cross-disciplinary program in innovation at the IAS.
C.F. is supported by FCT/Portugal (Grant No.~2023.01221.BD).
G.N.R. is supported by the James Arthur Postdoctoral Fellowship at New York University.

\vspace{\baselineskip}

\appendix

\section{Kinematic Variables and the Kinematic Mesh}\label{app:mesh}
In this appendix, we review the kinematic variables we are dealing with as well as a particularly useful way of organizing them: the kinematic mesh. 

Amplitudes are Lorentz-invariant objects that depend exclusively on the product---as computed by contraction with the Minkowski metric---of the momenta of the particles scattering. However, because of momentum conservation, these dot products are not independent from each other. Instead, a basis for the kinematics is given by the \textit{planar variables}:
\begin{equation}
    X_{i,j} = (p_i + \dots +p_{j-1})^2.
\end{equation}
These are precisely the invariants that appear as propagators in planar diagrams and thus the variables on which the amplitudes we are considering depend. They are related to \textit{non-planar} Mandelstams in the following way,
\begin{equation}
    X_{i,j} + X_{i+1,j+1}-X_{i,j+1}-X_{i+1,j} = c_{i,j}, \quad \text{with } \quad c_{i,j} = - 2p_i \cdot p_j,
    \label{eq:ceqn}
\end{equation}
where $i$ and $j$ are non-adjacent. 

\begin{figure}[t]
    \centering \includegraphics[width=0.8\textwidth]{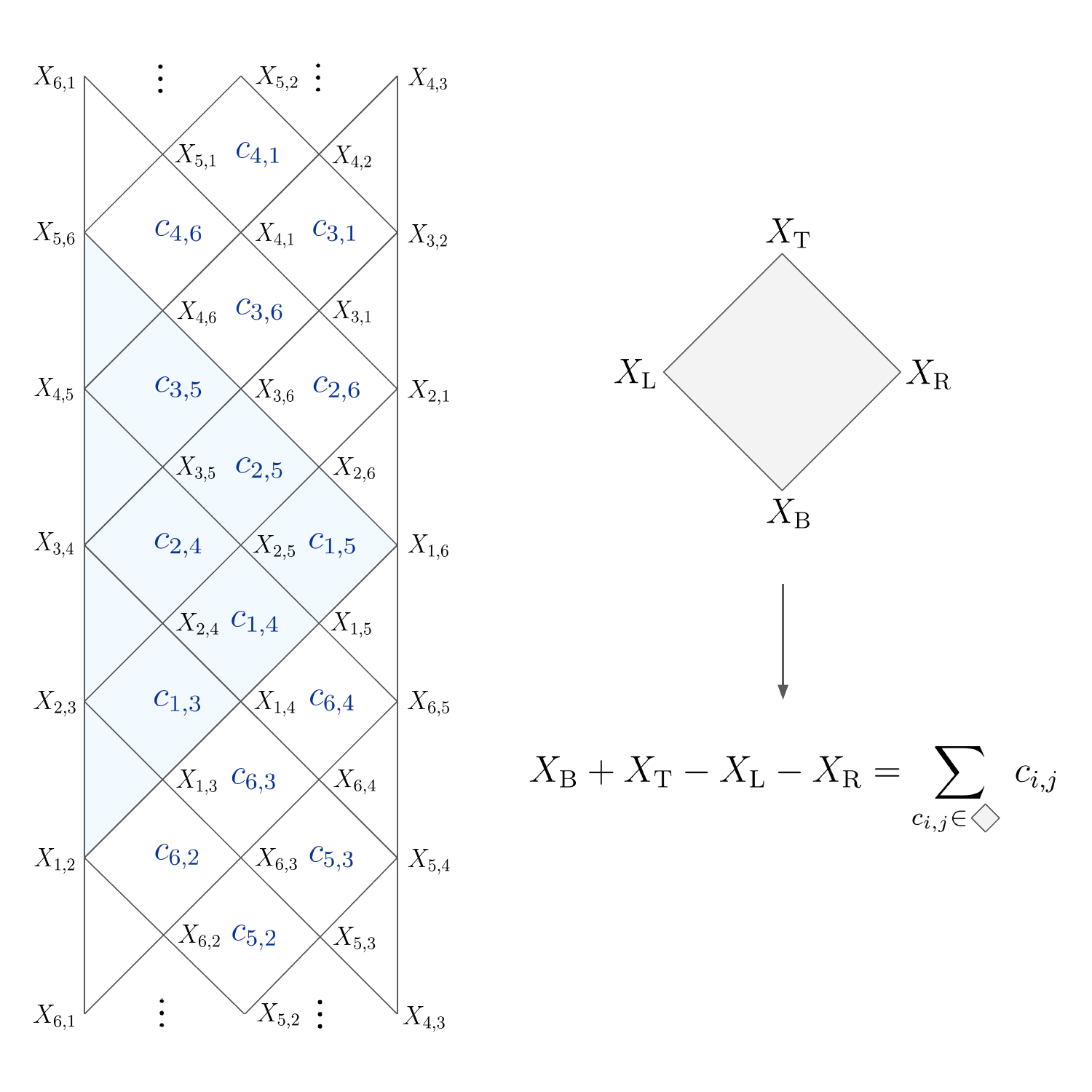}
    \caption{The six-point kinematic mesh described in text, providing a way of organizing Mandelstam variables for planar amplitudes.}
    \label{fig:mesh6pt}
\end{figure}

It is particularly useful to find a way of organizing the kinematic data that encodes these relations, and to do this we use the \textit{kinematic mesh}. In Fig.~\ref{fig:mesh6pt}, we present the kinematic mesh for six-point kinematics. 

We start by organizing the planar variables in a $45^\circ$ square grid. Going from left to right along a $45^\circ$ line we keep the first index fixed and add one for each grid point we hit in the second index, e.g., $X_{1,2}$, $X_{1,3}$, ..., $X_{1,n}$.  If, instead, we go from right to left along a $45^\circ$ line, we fix the second index and increase the first one. The boundary $X$ variables correspond to the momentum squared, which we set to zero for the case of massless particles, or otherwise the respective mass squared. Finally, to move vertically from one line to the one above, we add $1$ in both indices. This produces a configuration that naturally lets us associate a non-planar variable to each square, such that Eq.~\eqref{eq:ceqn} can be automatically read off from the mesh as explained in Fig.~\ref{fig:mesh6pt}.

As mentioned previously, the set of $n(n\,{-}\,3)/2$ planar variables forms a basis of the kinematic invariants. However, we can get a different basis by trading some of the $X$ variables for some $c$ variables. One way of doing this is by considering a subregion of the mesh that contains all the different planar variables, such as the blue triangle in Fig.~\ref{fig:mesh6pt}. Given such a subregion, we have that a possible basis for the kinematics is given by the $c_{i,j}$ inside and $n-3$ $X$ variables living on the boundary of this region. For example, for the blue triangle in Fig.~\ref{fig:mesh6pt}, we have the basis $\{X_{1,3},X_{1,4},X_{1,5},c_{1,3},c_{1,4},c_{1,5},c_{2,4},c_{2,5},c_{3,5}\}$. 

A systematic way of generating all possible subregions is by considering the triangulations of the $n$-gon. For each such collection of $n-3$ chords $ \mathcal{T}$ we consider the region of the mesh given by the $c_{i,j}$ with $(i,j) \notin \mathcal{T}$. Since the mesh is infinite, this produces an infinite region, so we further need to extract a finite subregion that contains all the $X_{i,j}$ exactly once. Under this construction, the blue region in Fig.~\ref{fig:mesh6pt} is then associated with the triangulation of the hexagon with chords $\{(2,6),(3,6),(4,6)\}$. For most of the text, we care about \textit{ray-like} triangulations, which are precisely of this type, so that the effective regions of the mesh we use are always triangular regions just like the one highlighted in Fig.~\ref{fig:mesh6pt}.

\vspace{\baselineskip}

\bibliographystyle{utphys-modified}
\bibliography{stringamps}

\end{document}